\documentclass[twocolumn,showpacs,superscriptaddress,reprint,longbibliography]{revtex4-2}
\usepackage{amsmath, amsfonts, amssymb}
\usepackage{enumerate}
\usepackage{graphicx}
\usepackage{color}
\usepackage{float}
\usepackage[breaklinks]{hyperref}
\usepackage[hyphenbreaks]{breakurl}
\usepackage{paralist}
\usepackage{multirow}
\usepackage{diagbox}
\usepackage{makecell}
\usepackage{soul}
\usepackage{cancel}
\begin{document}

\title{Long-range Cooper pair splitting by chiral Majorana edge states}

\newcommand{\bogota}{Departamento de F\'{\i}sica,
	Universidad Nacional de Colombia, Bogot\'a, Colombia}
\newcommand{\bosque}{Departamento de F\'{\i}sica,
	Universidad el Bosque, Bogot\'a, Colombia}
\newcommand{\uam}{Departamento de F\'{\i}sica Te\'orica de la Materia Condensada, Condensed Matter Physics Center (IFIMAC) and Instituto Nicol\'as Cabrera, Universidad Aut\'onoma de Madrid, Spain}
\newcommand{\order}{Preliminary author position}

\newcommand{\Tunja}{Escuela de F\'{\i}sica,
	Universidad Pedagógica y Tecnológica de Colombia (UPTC), Tunja, Colombia}

\author{Oscar Casas-Barrera}
\affiliation{\Tunja}
\affiliation{\bogota}

\author{Shirley G\'omez P\'aez}
\affiliation{\bogota}

\author{William J. Herrera}
\affiliation{\bogota}

\date{\today }

\begin{abstract}
We analyze the transport properties of a Cooper pair splitter device composed of two-point electrodes in contact with a ferromagnetic/superconductor (F/S) junction constructed on the surface of a topological insulator (TI). For the pair potential in the S region, we consider $s$- and $d$-wave symmetries, while for the F region, we focus on a magnetization vector normal to the TI surface. Non-local transport along the F/S interface is mediated by chiral Majorana edge states, with chirality controlled by the polarization of the magnetization vector. We demonstrate that crossed Andreev reflections slowly decays with the separation of the electrodes in standard clean samples. Our system exhibits a maximum Cooper pair-splitting efficiency of 80\% for a symmetrical voltage configuration, even in high-temperature superconductor devices.
\end{abstract}

\maketitle

\affiliation{\Tunja}

\affiliation{\bogota} \affiliation{\bosque}

\affiliation{\bogota}

\section{Introduction}

In recent years, various solid-state
systems with entangled electrons have been proposed for potential applications in quantum teleportation and computing \cite{Burkard_2007}. Among these standout Cooper pairs splitter devices, in which the electron pairs are stretched and extracted from the superconductor through non-local processes known as crossed Andreev reflections (CAR) \cite{Byers_1995,Deutscher_2000,Falci_2001}. The basic structure of such devices consists of a superconducting region (S) connected to two normal electrodes (N) separated by a distance of the order of the superconducting coherence length $\xi _{0}$ \cite
{Byers_1995,Deutscher_2000,Falci_2001,Lesovik_2001,Chtchelkatchev_2002,Bena_2002,Feinberg_2003,Melin_2003,Yamashita_2003,Stefanakis_2003,Melin_2004,Prada_2004,Beckmann_2004,Russo_2005,Brinkman_2006,Cadden_2006,Yeyati_2007,Beckmann_2007,Kalenkov_2007,Kalenkov_2007b,Benjamin_2008,Golubev_2009,Cadden_2009,Haugen_2010,Burset_2011,Gomez_2012,Beiranvand_2017,Beconcini_2018,Golubev_2019,Wu_2020,Jakobsen_2021}. Various designs incorporate intermediate quantum dots
\cite{Recher_2001,Recher_2003,Hofstetter_2009,Herrmann_2010,Schindele_2012,Cottet_2012,Michalek_2013,Michalek_2015,Michalek_2017,Bocian_2018,Golubev_2019}, anisotropic superconductivity \cite{Byers_1995,Stefanakis_2003,Takahashi_2006,Herrera_2009,Celis_2017}, and indeed, some of these have been realized experimentally \cite{Beckmann_2004,Russo_2005,Cadden_2006,Beckmann_2007,Hofstetter_2009,Cadden_2009,Herrmann_2010,Schindele_2012}. For subgap voltages, CARs compete with local Andreev reflections (AR) and elastic co-tunneling (EC) between electrodes.
Then, the pair-splitting efficiency is highly dependent on suppressing these secondary processes by appropriately setting the system parameters. While some systems could achieve an efficiency of 100\% under ideal conditions, impurities or defects present in samples can lead to quantum noise and decoherence processes \cite{Zhang_2015,Hou_2016}.

Topological superconductors (TS) offer a potential solution to this problem by presenting surface Andreev bound states (SABS) topologically protected against perturbations preserving the discrete symmetries of the system \cite%
{Volovik,Volovik_1997,Kane_2010,Zhang_2011,Kallin_2014,Schnyder_2015,Sato_2017,Oreg_2018}. Zero-energy SABS are Majorana modes, quasiparticles that are their own antiparticle and exhibit non-trivial statistics. This characteristic makes them promising candidates for implementation in topological quantum computing devices \cite{Wilczek_2009,Alicea_2011,Pachos,Elliott,Beenakker_2013,Sarma_2015}. Additionally, placing a conventional superconductor in contact with the surface of a topological insulator(TI) could give rise to an artificial chiral $p$-wave TS phase accompanied by chiral Majorana edge states (CMES) at the interface with a magnetic domain or ferromagnetic region (F) \cite{Kane_2008,Tanaka_2009,Stanescu_2010,Nagaosa_2010,Zareapour2012,Koren_2013,Xu_2015,Sun_2016_TI,Sun_2017,Dai_2017,Casas_2019_2,Casas_2020,science_2021,Kiphart2021,Bai2022,Li2023}.

In the last decade, several Cooper pairs splitter devices on the surface of a TI have been proposed. Most of these are F/S/F planar junctions where exchange fields in the F regions influence the efficiency of CAR processes, their doping levels, interfaces transparency, or the pair potential symmetry ($s$ or $d$) 
\cite{Niu_2010,Vali_2014,Vali_2015,Islam_2017,Zhang_K_2018}. However, transport across interfaces in this configuration results in oscillating CAR conductance that decays rapidly over distances on the order of $\xi _{0}$ due to the mediation of evanescent states. Nevertheless, numerical calculations have revealed that the presence of a TS phase with chiral $p$-wave symmetry, as theorized for Sr$_{2}$RuO$_{4}$, may lead to non-local, unidirectional transport that is independent of the electrode spacing \cite{Ikegaya_2019}. 

In this work, we investigate the spatial dependence and Cooper pair-splitting efficiency of CMES-mediated CAR processes at the interface of an F/S planar junction on the surface of a TI. In the F region, the TI is in contact with a ferromagnetic insulator with a magnetization vector normal to the surface. In the S region, the TI is proximitized by an intrinsic s- or d-wave superconductor.  For $s$ symmetry, we observe that the CMES lead to non-oscillating, long-range unidirectional CAR transport between the electrodes. This transport has a decay length of the order of $10^{2}\xi _{0}$ for samples with a typical degree of impurities, similar to that found in \cite{Ikegaya_2019}. However, in our system, the chirality can be controlled by the magnetization polarization of the F region, and the superconductivity is more robust against impurities compared to other potential chiral TS like Sr$_{2}$RuO$_{4}$ \cite{Mackenzie_1998,Anwar_2021}. In a Cooper pairs splitter configuration, we suppress EC processes by applying symmetrical bias voltages to the electrodes, thereby achieving a maximum splitting efficiency of $80\%$. An analogous behavior is expected for $d_{x^{2}-y^{2}}$ symmetry; however in this case, the samples must be exceptionally clean to avoid altering the superconducting state \cite{Balatsky_2006}.

This article is structured as follows: Section II discusses the model of the system and contains a derivation of each region's Hamiltonian and transport observables. Section III examines a toy model with infinite magnetization, analyzes the dispersion relations of CMES, and derives expressions for the conductance and the splitting efficiency as a function of the electrodes separation. Section IV presents the numerical calculations of observables for a system with finite magnetization and analyzes the dependence of CAR conductance on both the electrodes separation and the magnetization value. Finally, section V presents the conclusions of this work.

\section{Theoretical framework and transport observables}

\begin{figure}[!]
\centering
\includegraphics[width=1\columnwidth]{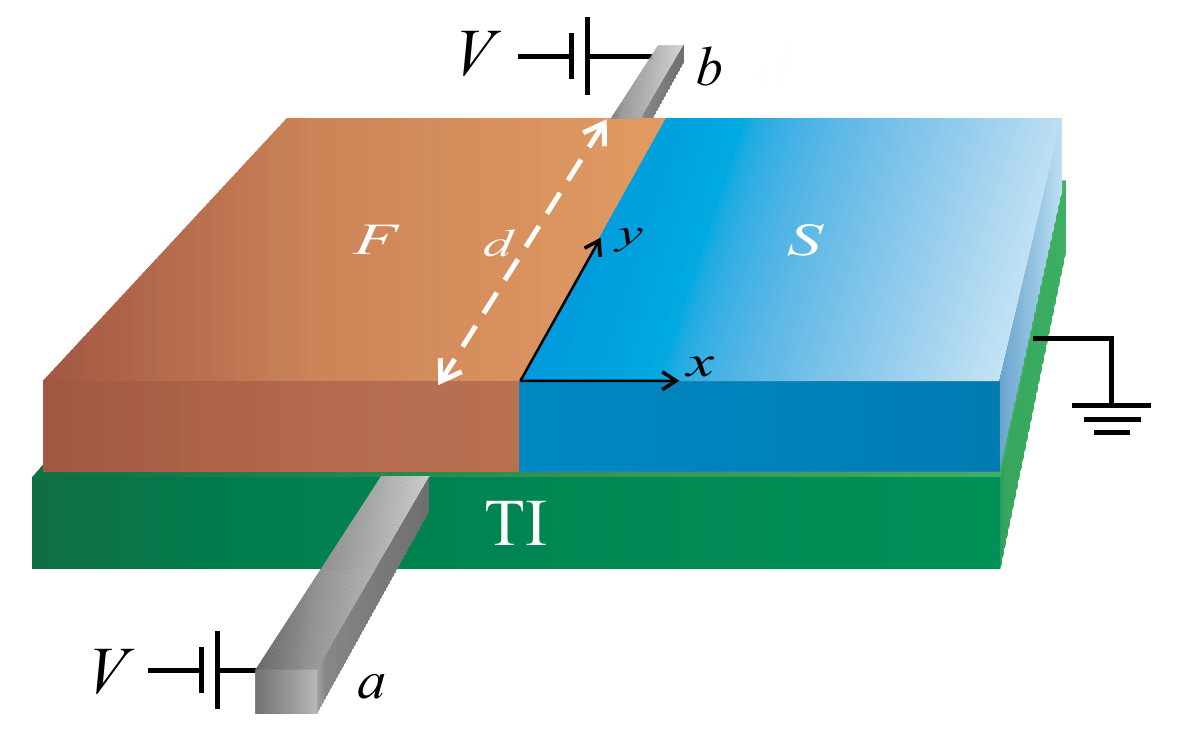}
\caption{Cooper pairs splitter device consisting of an F/S junction on the surface of a TI and two metallic electrodes separated by a distance $d$. The F/S interface is formed through the proximity effect on the TI. CAR transport between electrodes is primarily mediated by the CMES at the F/S interface.}
\label{fig:sistema}
\end{figure}

The Cooper pairs splitter device
analyzed in this work is illustrated in Fig. \ref{fig:sistema}. The system consists of an F/S planar junction constructed on the surface of a TI, in contact with two thin metallic electrodes. The F/S junction is formed through the proximity effect on the TI surface as follows: the left region is brought into contact with a ferromagnetic insulator (F), while the right region with a spin-singlet intrinsic superconductor (S), which can be either conventional ($s$-wave) or high-temperature ($d$-wave). The two thin metal electrodes, denoted by $a$ and $b$, are subjected to the same voltage bias $V$ and are separated by a certain distance $d$ on the TI surface within region F. 

The elementary excitations of the superconducting state induced in the S region are described by the Bogoliubov-de Gennes (BdG) Hamiltonian \cite{DeGennes,Kane_2008,Kane_2010,Stanescu_2010,Zhang_2011,Zhu2016}
\begin{equation}
\hat{H}_{BdG} =\left( 
\begin{array}{cc}
\hat{H}_{s} & \hat{\Delta} \\ 
\hat{\Delta}^{\dag } & -\hat{H}_{s}^{\dagger}
\end{array}\right) \text{,}  \label{HBdG}
\end{equation}
where $\hat{H}_{s} =v_{F}\left( \hat{\boldsymbol{%
\sigma }}\times \mathbf{\hat{p}}\right) _{z}-E_{F}\hat{\sigma_0}$ is the effective Hamiltonian of the TI surface around the $\Gamma $ point. Here, $v_{F}$ represents the speed of the charge carriers at the Fermi level of the normal TI, $%
\mathbf{\hat{p}}=-i\mathbf{\hbar \boldsymbol{\nabla _{\mathbf{r}}}}$ is the momentum operator, $\hat{\boldsymbol{\sigma }}=(\hat{\sigma}_{x},\hat{\sigma }_{y},\hat{\sigma}_{z}
)$ is the vector of Pauli matrices in the spin subspace, and $\hat{\sigma}_0$ is the corresponding identity matrix. $E_{F}$ denotes the Fermi level of the system \cite%
{Zhang_2009,Silvestrov_2012}. The induced superconducting order parameter is given by $\hat{\Delta}=\Delta _{0}$\textrm{cos}$\left[\beta \left( 2\theta
+\alpha \right)\right] i\hat{\sigma}_y$, where $\theta $ is the polar angular coordinate ($\beta=0$ for $s$-wave symmetry, $\beta=1$ for
$d$-wave symmetry, $\alpha =0$ for $d_{x^{2}-y^{2}}$ and $\alpha =\pi /2$ for $d_{xy}$). In the weak coupling limit ($E_F\gg \Delta _{0}$), the system's properties are predominantly determined by the states near the Fermi surface ($\hat{\psi} _{e}\simeq\left( 1,-i\mathrm{e}^{i\theta },0,0\right) ^{T}/\sqrt{2}$
and $\hat{\psi}_{h}\simeq\left( 0,0,-1,i\mathrm{e}^{-i\theta }\right) ^{T}/\sqrt{2}$). When projecting $\hat{H}_{BdG}$ onto this basis of states, we obtain the following effective spinless anisotropic order parameter with mixed symmetry
\begin{equation}
\Delta_{eff}=\left\langle \hat{\psi} _{e}\left\vert \hat{H}_{BdG} \right\vert \hat{\psi} _{h}\right\rangle =i\Delta_{0}\cos\left[\beta\left( 2\theta
+\alpha \right)\right]\mathrm{e}
^{-i\theta }\text{.}  \label{delta_efectivo}
\end{equation}

To describe the normal left region F with an induced exchange field perpendicular to the surface, $\mathbf{M}=M\mathbf{\hat{z}}$, we set $\hat{\Delta}=$$\hat{0}$ in (\ref{HBdG}) and introduce the corresponding Zeeman-type term $\hat{H}_{Z}=M\hat{\sigma} _{z}$ to the $\hat{H}_{s}$ Hamiltonian. This term breaks the time-reversal symmetry of (\ref{HBdG}) at the F/S interface, inducing topologically protected CMES \cite{Kane_2008,Tanaka_2009,Nagaosa_2010}. It will be assumed that $E_F=0$ for region F to ensure that transport occurs only on the TI surface \cite{Nagaosa_2010}.

The transport properties of this kind of 2D junction can be described using a Hamiltonian approach \cite{Yeyati_1996,Cuevas}, where adjacent regions of the system are coupled through a tight-binding Hamiltonian (refer to appendix \ref{sec:app1D}). The equilibrium Green's functions for each isolated region are derived analytically by the asymptotic solutions method \cite{McMillan_1968,Herrera_2010,Casas_2019_1,Casas_2020} (see appendices \ref{sec:app1A} to \ref{sec:app1C} for technical details). Then, the equilibrium Green's functions for the entire system are computed non-perturbatively in the hopping parameters by solving an algebraic Dyson equation with the equilibrium Green's functions of the regions. Finally, the system's transport properties are calculated in terms of the non-equilibrium (or Keldysh) Green's functions (NEGF), which can be expressed in relation to their equilibrium counterparts under stationary regime.

For low voltages ($eV\ll \Delta_0 $), an electron incident on an electrode can either be reflected as a hole within the same electrode (AR process), transmitted as an electron to the second electrode (EC process), or reflected as a hole at the second electrode (CAR process), which corresponds to Cooper pair-splitting under time reversal \cite{Byers_1995,Deutscher_2000,Falci_2001,Alfredo_2010,Herrmann_2010,Burset_2011,Michalek_2017}. For a configuration with symmetrical voltajes ($V_a=V_b=V$), only AR and CAR processes contribute to the current of the system (e.g. as evaluated at electrode $a$)
\begin{equation}
I_a=I_{AR,a}+I_{CAR}\text{,} \label{I}
\end{equation}
where the contribution of AR processes to the current is given by
\begin{equation}
I_{AR.a}=\frac{2e}{h}\int R_{AR,a}\left(E\right) \left(n_{aN,e}-n_{aN,h}\right) dE\text{,} \label{IAR}
\end{equation}
whith $R_{AR,a}$ the local Andreev reflection coefficient given by
\begin{equation}
R_{AR,a}\left(E\right) =\mathrm{Tr}\left( \hat{\Gamma}_{aN,e}\hat{G}_{aaS,eh}^{r}\hat{\Gamma}_{aN,h}\hat{G}_{aaS,eh}^{a}\right) \text{.}
\end{equation} 

Here, $\hat{G}_{aa'S}^{a/r}$ represents the advanced/retarded equilibrium Green's functions of the system, evaluated at the coordinates of the point contact electrodes inside $S$ region, $\hat{\Gamma}_{aN,\mu} =2\pi [\hat{t}_{a}^{\dag }\hat{\rho}_{aN}\hat{t}_{a}]_{\mu\mu}$ the level widths matrices with $\mu=(e,h)$ a Nambu index, $\hat{t}_{a}$ the hopping matrix in spin space between electrode $i$ and F/S junction, $\hat{\rho}_{aN}=- \mathrm{Im}(\hat{g}_{aN}^{r})/\pi$, the density of states matrix of electrode $a$ with the retarded/advanced equilibrium Green function $\hat{g}_{aN}^{r(a)}$, and $n_{aN,e/h}\left( E\right) =f_a\left(E\mp eV_a\right)$ represents the corresponding Fermi-Dirac occupancy functions. On the other hand, the CAR contribution to the current is given by 
\begin{equation}
I_{CAR}=\frac{2e}{h}\int T_{CAR,ab}\left(E\right)\left( n_{aN,e}-n_{bN,h}\right)dE
\text{,}  \label{ICAR}
\end{equation}
where the respective transmission coefficient $T_{CAR,ab}$ is given by
\begin{equation}
T_{CAR,ab}\left(E\right)=\mathrm{Tr}\left( \hat{\Gamma}_{aN,e}\hat{G}_{abS,eh}^{r}\hat{\Gamma}_{bN,h}\hat{G}_{baS,eh}^{a}\right) \text{,} \label{TCAR}
\end{equation}	

Despite the formal resemblance of the last expressions to the more traditional BTK formalism, they are applicable beyond the tunnel limit approximation ($[\hat{t}_a]_{\sigma\sigma'}\ll \hbar v_F$) and involve all perturbation orders with respect to the hopping parameters between regions \cite{Yeyati_1996}. In our system, there is only competition between CAR and AR processes, which is reflected in the expression for the total differential conductance across electrode $a$ ($\sigma_{a}=dI_{a}/dV$) at $T=0$,
\begin{eqnarray}
\sigma_{a}(V)&=&2\sigma_{0}(T_{CAR}\left( V\right)+T_{CAR}\left( -V\right))+2R_{AR,a}\left( V\right))\text{,}  \label{Cond}
\end{eqnarray}
where $\sigma_{0}=2e^{2}/h$ is the quantum of conductance and the Cooper pair-splitting efficiency is defined as \cite{Herrmann_2010,Burset_2011,Michalek_2017}
\begin{equation}
	\eta =4\sigma_{0}\frac{T_{CAR}\left( V\right)+T_{CAR}\left( -V\right)}{\sigma_{a}\left( V\right)+\sigma_{b}\left( V\right)}\text{.}  \label{eficiencia}
\end{equation}
For $\eta$ to be equal to $1$, the local Andreev reflections are completely suppressed. It is convenient to compare the CAR conductance (\ref{Cond}) with the electronic conductance of the system in its normal state and null magnetization for $d=0$.
\begin{equation}
\sigma_{ee,0} =\sigma_{0}T_{EC}\left(0\right) \text{.}  \label{Cond0}
\end{equation}

\section{Toy Model}

First, we consider the system depicted in Fig. \ref{fig:sistema} under conditions of high doping in the S region ($E_F\gg \Delta _{0}$, $k_{e/h} \simeq k_F$), 
high magnetization of F region ($M\gg\Delta_{0}$) and in the tunnel limit for the electrode couplings (refer to appendices \ref{sec:app1A} to \ref{sec:app1C} for further details). Under these conditions, the Green function of the system, evaluated at the point contact of the electrodes with the F/S interface, is represented by (\ref{Fourier2}) with $x_{a}=x_{b}=0$, $%
y_{b}-y_{a}=d$, where the integrand $\hat{G}_{RR}^{r,a}\left(
E,0,0,q\right) \simeq\hat{g}_{S0}^{r,a}\left( E,q\right) $ corresponds to the equilibrium Green function of the uncoupled S region (\ref{GreenZZ2})

\begin{equation}
\hat{g}_{S0}^{r,a}\left( E,q\right) =\frac{-1}{\hbar v_{F}}\left( 
\begin{array}{cccc}
K & 0 & N\mathrm{e}^{-i\phi _{+}} & 0 \\ 
1 & 0 & 0 & 0 \\ 
N\mathrm{e}^{i\phi _{-}} & 0 & K & 0 \\ 
0 & 0 & -1 & 0%
\end{array}%
\right) \text{,} \label{GS0}
\end{equation}

with $q\simeq k_F \sin\theta$ the conserved wave vector along the interface,%
\begin{eqnarray}
K &=&\frac{i}{D}\left( 1-\gamma_0 ^{2}\mathrm{e}^{-i\Delta \varphi }\right) 
\text{, \ \ \ }N=-\frac{\gamma_0 }{D}\left( \mathrm{e}^{i\theta }+\mathrm{e}%
^{-i\theta }\right) \text{,}  \notag \\
D &=&\mathrm{e}^{i\theta }+\gamma_0 ^{2}\mathrm{e}^{-i\theta }\mathrm{e}%
^{-i\Delta \varphi }\text{,} \\
\gamma_0 &=&\sqrt{\frac{E-\Omega }{E+\Omega }}\text{, \ \ \ }\Omega =\sqrt{%
E^{2}- \Delta \left( \theta \right)  ^{2}}\text{.}
\end{eqnarray}

\begin{figure}[!]
	\centering
	\includegraphics[width=1.0\columnwidth]{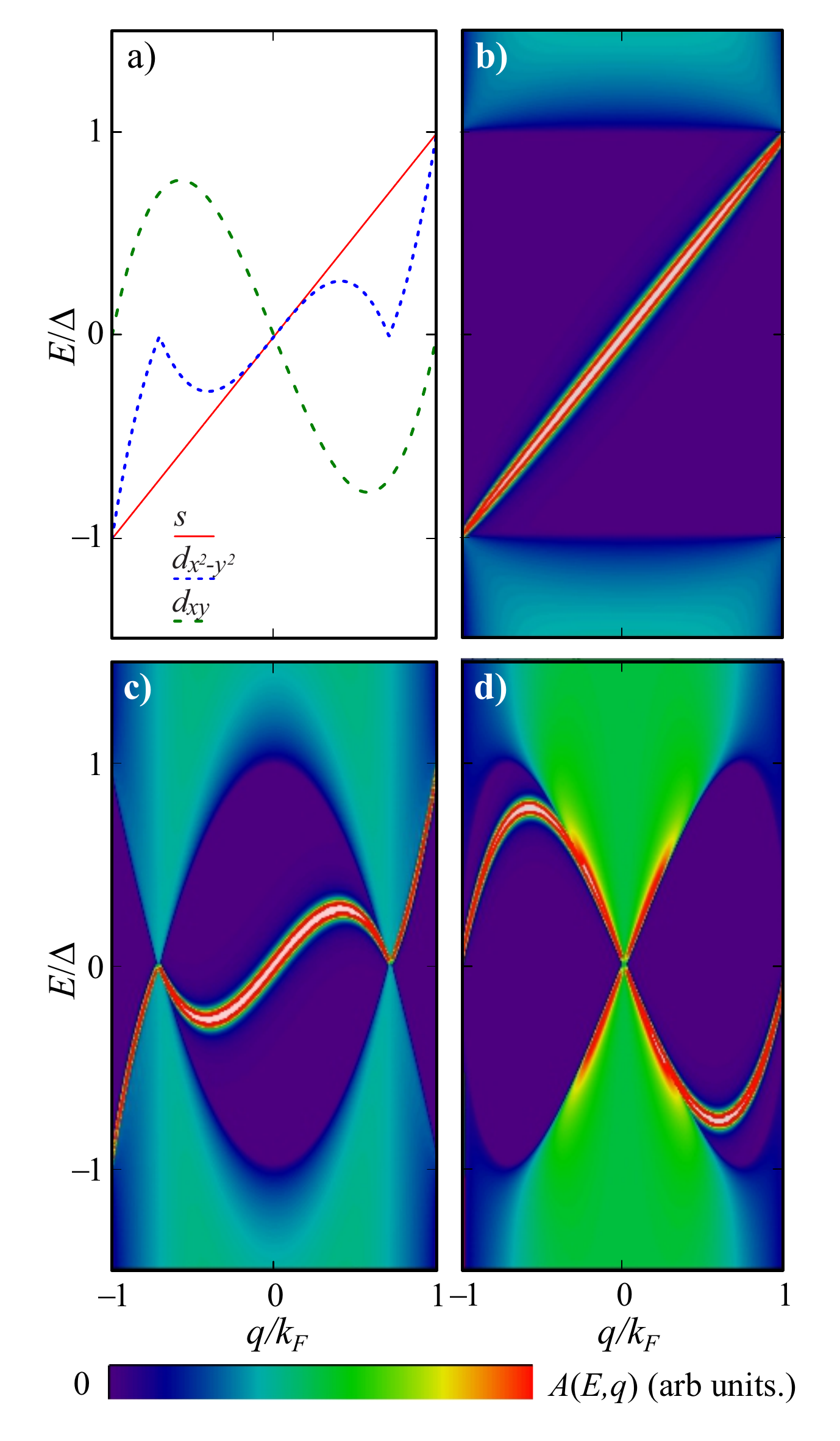}
	\caption{a) Surface Andreev bound states at the F/S interface with infinite magnetization obtained from expression (\ref{SABS}) ($M$ $\rightarrow -\infty $). The opposite polarization of $M$ changes the functions parity. Panels (b) - (d) spectral density $A$ evaluated at the F/S junction interface for a superconductor with b) $s$-wave, c) $%
	d_{x^{2}-y^{2}}$, and d) $d_{xy}$ symmetry. In all three cases, the presence of chiral Majorana edge states is observed.}
	\label{fig:estados}
\end{figure}

For the case of a strong magnetic barrier, the dispersion relation of the SABS is obtained from the poles of the Green function (\ref{GS0}) and can be expressed as
\begin{eqnarray}
E &=&\pm \left\vert \Delta _{eff}\right\vert \text{\textrm{cos}}\left(
\Delta \varphi /2\right) \text{,}  \label{SABS} \\
\Delta _{+} &=&\Delta _{eff}\left( \theta \right) \text{, \ \ \ \ }\Delta
_{-}=\Delta _{eff}\left( \pi -\theta \right) \text{,} \\
\Delta \varphi &=&\varphi _{+}-\varphi _{-}\text{, \ \ \ \ }\varphi _{\pm }=%
\text{\textrm{arg}}\left( \Delta _{\pm }\right) \text{,}  \notag
\end{eqnarray}
with $\Delta_{eff}$ is the effective surface gap given by (\ref{delta_efectivo}). It is important to note that our model aligns with the general expression for SABS \cite{Lofwander_2001} and is consistent with those found for planar F/S junctions in TIs with finite magnetizations \cite{Kane_2008,Tanaka_2009,Nagaosa_2010}. In Fig. \ref{fig:estados} a), CMES dispersion curves are shown for $s$-wave, $d_{x^{2}-y^{2}}$ and $%
d_{xy}$ symmetries. These states are also depicted in panels (b)-(c) of Fig. \ref{fig:estados}, where the spectral density $A(q,E)=\textrm{Tr}[\hat{\rho}_{F/S}(q,E)]$ evaluated at the interface of an F/S junction with infinite magnetization is shown for different symmetries of the pair potential.

\begin{figure}[!]
	\centering
	\includegraphics[width=0.9\columnwidth]{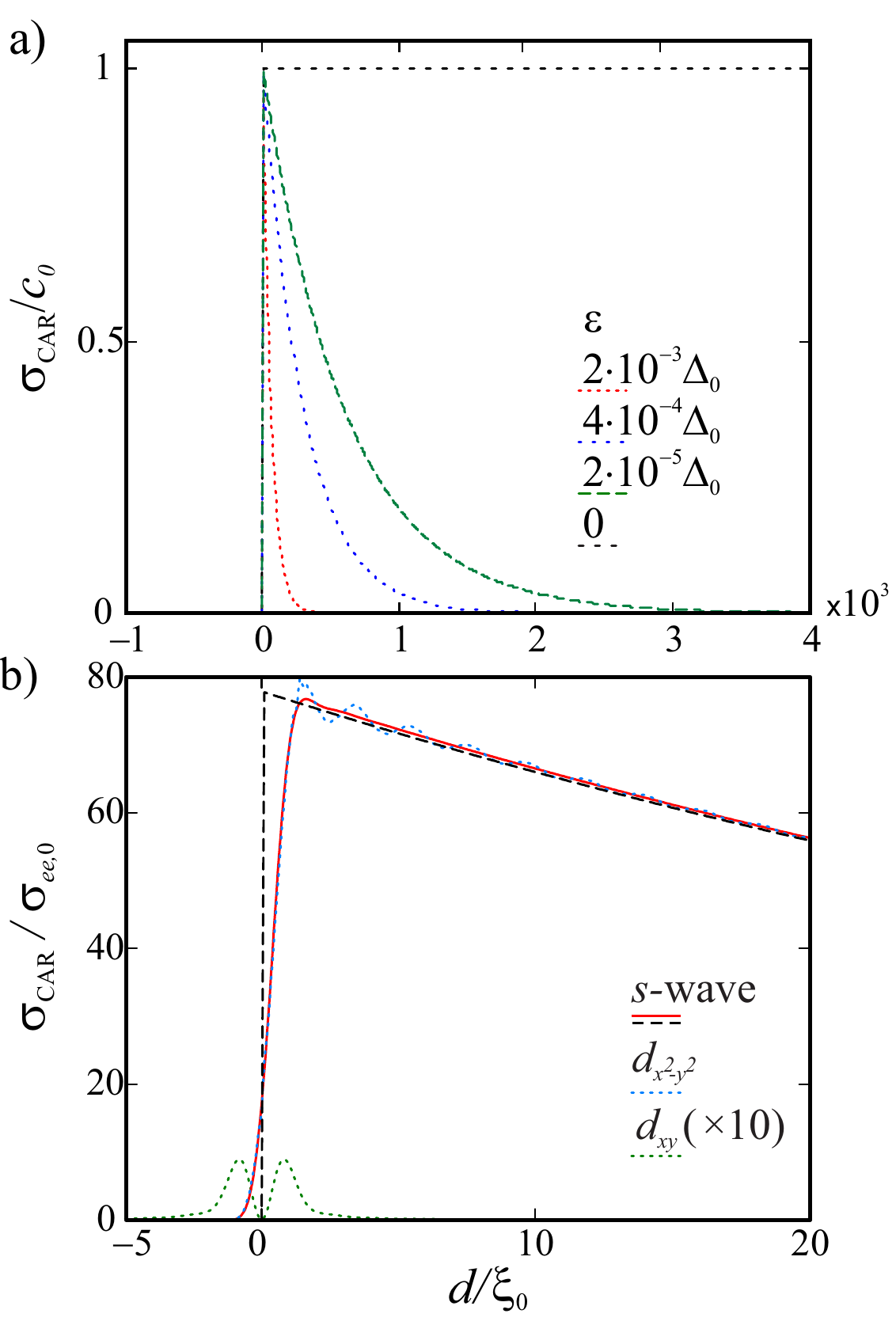}
	\caption{a) Crossed Andreev reflection conductance $\sigma_{CAR}\left(V=0\right) $ along the F/S interface as a function of distance $d$ for $s$-wave symmetry with infinite magnetization (\ref{TCARS}), $E_F=13 \Delta _{0}$ and different values of $\protect\varepsilon$ in the tunnel limit. All curves in panel a) are normalized to $c_0$ (\ref{C}). b) Curves of $\sigma_{CAR}(V=0)$ for the different symmetries with infinite magnetization, $E_F=13 \Delta _{0}$, and $\protect\varepsilon =0.002$ $\Delta _{0}$. The dashed black line represents the fit with the analytic function (\ref{TCARS}) for $s$-wave symmetry. All curves in panel a) are normalized to $\sigma_{ee,0}$ (\ref{Cond0}).}
	\label{fig:Tcar}
\end{figure}

	The dispersion relation is linear in $q$ for $s$-wave symmetry, while undulated for $d$-wave symmetries. Note that the density of states for both $s$-wave and $d_{x^{2}-y^{2}}$ symmetries presents a high value and maintains the same chirality near $q=0$, while for $d_{xy}$ becomes symmetrical and diffuse. In all instances, the dispersion relations are odd functions of $q$, and the polarization of magnetization in the F region defines their chirality. The following section shows that these states account for the solid chiral character of non-local transport processes along the interface.

For $s$-wave symmetry and $V=0$, the calculation of (\ref{TCAR}) can be performed analytically, resulting in
\begin{eqnarray}
\sigma_{CAR}\left( d\right) &=&c_0\left[ \Theta \left( sd\right)
\left( 1+\frac{\varepsilon }{\Delta _{0}}\right) \right] ^{2}\mathrm{e}^{-d/\lambda }\text{,} \label{TCARS}\\
c_0 &=&\sigma_0\tilde{\Gamma}_{e,b}\tilde{\Gamma}_{h,a}\left( 2\pi iL/\hbar v_{F}\right)
^{2}\left( k_{F}/\pi \right) ^{6}\text{,} \label{C}
\end{eqnarray}%
where $\Theta (x)$ is the Heaviside step function, $s=-%
\mathrm{sgn}(M)$, $\tilde{\Gamma}_{\mu,iN}\equiv\text{Tr}[{\hat{\Gamma}}_{\mu,iN}]$, $\varepsilon$ is the imaginary part of the excitation energy, and $\lambda =\Delta _{0}/2\varepsilon k_{F}=\xi _{0}(\hbar
v_{F}/2\pi \varepsilon k_{F})$ represents the characteristic decay length of CAR processes along the interface. Here $\xi
_{0}=\pi \Delta _{0}/\hbar v_{F}$ is the coherence length of the conventional superconductor. The curves in Fig. \ref{fig:Tcar} (a) show the behavior of $\sigma_{CAR}\left( d\right) $ for different values of $\varepsilon $. 

\begin{figure*}[!]
	\centering
	\includegraphics[width=0.8\textwidth]{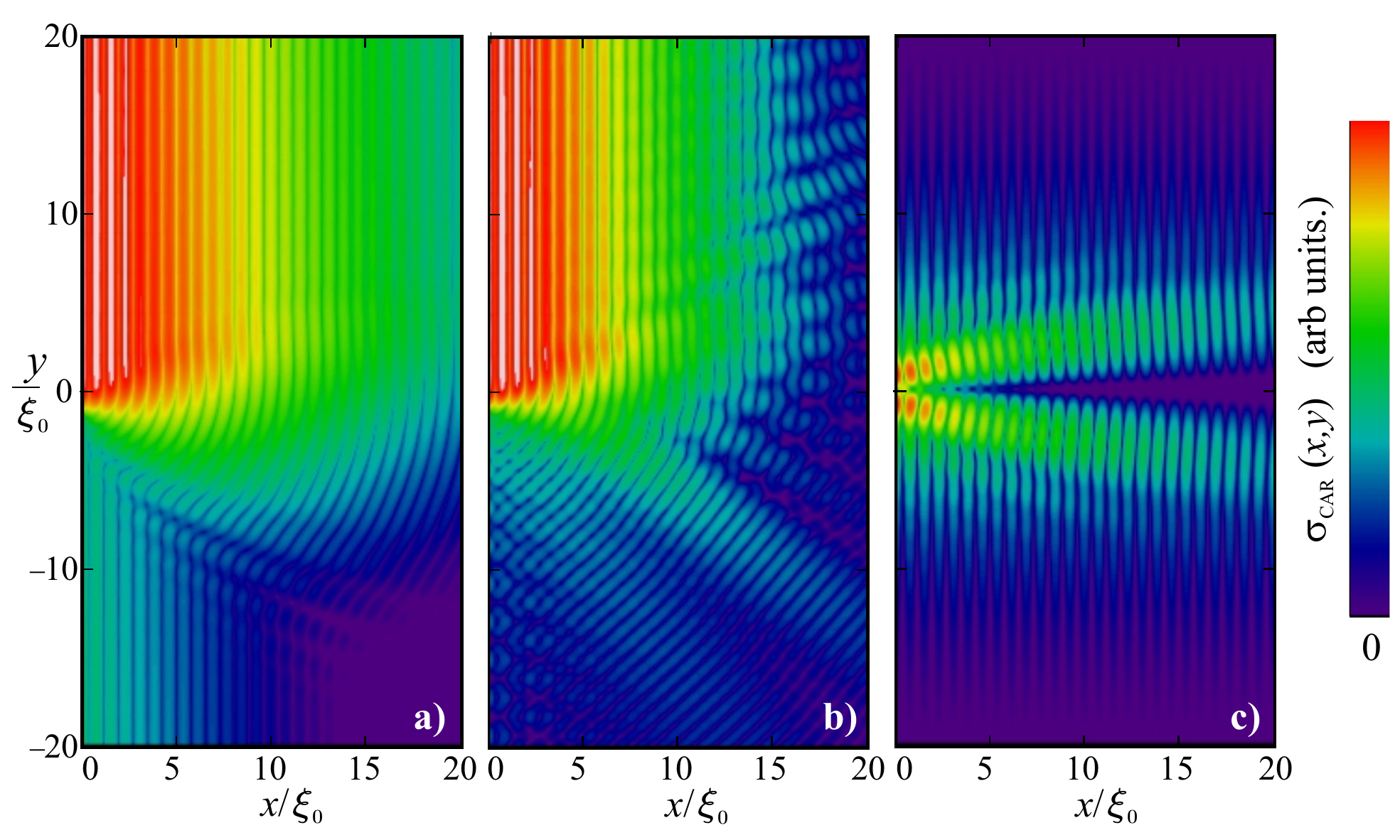}
	\caption{Maps of crossed Andreev reflection conductance $\sigma_{CAR}(V=0)$ in the tunnel limit as a function of the $b$ electrode coordinates over the S region with the $a$ electrode fixed at the F/S interface ($\mathbf{r}_a=(0,0)$, $\mathbf{r}_b=(x,y)$), for infinite magnetization of the F region and a) $s$-wave, b) $d_{x^{2}-y^{2}}$ c) $d_{xy}$ symmetry for the S region. A logarithmic scale is used for $\sigma_{CAR}$.}
	\label{fig:FIG 5}
\end{figure*}

Note that the electron-hole component of the Green function exhibits unidirectional behavior (odd with distance) and approaches the step function in the limit $\varepsilon \rightarrow 0$. Since $\varepsilon \sim \hbar /2\Delta\tau$ with $\Delta\tau$ representing the quasiparticle lifetime, it is expected that for weak-disorder junctions, these processes have a long-range (decay length on the order of $10^{2}\xi _{0}$), similar to that observed in a 2D intrinsic $p$-wave chiral superconductor (appendix \ref{sec:app1E}) or a 3D one \cite{Ikegaya_2019}. In this case, the propagating CMES present along the F/S interface favorably mediates the CAR processes, and confers their chirality and topological protection. It is essential to highlight that this distance-dependent behavior contrasts with that obtained for conventional superconductors with $s$-wave ($\sigma_{CAR}\left( d\right) \propto e^{-2d/\pi \xi _{0}}/d^{3}$) or $d$-wave ($\sigma_{CAR}\left( d\right) \propto 1/d^{2}$) superconductivity \cite{Herrera_2009}.

Although the conductance increase with the Fermi level of the S region, the direct AR conductance is approximately a quarter of that of the CAR. According to formula (\ref{eficiencia}), the splitting efficiency percentage would reach a maximum of $80\%$ for symmetrical voltages ($\Theta (0)=1/2$). In Fig. \ref{fig:Tcar} (b), the behavior of $\sigma_{CAR}(V=0)$ for infinite magnetization across different symmetries of the pair potential is observed. For $d_{x^{2}-y^{2}}$ symmetry case, the results are qualitatively similar to those of the $s$-wave symmetry due to the mediation of CMES around $q=0$ [Fig. \ref{fig:estados} (c)], and because $\Delta \varphi =0$ for most of the integration interval (except in the neighborhood of $\theta =\pm \pi /4$  nodes). However, there are slight additional oscillations due to the angular dependence of $\Delta
\left( \theta \right) $. Both symmetries, $s$-wave and $d_{x^{2}-y^{2}}$
are closely align with the analytical result (\ref{TCARS}). 

For the $d_{xy}$ case, the overall amplitude is lower than in the previous cases. Here, $\sigma_{CAR}(V=0)$ is an even function of $d$ with two small maxima flanking $d=0$. This behavior is attributed to the presence of two low-density counter-propagating CMES around $q=0$ [Fig. \ref{fig:estados}(d)]. Mathematically because $\Delta \varphi =\pi $ for the integration interval, resulting in real roots for $D$ and making analytical integration impossible. Furthermore, numerical computation shows an even behavior with distance, decreasing proportionally to $\sim 1/d^{2}$ as in the conventional $d_{xy}$ case \cite{Herrera_2009}. The mixed symmetry of the effective surface pair potential (\ref{delta_efectivo}), and the corresponding nodal dispersion relation for SABS with counter propagating CMES (\ref{SABS}, fig:FIG 2 (d)), prevents the $d_{xy}$ symmetry from presenting the characteristic ZBCP associated with the zero-energy flat SABS of the conventional $d_{xy}$ case.

All these behaviors can also be visualized in Fig.  \ref{fig:FIG 5}, where maps of $\sigma_{CAR}\left( V=0\right)$ are shown for the S region as a function of $\mathbf{r}_{b}\mathbf{=}(x,y)$, with $%
\mathbf{r}_{a}$ fixed at the F/S interface with infinite magnetization. As observed in panels a) and b) for $s$-wave and $d_{x^{2}-y^{2}}$ symmetries respectively, the strong chirality of the non-local transport arise from the presence of CMES around $\mathbf{k}=0$. In contrast, while for the $d_{xy}$ case, a symmetrical interference pattern is observed due to the low spectral density of these states at $q=0$.

\section{Cooper pair splitter device with finite magnetization in the F region}
\begin{figure}[!]
\centering
\includegraphics[width=0.9\columnwidth]{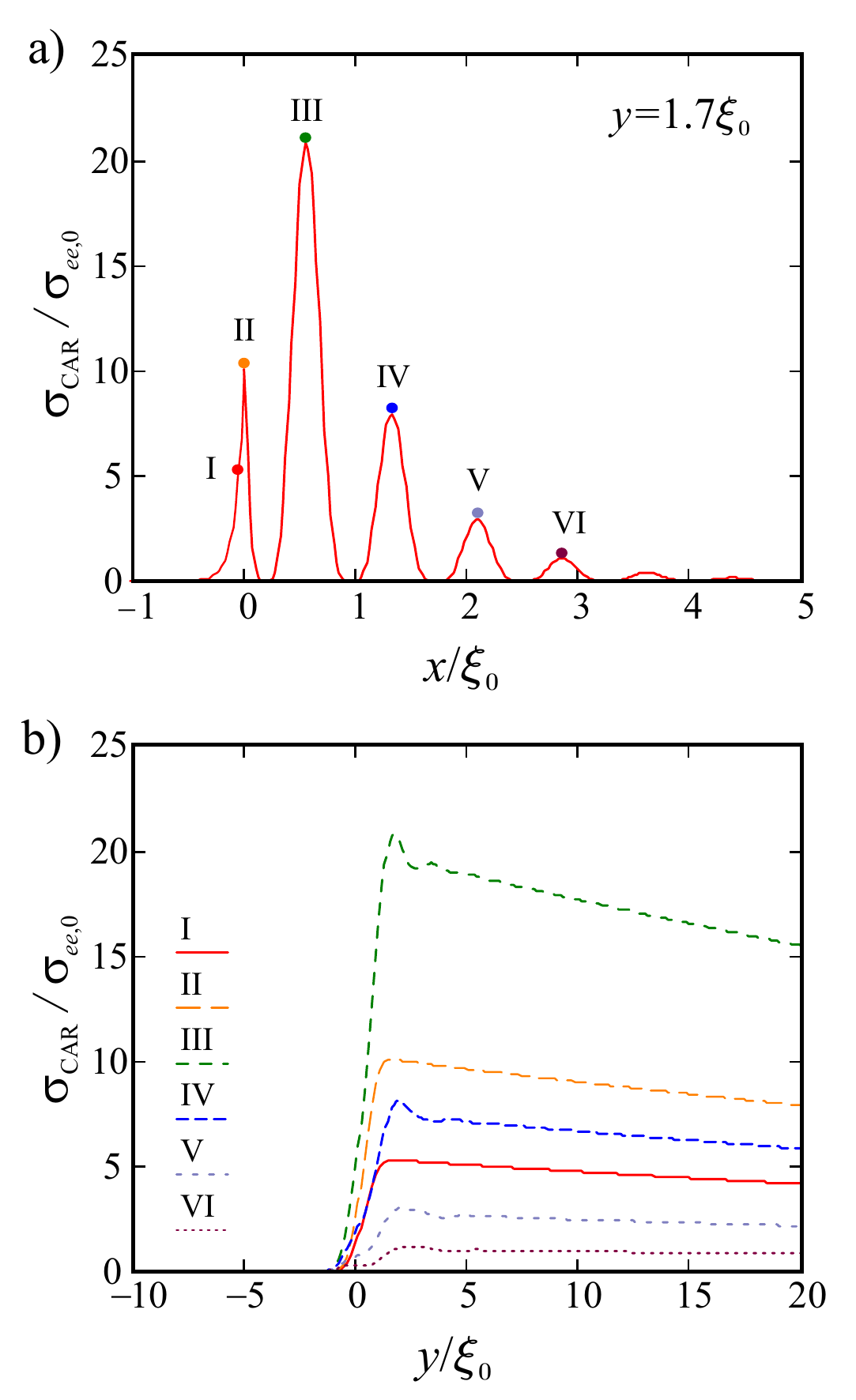}
\caption{Crossed Andreev reflection conductance $\sigma_{CAR}(V=0)$ for $s$-wave symmetry in the tunnel limit: a) as a function of the perpendicular distance to the interface F/S interface $x$ for a constant electrode separation $d=y=1.7\protect\xi_{0}$, and b) as a function of $y$ for the points labeled in panel a). In both figures, $M=-10$ $\Delta_{0}$. All curves are normalized as in Fig. \ref{fig:Tcar} b).}
\label{fig:FIG 6}
\end{figure}

In this section, we examine the system depicted in Fig. \ref{fig:sistema} with finite magnetization in region F. As shown in panel a) of Fig. \ref{fig:FIG 6}, $\sigma_{CAR}$ exhibits an oscillatory decay towards the interior of the region S over distances approximately of $5\xi_0$. In contrast, it decays smoothly towards region F ($x<0$), with a characteristic length that decreases with the value of $\left\vert M\right\vert$. Across the interface, $\sigma_{CAR}$ exhibits the behavior discussed above for the F/S interface for $s$-wave, as illustrated in panel b) corresponding to the maxima of panel a) in Fig. \ref{fig:FIG 6}. The results for the $s$-wave symmetry, show that the propagating CMES facilitate long-range CAR transport along the F/S interface. Similarly, for the $d_{x^{2}-y^{2}}$ symmetry, analogous results were obtained despite the presence of nodes in $\Delta(\theta)$. This is because the longitudinal transport predominantly occurs along the lobes of $\Delta(\theta)$. This observation suggest the potential for achieving long-range CAR in high-temperature superconductors (HTS).
\begin{figure}[!]
	\centering
	\includegraphics[width=0.9\columnwidth]{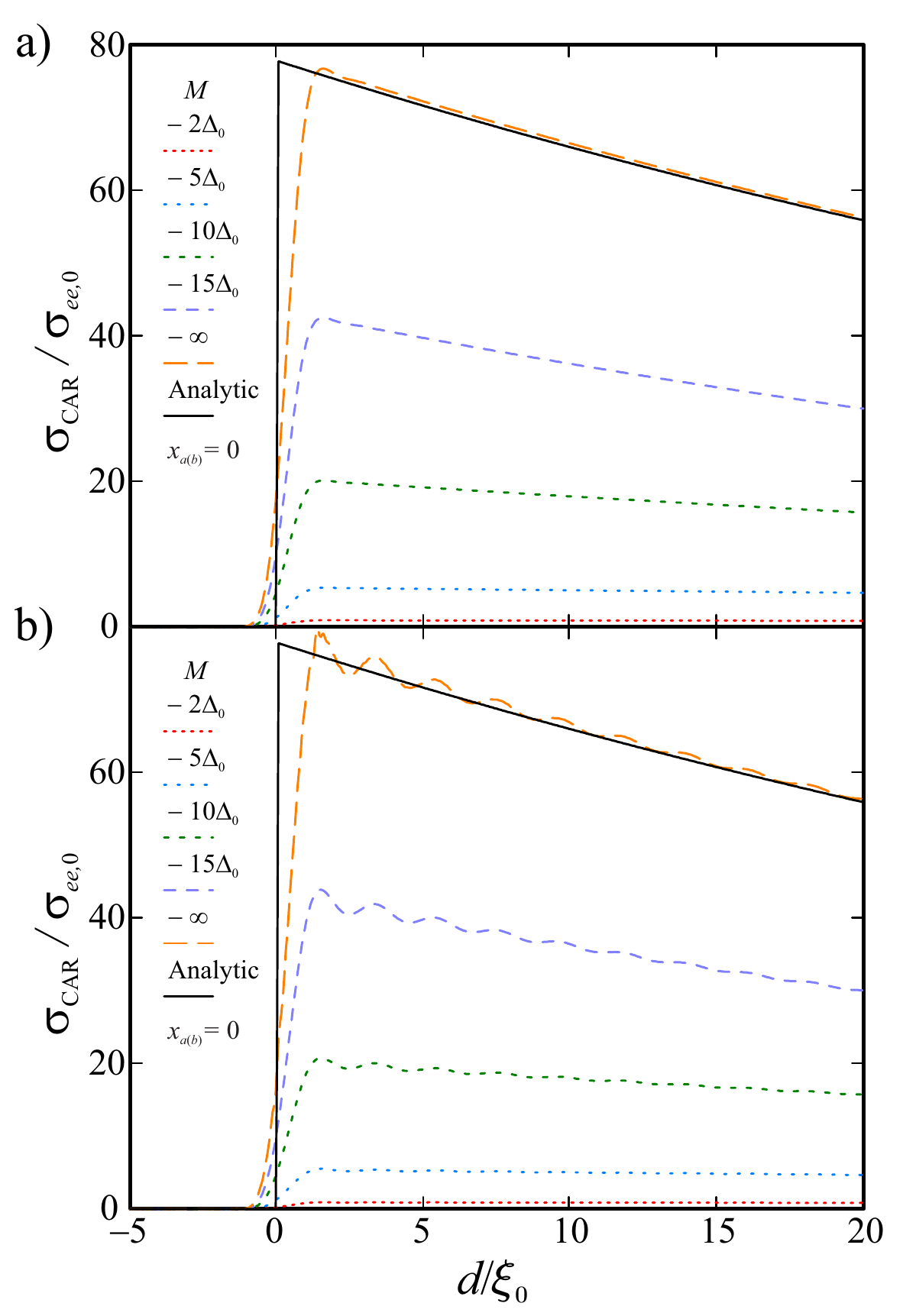}
	\caption{Crossed Andreev reflection conductance $\sigma_{CAR}(V=0)$ in the tunnel limit: a) for $s$-wave symmetry, and b) for $d_{x^{2}-y^{2}}$ and different values of magnetization ($E_{FR}=13$ $\Delta _{0}$, $\protect\varepsilon =0.002$ $\Delta _{0}$).  All curves are normalized as in Fig. \ref{fig:Tcar} b).}
	\label{fig:FIG 7}
\end{figure}
\begin{figure}[!]
	\centering
	\includegraphics[width=0.9\columnwidth]{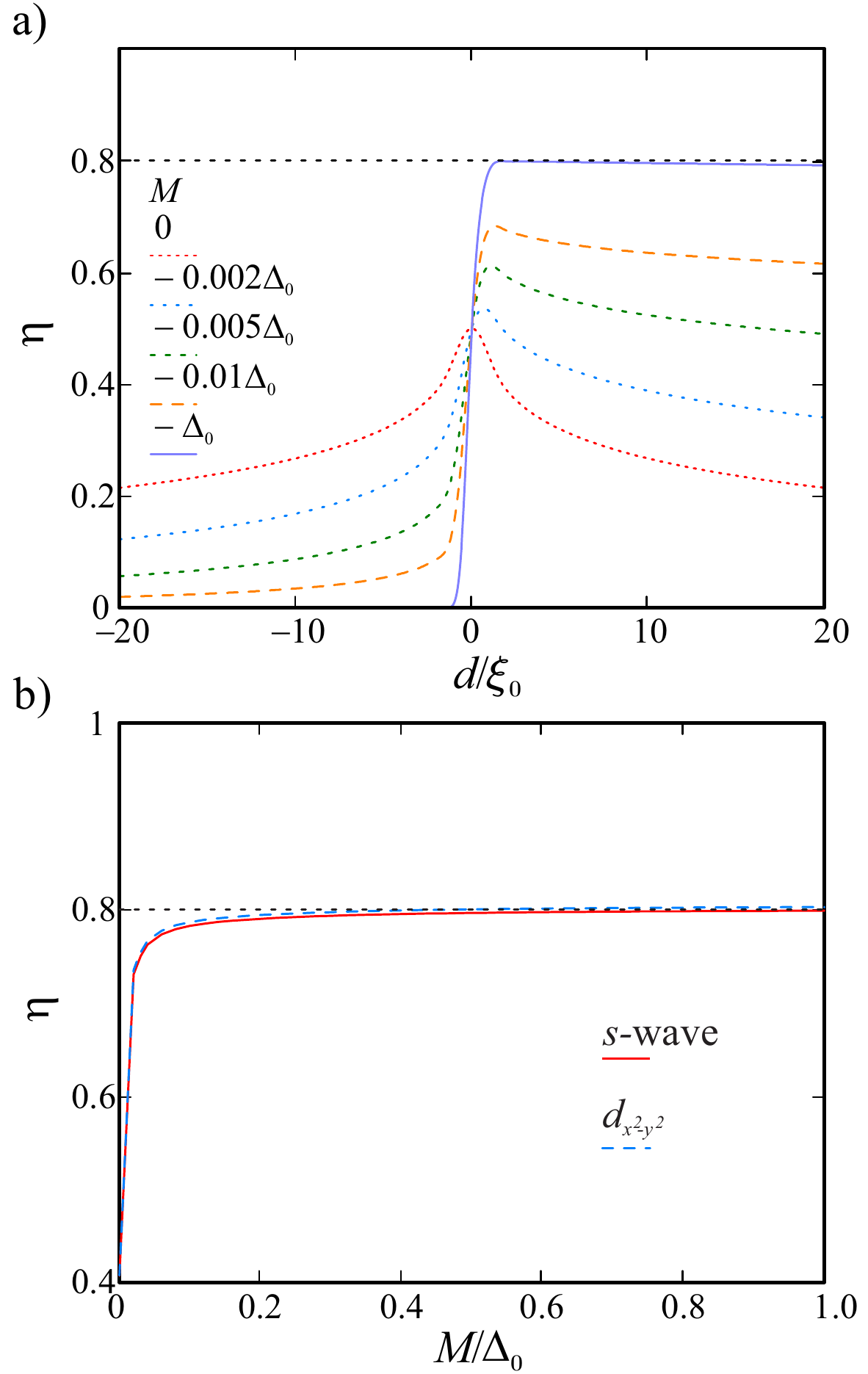}
	\caption{Cooper pair-splitting efficiency $\eta$ as a function of a) the electrode separation $d$ for $s$-wave symmetry and different values of $M$; b) as a function of magnetization for both $s$- and $d$-wave symmetries with $d=1.6$ $\protect\xi _{0}$.}
	\label{fig:FIG 8}
\end{figure}
In Fig. \ref{fig:FIG 7}, we analyze the effect of the magnitude of $\mathbf{M}$ on the CAR conductance at the interface. Panel a) displays the curves of $\sigma_{CAR}(V=0)$ for $s$-wave symmetry as a function of $d$ for various magnetization values. As seen in this figure, the maximum CAR conductance increases with the value of $\left\vert M\right\vert $, but at a decelerating rate until reaching a limit value for $\left\vert
M\right\vert =\infty$. A similar trend was observed for $d_{x^{2}-y^{2}}$ symmetry in panel b), although with the same undulations in Fig. \ref{fig:Tcar}. Another important aspect is that the decay length of $\sigma_{CAR}$ decreases with magnetization until reaching the minimum value of $\lambda =\Delta
_{0}/2\varepsilon k_{F}$ for the infinite case. 

Additionally, the Fig. \ref{fig:FIG 8} a) shows the Cooper pair-splitting efficiency as a function of the distance along the F/S interface for various $M$ values. As can be seen, the efficiency becomes increasingly chiral as the magnetization increases, at the same time as its maximum value between $0<d<\xi _{0}$ gradually increases from $0.5$ to $\sim 
$ $0.8$. Panel b) further reveals that efficiency rapidly reaches its peak for magnetization values on the order of $\Delta _{0}$. Beyond this point, it remains constant for both $s$-wave and $d_{x^{2}-y^{2}}$ symmetries.

\begin{figure}[!]
	\centering
	\includegraphics[width=0.9\columnwidth]{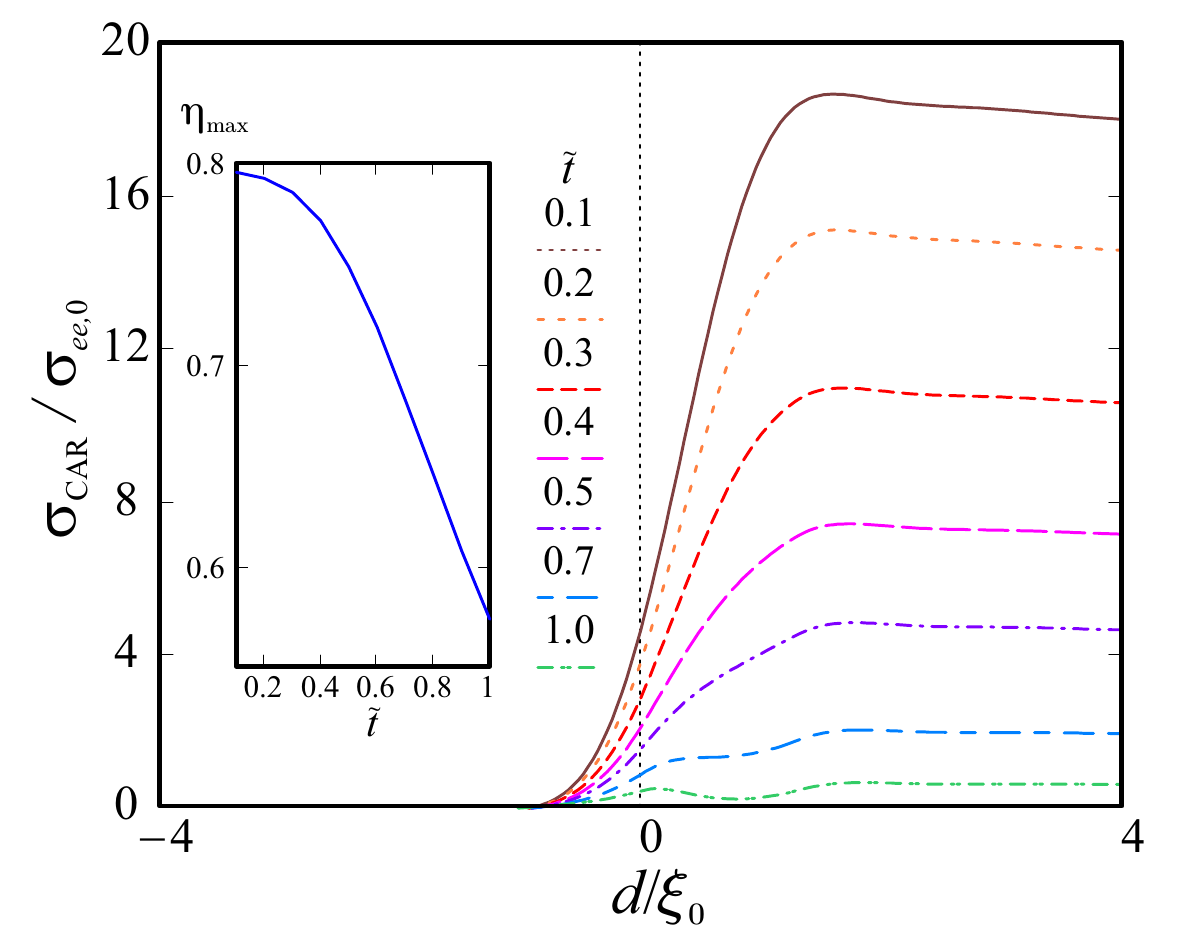}
	\caption{Crossed Andreev reflection conductance $\sigma_{CAR}(V=0)$ for $s$-wave symmetry, with $M=-10\Delta_0$ and  different values of the normalized hopping amplitude $\tilde t$ ($\protect\varepsilon =0.002$ $\Delta _{0}$). The inset shows the dependence of the maximum efficiency $\eta_{max}$ for different values of $\tilde t$. (All curves are normalized as in Fig. \ref{fig:Tcar} b).)}
		\label{fig:FIG 9}
\end{figure}

Finally, figure \ref{fig:FIG 9} shows the effect of the transparency of the coupling between the normal electrodes and the F/S junction (see appendix \ref{sec:app1B}). As can be seen in the plot, the normalized CAR conductance decreases globally and exhibits the characteristic behavior of the tunnel limit for values of the normalized hopping parameter $\tilde{t}_i$ less than $0.7$. However, while for higher values of $\tilde{t}_i$ approaching the transparent limit, the CAR conductance presents a minimum around $d\sim \protect\xi _{0}$. This is associated with the momentary increase in the local Andreev reflections at the electrode point contacts. Despite this variation, the CAR conductance still preserves its long-range chiral character and efficiency above $0.58$ for $d>1.5\protect\xi _{0}$, indicating that this behavior is independent of the contact barrier transparency.

\section{Conclusions}

We have examined the efficiency of crossed Andreev reflection processes in a Cooper pair splitter device composed of two thin metal electrodes in contact with an F/S junction formed on the surface of a TI through the proximity effect. We considered a magnetization vector normal to the TI surface for the F region and an induced $s$- and $d$-wave superconducting order parameter for the S region. The compelling spinless chiral $p$-wave symmetry at the F/S interface presents chiral Majorana edge states. The chirality of these states can be controlled by the magnetization polarization. Using a simple analytical model, we demonstrate that the conductance of crossed Andreev reflection mediated by chiral Majorana edge states along the F/S interface does not oscillate with electrode separation and exhibits slow decay in low-disorder samples (long-range behavior). Under a symmetrical voltages configuration, our system could achieve a maximum splitting efficiency of 80\% and remains stable with electrode separation for $M\sim \Delta _{0}$. Our results are also valid for the $d_{x^{2}-y^{2}}$ symmetry and could be easily extrapolated to finite temperatures, paving the way for potential experimental realization in High Tc supercondutors.

\acknowledgements

We wish to acknowledge the support of Universidad Nacional de Colombia, DIEB,
Código Hermes 57739.

\appendix

\section{Hamiltonian Approach and Calculation of the Non-local Current of the System \label{sec:app1D}}

For a superconducting system with two contacts (as in a Cooper pairs-splitter device), the Hamiltonian has the form \cite{Yeyati_1996,Cuevas,Casas_2019_1,Casas_2019} 
\begin{equation}
	\hat{H}=\hat{H}_{a}+\hat{H}_{b}+\hat{H}_{s}+\hat{H}_{as}+\hat{H}_{bs}\text{,}  \label{AHTI}
\end{equation}%
with $\hat{H}_{a,b,s}$ the Hamiltonians of normal electrodes $a$, $b$ and the superconducting region $s$, and $\hat{H}_{is}$ the hopping Hamiltonians between the point electrode $i$ and region $s$%
\begin{equation}
	\hat{H}_{is}=t_{i}\sum_{\sigma ,\sigma ^{\prime }}\mathrm{e}^{i\phi _{i}(\tau
		)/2}\hat{c}_{i\sigma }^{\dagger }\hat{b}_{i\sigma ^{\prime }}+\text{h.c.,}
	\label{eq:hop3}
\end{equation}
where $\sigma ,\sigma ^{\prime }=\uparrow ,\downarrow $ are the spin projection indices in the $z$ direction, $\phi _{i}(\tau
)=\phi _{0}+2(E_{F,i}-E_{F,s})\tau /\hbar $ are the time-dependent gauge phases induced by the gradient of chemical potential in the vicinity of the electrode $i=a,b$ with region S. Here $t_{i}$ is the hopping amplitude at this contact point, the $\hat{c}_{i\sigma }$ are the annihilation operators for the electrode $i$, and $\hat{b}_{i\sigma }$ are the annihilation operators at the point of the region S in contact with the electrode $i$. In the Heisenberg picture the average current in contact $i$ is given by
\begin{gather}
	I_{i}(\tau )=-e\left\langle \frac{d}{d\tau }\hat{N}_{i}(\tau )\right\rangle \\
	=it_{i}\frac{e}{\hbar }\sum_{\sigma ,\sigma ^{\prime }}\left[ \left\langle 
	\hat{c}_{i\sigma }^{\dag }(\tau )\hat{b}_{i\sigma ^{\prime }}(\tau
	)\right\rangle -\left\langle \hat{b}_{i\sigma ^{\prime }}^{\dagger }(\tau )%
	\hat{c}_{i\sigma }(\tau )\right\rangle \right] \text{,}  \notag
\end{gather}%
which can be expressed in terms of the Keldysh Green functions as%
\begin{gather}
	\hat{G}_{ijkl}^{\alpha \beta }\left( \tau _{\alpha },\tau _{\beta }^{\prime
	}\right) =-i\left\langle \hat{T}_{c}\left[ \hat{D}_{q,ki}\left( \tau _{\alpha
	}\right) \hat{D}_{q,lj}^{\dag }\left( \tau _{\beta }^{\prime }\right) \right]
	\right\rangle \text{,}  \\
	\hat{D}_{ki}\left( \tau \right) =\left( \hat{d}_{ki\uparrow }\left( \tau
	\right) ,\hat{d}_{ki\downarrow }\left( \tau \right) ,\hat{d}_{ki\uparrow
	}^{\dag }\left( \tau \right) ,\hat{d}_{ki\downarrow }^{\dag }\left( \tau
	\right) \right) ^{T} \text{,}
\end{gather}%
with $i,j=a,b$ representing the contact indices, $k,l=N,S$ the region indices ($\hat{d}_{ni\sigma}=\hat{c}_{i\sigma}$ , $\hat{d}_{si\sigma}=\hat{b}_{i\sigma}$), $\alpha ,\beta =+,-$ the indices of the Keldysh temporal contour and $\hat{T}_{c}$ is the Keldysh temporal ordering operator. Assuming normal electrodes on the surface of the TI, the hopping Hamiltonian \ref{eq:hop3} takes the form \ref{H_hopping_transparente}. Considering the two possible choices of the boundary conditions, the average current acquires a factor of $2$ \cite{Casas_2019,Casas_2020}. In a stationary situation, the average current in the electrode $i$ can be expressed in the energy space as
\begin{equation}
	I_{i}=\frac{e}{h}\int dE\mathrm{Tr}\left(  \hat{\tau} _{z}\left[ \hat{t}%
	_{i}\hat{G}_{iSN}^{+-}(E)-\hat{t}_{i}^{\dag }\hat{G}_{iNS}^{+-}(E)\right]
	\right) \text{,}
\end{equation}
with $\hat{\tau}_k$ representing the Pauli matrices in Nambu space, and $\hat{t}_{i}$ the self-energy matrix associated with \ref{eq:hop3}, we employ the shorthand notation for repeated indices ($ii\rightarrow i$). Using the following Dyson equations \cite{Cuevas} 
\begin{eqnarray}
	\hat{G}_{iSN}^{+-}(E) &=&\hat{G}_{iS}^{+-}(E)\hat{t}_{i}^{T}\hat{g}%
	_{iN}^{a}(E)+\hat{G}_{iS}^{r}(E)\hat{t}_{i}^{T}\hat{g}_{iN}^{+-}(E)\text{,}
	\label{Dy3} \\
	\hat{G}_{iNS}^{+-}(E) &=&\hat{g}_{iN}^{+-}(E)\hat{t}_{i}\hat{G}_{iS}^{a}(E)+%
	\hat{g}_{iN}^{r}(E)\hat{t}_{i}\hat{G}_{iS}^{+-}(E)\text{,}  \label{Dy4}
\end{eqnarray}%
the average current can be written in terms of Green functions evaluated over a single type of region as
\begin{eqnarray}
	I_{i} &=&\frac{e}{2h}\int dE\mathrm{Tr}\left( \hat{\tau}_{z}\hat{t}%
	_{i}^{\dag }\left[ \hat{g}_{iN}^{+-}\left( E\right) \hat{t}_{i}\hat{G}%
	_{iS}^{-+}\left( E\right) \right. \right. \\
	&&\left. \left. -\hat{g}_{iN}^{-+}\left( E\right) \hat{t}_{i}\hat{G}%
	_{iS}^{+-}\left( E\right) \right] \right) \text{,}  \notag
\end{eqnarray}%
where the unperturbed Keldysh Green functions of the electrodes are given by
\begin{eqnarray}
	\hat{g}_{iN}^{+-}\left( E\right) &=&2\pi i\hat{\rho}_{iN}\left( E\right) 
	\hat{n}_{iN}\left( E\right) \text{,}  \label{Kmn} \\
	\hat{g}_{iN}^{-+}\left( E\right) &=&-2\pi i\hat{\rho}_{iN}\left( E\right)
	\left( \hat{\tau}_0\otimes\hat{\sigma}_0-\hat{n}_{iN}\left( E\right) \right) \text{,}  \label{Knm}
\end{eqnarray}
with $\hat{n}_{iN}\left( E\right)=\mathrm{diag}%
(n_{i,e \uparrow }(E),n_{i,e \downarrow }(E),n_{i,h \uparrow }(E),n_{i,h \downarrow }(E))$ the occupation matrix of electrode $i$.

By considering the following Dyson equation, the non-equilibrium Green functions can be expressed in terms of the local and non-local equilibrium Green functions of the system ($\gamma =+-,-+$) \cite%
{Casas_2019,Casas_2020} 
\begin{equation}
	\hat{G}_{iS}^{\gamma }=\hat{G}_{iS}^{r}\hat{t}_{i}^{\dag }\hat{g}%
	_{iN}^{\gamma }\hat{t}_{i}\hat{G}_{iS}^{a}+\hat{G}_{ijS}^{r}\hat{t}_{i}%
	\hat{g}_{jS}^{\gamma }\hat{t}_{i}^{\dag }\hat{G}_{jiS}^{a}\text{,}
	\label{DyK3}
\end{equation}
which in turn are obtained from the equilibrium Green function of the isolated regions $\hat{g}_{ijk}^{r/a}$ by a Dyson equation of the form
\begin{equation}
	\hat{G}_{ijk}^{r/a}=\hat{g}_{ijk}^{r/a}+\hat{g}_{imk}^{r/a}\hat{\Sigma}_{mn}%
	\hat{G}_{njk}^{r/a}\text{,}\label{Dyson_equation}
\end{equation}
where the coupling self-energies between regions $\hat{\Sigma}_{mn}=\hat{\Sigma}%
	_{nm}^{T}\equiv \hat{t}$ ($m,n=L,S$) correspond to the matrix form of $\hat{H}_{is}$.
By defining the expression
	\begin{equation}
		I_{ij}=\frac{e}{h}\int dE\mathrm{Tr}\tau _{z}\bar{\rho}_{iN}\left[
		\hat{n}_{iN}\hat{G}_{ijS}^{r}\bar{\rho}_{jN}-\hat{G}_{ijS}^{r}\bar{\rho}_{jN}\hat{%
			n}_{jN}\right] \hat{G}_{ijS}^{a}\text{,}  \label{corriente} 
\end{equation}
the total average current induced in the electrode $i$ takes the form
\begin{equation}
		I_{i}=I_{ii}+I_{ij}, 
\end{equation}
with $j\neq i$. Expressing (\ref{corriente}) in terms of matrices in the Nambu subspace we obtain the expressions (\ref{IAR}), (\ref{ICAR}) and (\ref{I}) for a symmetrical voltajes configuration. For independent (or different) voltajes, $I_{ij}$ presents an EC contribution given by 
\begin{eqnarray}
		I_{EC}=\frac{2e}{h}\int T_{EC,ij}\left(E\right)\left( n_{iN,e}-n_{jN,e}\right)dE
		\text{,}  \label{IEC}  \\
		T_{EC,ij}\left(E\right) =\mathrm{Tr}\left( \hat{\Gamma}_{iN,e}\hat{G}_{ijS,ee}^{r}\hat{\Gamma}
		_{jN,e }\hat{G}_{jiS,ee}^{a}\right) \text{.}  
\end{eqnarray}

\section{Asymptotic Solutions Method for Green Functions in TIs \protect\cite{Casas_2020} \label{sec:app1A}}

For a region in a planar junction with translational invariance along the $y$ axis, the advanced ($a$) and retarded ($r$) Green functions can be expressed as
\begin{equation}
\hat{g}^{r,a}\left( E,x,x^{\prime },y-y^{\prime }\right) =\int dq\mathrm{e}%
^{iq\left( y-y^{\prime }\right) }\hat{g}^{r,a}\left( E,x,x^{\prime
},q\right) \text{,}  \label{Fourier}
\end{equation}%
where the integrand satisfies the inhomogeneous equation
\begin{equation}
\left[ \left( E\pm i\varepsilon \right) -\hat{H}\left( x,q\right) %
\right] \hat{g}^{r/a}\left( E,x,x^{\prime },q\right) =\delta \left(
x-x^{\prime }\right) \text{,}  \label{GreenTI}
\end{equation}%
with $E$ being the excitation energy of the system, $q$ the conserved wave vector along the interface, and $\varepsilon $ an infinitesimal scalar. These Green functions can be expressed in terms of the asymptotic solutions of the system as
\begin{equation}
\hat{g}(x,x^{\prime })=\left\{ 
\begin{array}{cc}
\sum\limits_{\mu ,\nu =e,h}\hat{C}_{\mu \nu }\hat{\Psi} _{<}^{\mu }\left( x\right)
\hat{\Psi} _{>}^{\nu T}\left( x^{\prime }\right) & x<x^{\prime } \\ 
\sum\limits_{\mu ,\nu =e,h}\hat{C}_{\mu \nu }^{\prime }\hat{\Psi} _{>}^{\nu
}\left( x\right) \hat{\Psi} _{<}^{\mu T}\left( x^{\prime }\right) & x>x^{\prime }%
\end{array}%
\right. \text{,}  \label{GreenZZ2}
\end{equation}%
where $\hat{C}_{\mu \nu }$ ($\nu =e,h$) are matrix coefficients determined by equation (\ref{GreenTI}), and $\hat{\Psi}
_{<,>}^{\mu }\left( x\right) $ represents the asymptotic solutions of $\hat{H}$. These solutions obey specific boundary conditions at the left ($<$) and right ($>$) edges of region $i$. For the right semi-infinite superconducting region S, these solutions involve processes as conventional reflections ($\mu =\nu $) and branch exchange $e-h$ ($\mu \neq \nu $) at the interface

\begin{eqnarray}
\hat{\Psi} _{<}^{e}\left( x\right) &=&\hat{\psi} _{-}^{e}\left( x\right) +r_{L}^{ee}\hat{\psi}
_{+}^{e}\left( x\right) +r_{L}^{eh}\hat{\psi} _{-}^{h}\left( x\right) \text{,}
\label{asymp} \\
\hat{\Psi} _{<}^{h}\left( x\right) &=&\hat{\psi} _{+}^{h}\left( x\right) +r_{L}^{hh}\hat{\psi}
_{-}^{h}\left( x\right) +r_{L}^{he}\hat{\psi} _{+}^{e}\left( x\right) \text{,} 
\notag \\
\hat{\Psi} _{>}^{e}\left( x\right) &=&\hat{\psi} _{+}^{e}\left( x\right) \text{, \ }\hat{\Psi}
_{>}^{h}\left( x\right) =\hat{\psi} _{-}^{h}\left( x\right) \text{,}
\end{eqnarray}
where $\hat{\psi} _{h}^{\mu }\left( x\right) $ are eigensolutions of $%
\hat{H}_{BdG}\left(x,q\right) $ that propagate in the $%
\eta \mathbf{\hat{x}}$ ($h =\pm 1$) direction, and the $r_{i}^{\mu \nu }$ are the reflection coefficients on the left ($L$) or right ($R$) edge defined by the chosen boundary conditions. For the left magnetic normal region F, there are no $e-h$ conversion processes ($r_{R}^{eh}=r_{R}^{he}=0$)

\begin{eqnarray}
\hat{\Psi} _{>}^{e}\left( x\right) &=&\hat{\psi} _{+}^{e}\left( x\right) +r_{R}^{ee}\hat{\psi}
_{-}^{e}\left( x\right) \text{,} \\
\hat{\Psi} _{>}^{h}\left( x\right) &=&\hat{\psi} _{-}^{h}\left( x\right) +r_{R}^{hh}\hat{\psi}
_{+}^{h}\left( x\right) \text{,}  \notag \\
\hat{\Psi} _{<}^{e}\left( x\right) &=&\hat{\psi} _{-}^{e}\left( x\right) \text{, \ }\hat{\Psi}
_{<}^{h}\left( x\right) =\hat{\psi} _{+}^{h}\left( x\right) \text{.}
\end{eqnarray}

The surface of a TI lacks open boundaries, allowing for the selection of artificial boundary conditions as long as the corresponding $\hat{H}_{T}$ ensures a transparent coupling between the regions on the TI surface. In this work, we adopt the boundary conditions $\hat{\Psi} _{<,\downarrow }^{\mu }\left( x_{L}\right) =\hat{\Psi} _{>,\uparrow
}^{\mu }\left( x_{R}\right) =0$ for the spin components of spinors at the left  $x_{L}$ and right $x_{R}$ edge of each region. Consequently, the coupling Hamiltonian $\hat{H}_{T}$ assumes the specific form
\begin{equation}
	\hat{H}_{T}=t\int \mathrm{d}q\hat{c}_{q,L\downarrow }^{\dagger }\hat{c}%
	_{q,R\uparrow }\text{+h.c},  \label{H_hopping_transparente}
\end{equation}%
with $t=\hbar v_{F}$ the transparent hopping amplitude between adjacent 
regions. Thus, the equilibrium Green functions of the coupled system $\hat{G}_{kl}=\hat{G}^{r/a}(x_{k},x_{l}^{\prime })$ ($x_{k}$ in region $k$) can be determined by a Dyson of the form (\ref{Dyson_equation})
\begin{equation}
	\hat{G}_{kl}=\hat{g}_{k}\delta _{kl}+\hat{g}_{k}\delta _{km}\hat{\Sigma}_{mn}%
	\hat{G}_{nl}\text{,}  
\end{equation}
where the self-energies $\hat{\Sigma}_{LR}=\hat{\Sigma}%
_{RL}^{T}=t\tau _{z}(\sigma _{x}-i\sigma _{y})\equiv \hat{t}$ correspond to the matrix form of $\hat{H}_{T}$. This boundary condition corresponds physically to a high-intensity magnetic barrier ($M$ $\rightarrow \pm \infty $), where the spinors components with spin projection opposite to the magnetization direction become null. Therefore, to model large magnetization values and doping levels for the left region, it suffices to exclude the coupling with the F region and only consider the right semi-infinite region with an open boundary condition. 

\section{Parameters of the magnetic region's Green Function\label{sec:app1B}}

The spectrum of Hamiltonian (\ref{HBdG}) for the magnetic region in a TI surface is
\begin{equation}
E_{e/h}=\pm \left( \sqrt{\left( \hbar v_{F}\left\vert \mathbf{k}\right\vert
\right) ^{2}+M^{2}}-E_F\right) \text{,}  \label{espectroM}
\end{equation}%
with eigenspinors
\begin{equation}
\hat{\psi} _{\eta }^{e}\left( \mathbf{r}\right) =\mathrm{e}^{iqy}\mathrm{e}%
^{\eta ik_{e}x}\left( \hat{\varphi}_{\varepsilon }^{e},0\right) ^{T}%
\text{, \ \ \ }\hat{\psi} _{\eta }^{h}\left( \mathbf{r}\right) =\mathrm{e}%
^{iqy}\mathrm{e}^{\eta ik_{h}x}\left( 0,\hat{\varphi}_{\varepsilon
}^{h}\right) ^{T}\text{,}
\end{equation}%
where 
\begin{eqnarray}
\hat{\varphi}_{\eta }^{e} &=&\left( M_{+}^{e},-\eta iM_{-}^{e}%
\mathrm{e}^{\eta i\theta _{e}}\right) ^{T}/\sqrt{2}\text{,} \\
\hat{\varphi}_{\eta }^{h} &=&\left( \eta iM_{+}^{h}\mathrm{e}%
^{\eta i\theta _{h}},M_{-}^{h}\right) ^{T}/\sqrt{2}\text{,} \\
M_{\pm }^{e} &=&\sqrt{E+E_F\pm M}/\sqrt{E+E_F}\text{,} \\
M_{\pm }^{h} &=&\sqrt{E_F-E\pm M}/\sqrt{E_F-E}\text{,} \\
\mathrm{e}^{i\theta _{\mu }} &\equiv &\hbar v_{F}\left( k_{\mu
}+iq\right) /\left\vert \mathbf{k}\right\vert \text{,}
\end{eqnarray}%
and with wave vector in $x$ direction
\begin{equation}
k_{e/h}=\mathrm{sgn}\left( E_F\pm E\right) \sqrt{\frac{\left( E_F\pm
E\right) ^{2}-M^{2}}{\left( \hbar v_{F}\right) ^{2}}-q^{2}}\text{,}
\end{equation}
where $sgn$ sets the correct sign for the valence band. For the chosen boundary conditions, the reflection coefficients are
\begin{equation}
r_{R}^{ee}=-1\text{, \ \ }r_{R}^{hh}=\mathrm{e}^{-2i\theta _{h}}\text{.}
\end{equation}

Integrating equation (\ref{GreenTI}) between $x^{\prime }-0^{+}$ and $
x^{\prime }+0^{+}$, we get the following constraint relation
\begin{equation}
\hat{g}\left( x^{\prime }+0^{+},x^{\prime }\right) -\hat{g}\left( x^{\prime
}-0^{+},x^{\prime }\right) =\frac{i}{\hbar v_{F}}\left( \tau _{z}\otimes
\sigma _{y}\right) \text{.}  \label{LigaduraTI}
\end{equation}

From this expression, we obtain the matrix coefficients of (\ref{GreenZZ2}) for this region
\begin{eqnarray}
\hat{C}_{\mu \nu } &=&\hat{C}_{\mu \mu }\delta _{\mu \nu }\text{, \ \ }\hat{C%
}_{\mu \mu }=\hat{C}_{\mu \mu }^{\prime }=\hat{C}_{ee}\text{,}
\label{CeeTIF} \\
\hat{C}_{ee} &=&\frac{-i}{\hbar v_{F}}\left( \frac{N_{e}}{\text{\textrm{cos}}%
\theta _{e}}\frac{\tau _{0}+\tau _{z} }{2}+\frac{N_{h}}{\text{%
\textrm{cos}}\theta _{h}}\frac{\tau _{0}-\tau _{z} }{2}\right) 
\text{,} \\
N_{e/h} &=&\frac{E_F\pm E}{\sqrt{\left( E_F\pm E\right) ^{2}-M^{2}}}%
\text{.}
\end{eqnarray}

For the studied system, the contact points of the electrodes with the F region are modeled as heavily doped magnetic leads
\begin{equation}
\hat{g}_{i}^{r(a)}\simeq \frac{-i}{\hbar v_{F}}\left( 
\begin{array}{cc}
\tilde{g} & 0 \\ 
0 & -\tilde{g}^{\ast }%
\end{array}%
\right) \text{, \ \ \ \ \ \ }\tilde{g}=\left( 
\begin{array}{cc}
0 & 1 \\ 
0 & i%
\end{array}%
\right) \text{.}
\end{equation}

The effect of finite width $L$ of electrodes is modeled by a weight factor $f\left( q\right) $ in the Fourier transform (\ref{Fourier}) of the equilibrium Green functions of the junction \cite{Takagaki_1999,Herrera_2009b}
\begin{eqnarray}
\hat{G}_{ij}^{r,a}\left( E\right) &\rightarrow & \\
\int&dq&
\left\vert f\left( q\right) \right\vert ^{2}\mathrm{e}^{iq\left(
	y_{j}-y_{i}\right) }\hat{G}^{r,a}\left( E,x_{j},x_{i},q\right)\text{,}
\label{Fourier2} \\
f\left( q\right) &=&\left\langle k_{1}|q\right\rangle \sqrt{k_{F}^{2}-q^{2}}%
\text{, \ \ } \\
\left\langle k_{1}|q\right\rangle &=&\sqrt{\frac{\pi }{L_{y}^{3}}}\frac{%
	\text{\textrm{cos}}(qL/2)}{k_{1}^{2}-q^{2}}\text{.}
\end{eqnarray}

This factor is proportional to the transverse wave vector of states at the superconducting region and depends only on the first transverse mode of the electrode. ($k_{1}=\pi /L$). The normalized hopping parameter $\tilde t$ between electrodes and the F/S region is related to this factor by the expression $\tilde t=t\sqrt{F_0}/\hbar v_F$ with
\begin{equation}
F_0=\int dq
\left\vert f\left( q\right) \right\vert ^{2}\text{.}
\end{equation}

\section{Parameters of the superconducting region's Green function  \label{sec:app1C}}

The excitation spectrum of Hamiltonian (\ref{HBdG}) for the superconducting region is
\begin{equation}
	E\left( \mathbf{k}\right) =\sqrt{\left(\hbar v_F\left\vert 
		\mathbf{k}\right\vert -E_F\right) ^{2}+|\hat{\Delta}|
		^{2}}\text{,}  \label{espectro}
\end{equation}
where $|\hat{\Delta}|=\textrm{Tr}(\hat{\Delta}^{\dagger}\hat{\Delta})/2$ \cite{Sigrist_2005}. The associated eigenspinors are given by
\begin{align}
\hat{\psi} _{\eta }^{e}\left( x\right) & =\mathrm{e}^{\eta
ik_{e}x}\left( u_{0}\hat{\varphi}_{\eta }^{e},-iv_{0}\text{\textrm{e}}%
^{-i\phi _{\eta }}\sigma _{y}\hat{\varphi}_{\eta }^{e}\right)
^{T}\text{,} \\
\hat{\psi} _{\eta }^{h}\left( x\right) & =\mathrm{e}^{\eta
ik_{h}x}\left( -iv_{0}\sigma _{y}\hat{\varphi}_{\eta }^{h},u_{0}\text{%
\textrm{e}}^{-i\phi _{\eta }}\hat{\varphi}_{\eta }^{h}\right)
^{T}\text{,}  \notag
\end{align}%
where the coherence factors are defined as
\begin{equation}
u_{0}=\sqrt{\frac{1}{2}\left( 1+\frac{\Omega }{E}\right) }\text{,}\,\,v_{0}=%
\sqrt{\frac{1}{2}\left( 1-\frac{\Omega }{E}\right) }\text{,}\,\,\Omega =%
\sqrt{E^{2}-\left\vert \Delta \right\vert ^{2}}\text{,}
\end{equation}%
and the spinors $\hat{\varphi}_{\eta }^{\mu }$ ($\mu =e,h$) by the expressions%
\begin{eqnarray}
\hat{\varphi}_{\eta }^{e} &=&\left( 1,-\eta i\mathrm{e}%
^{\eta i\theta _{e}}\right) ^{T}/\sqrt{2}\text{,}  \label{espinor1} \\
\hat{\varphi}_{\eta }^{h} &=&\left( \eta i\mathrm{e}%
^{\eta i\theta _{h}},1\right) ^{T}/\sqrt{2}\text{,}  \label{espinor2}
\end{eqnarray}%
with wave vector $x$ 
\begin{equation}
k_{e/h}=\sqrt{\frac{\left( E_F\pm \Omega \right) ^{2}}{\hbar ^{2}v_{F}^{2}}%
-q^{2}}\text{.}
\end{equation}

For the chosen boundary conditions, the reflection coefficients are:
\begin{eqnarray*}
r_{L}^{ee} &=&\frac{1}{Y}\left( \mathrm{e}^{-i\theta _{e}}-\gamma_0 ^{2}%
\mathrm{e}^{-i\theta _{h}}\right) \mathrm{e}^{-i\phi _{-}}\text{,} \\
r_{L}^{eh} &=&-\frac{1}{Y}\gamma_0 \left( \mathrm{e}^{-i\theta _{e}}\mathrm{e}%
^{-i\phi _{+}}+\mathrm{e}^{i\theta _{e}}\mathrm{e}^{-i\phi _{-}}\right) 
\text{,} \\
r_{L}^{he} &=&-\frac{1}{Y}\gamma_0 \left( \mathrm{e}^{-i\theta _{h}}\mathrm{e}%
^{-i\phi _{+}}+\mathrm{e}^{i\theta _{h}}\mathrm{e}^{-i\phi _{-}}\right) 
\text{,} \\
r_{L}^{hh} &=&-\frac{1}{Y}\left( \mathrm{e}^{i\theta _{e}}-\gamma_0 ^{2}%
\mathrm{e}^{i\theta _{h}}\right) \mathrm{e}^{-i\phi _{+}}\text{,} \\
Y &=&\gamma_0 ^{2}\mathrm{e}^{-i\theta _{h}}\mathrm{e}^{-i\phi _{+}}+%
\mathrm{e}^{i\theta _{e}}\mathrm{e}^{-i\phi _{-}} \text{, \ \ }\gamma_0
=v_{0}/u_{0}\text{.}
\end{eqnarray*}

In this case the matrix coefficients of (\ref{GreenZZ2}) are given by
\begin{eqnarray*}
\hat{C}_{ee} &=&-\frac{1}{\hbar v}\frac{1}{C^{2}+BF-AH}\left( 
\begin{array}{cccc}
H & 0 & -C & B \\ 
0 & H & -F & -C \\ 
C & B & -A & 0 \\ 
-F & C & 0 & -A%
\end{array}%
\right) \text{,} \\
\hat{C}_{hh} &=&\Lambda \hat{C}_{ee}\text{, \ \ }\hat{C}_{eh}=\hat{C}_{he}=0%
\text{,}
\end{eqnarray*}

where
\begin{eqnarray*}
A &=&i\left( \mathrm{e}^{-i\theta _{e}}+\mathrm{e}^{i\theta _{e}}\right)
-\Lambda i\gamma_0 ^{2}\left( \mathrm{e}^{-i\theta _{h}}+\mathrm{e}^{i\theta
_{h}}\right) \text{,} \\
B &=&i\gamma_0 \Lambda \left( \mathrm{e}^{i\theta _{h}}\mathrm{e}%
^{-i\phi _{+}}+\mathrm{e}^{-i\theta _{h}}\mathrm{e}^{-i\phi _{-}}\right) \\
&-&i\gamma_0\left( \mathrm{e}^{i\theta _{e}}\mathrm{e}^{-i\phi _{+}}+\mathrm{e}%
^{-i\theta _{e}}\mathrm{e}^{-i\phi _{-}}\right) \text{,} \\
C &=&\gamma_0 \left( \mathrm{e}^{-i\phi _{+}}-\mathrm{e}^{-i\phi _{-}}\right)
\left( \Lambda -1\right) \text{,} \\
F &=&i\gamma_0 \Lambda \left( \mathrm{e}^{-i\theta _{h}}\mathrm{e}%
^{-i\phi _{+}}+\mathrm{e}^{i\theta _{h}}\mathrm{e}^{-i\phi _{-}}\right)  \\
&-&i\gamma_0\left( \mathrm{e}^{-i\theta _{e}}\mathrm{e}^{-i\phi _{+}}+\mathrm{e}%
^{i\theta _{e}}\mathrm{e}^{-i\phi _{-}}\right)  \text{,} \\
H &=&i\gamma_0 ^{2}\left( \mathrm{e}^{-i\theta _{e}}+\mathrm{e}^{i\theta
_{e}}\right) \mathrm{e}^{-i\phi _{+}}\mathrm{e}^{-i\phi _{-}}  \\
&-&\Lambda i\left( \mathrm{e}^{-i\theta _{h}}+\mathrm{e}^{i\theta _{h}}\right) \mathrm{e%
}^{-i\phi _{+}}\mathrm{e}^{-i\phi _{-}}\text{,} \\
\Lambda &=&r_{L}^{eh}/r_{L}^{he}\text{.}
\end{eqnarray*}

\section{CAR processes in conventional chiral $p$-wave superconductivity \label{sec:app1E}}

For a conventional superconductor, the element $eh$ of the Green function evaluated at the edge of a semi-infinite region is given by
\begin{eqnarray}
\hat{g}_{S0,eh}^{r,a}\left( E,q\right) =\frac{-2im}{\hbar ^{2}D}\gamma_0 
\mathrm{e}^{i\varphi _{-}} \\
\times\left( \frac{1}{k_{e}}+\frac{1}{k_{h}} +\frac{D\left( 1-%
\mathrm{e}^{i\Delta \varphi }\right) }{2\left( 1-\gamma ^{2}\right) }\left( 
\frac{1}{k_{e}}-\frac{1}{k_{h}}\right) \right) \text{,} \\
D =\left( 1-\gamma_0 ^{2}\mathrm{e}^{-i\Delta \varphi }\right) \text{,}
\end{eqnarray}
with $m$ the mass of the electron. For chiral $p$-wave symmetry $\Delta
\left( \theta \right) =\Delta _{0}e^{i\theta }$, then $e^{i\varphi
_{+}}=e^{i\theta }=-e^{-i\varphi _{-}}$, and for $E=0+i\varepsilon $ and $E_F\gg \Delta _{0}$ ($k_{e}\approx k_{h}$) we get
\begin{eqnarray}
\hat{g}_{S0,eh}^{r,a}\left( 0,\theta \right) &\approx &-\frac{4m}{\hbar ^{2}}%
\frac{\delta \mathrm{e}^{-i\theta }}{\left( 1-\delta ^{2}\mathrm{e}%
^{-2i\theta }\right) }\frac{1}{k_{F}\text{\textrm{cos}}\theta }\text{,} \\
\delta &=&\sqrt{\frac{\zeta -\varepsilon }{\zeta +\varepsilon }}\text{, \ \ \ }%
\zeta =\sqrt{\varepsilon ^{2}+\Delta _{0}^{2}}\text{.}  \notag
\end{eqnarray}

Expressing the corresponding integral (\ref{Fourier2}) in terms of $q\approx k_{F}\mathrm{sin}\theta$, we obtain for  (\ref{Cond})
\begin{equation}
\sigma_{CAR}\left( d\right) =\sigma_0\tilde{\Gamma}_{e,b}\tilde{\Gamma}_{h,a}\left( \frac{4mL%
}{\hbar ^{2}}\right) ^{2}\left( \frac{k_{F}}{\pi }\right) ^{4}\Theta
^{2}\left( sd\right) \mathrm{e}^{-d/\lambda}\text{,}
\end{equation}
which has a dependence on the electrode separation $d$ identical to that of the case studied above for a TI in contact with a conventional $s$-wave superconductor.

\bibliography{article_OCasas_SGomez_WHerrera_2023}

\begin{thebibliography}{101}%
\makeatletter
\providecommand \@ifxundefined [1]{%
 \@ifx{#1\undefined}
}%
\providecommand \@ifnum [1]{%
 \ifnum #1\expandafter \@firstoftwo
 \else \expandafter \@secondoftwo
 \fi
}%
\providecommand \@ifx [1]{%
 \ifx #1\expandafter \@firstoftwo
 \else \expandafter \@secondoftwo
 \fi
}%
\providecommand \natexlab [1]{#1}%
\providecommand \enquote  [1]{``#1''}%
\providecommand \bibnamefont  [1]{#1}%
\providecommand \bibfnamefont [1]{#1}%
\providecommand \citenamefont [1]{#1}%
\providecommand \href@noop [0]{\@secondoftwo}%
\providecommand \href [0]{\begingroup \@sanitize@url \@href}%
\providecommand \@href[1]{\@@startlink{#1}\@@href}%
\providecommand \@@href[1]{\endgroup#1\@@endlink}%
\providecommand \@sanitize@url [0]{\catcode `\\12\catcode `\$12\catcode
  `\&12\catcode `\#12\catcode `\^12\catcode `\_12\catcode `\%12\relax}%
\providecommand \@@startlink[1]{}%
\providecommand \@@endlink[0]{}%
\providecommand \url  [0]{\begingroup\@sanitize@url \@url }%
\providecommand \@url [1]{\endgroup\@href {#1}{\urlprefix }}%
\providecommand \urlprefix  [0]{URL }%
\providecommand \Eprint [0]{\href }%
\providecommand \doibase [0]{https://doi.org/}%
\providecommand \selectlanguage [0]{\@gobble}%
\providecommand \bibinfo  [0]{\@secondoftwo}%
\providecommand \bibfield  [0]{\@secondoftwo}%
\providecommand \translation [1]{[#1]}%
\providecommand \BibitemOpen [0]{}%
\providecommand \bibitemStop [0]{}%
\providecommand \bibitemNoStop [0]{.\EOS\space}%
\providecommand \EOS [0]{\spacefactor3000\relax}%
\providecommand \BibitemShut  [1]{\csname bibitem#1\endcsname}%
\let\auto@bib@innerbib\@empty
\bibitem [{\citenamefont {Burkard}(2007)}]{Burkard_2007}%
  \BibitemOpen
  \bibfield  {author} {\bibinfo {author} {\bibfnamefont {G.}~\bibnamefont
  {Burkard}},\ }\bibfield  {title} {\bibinfo {title} {Spin-entangled electrons
  in solid-state systems},\ }\href
  {http://stacks.iop.org/0953-8984/19/i=23/a=233202} {\bibfield  {journal}
  {\bibinfo  {journal} {Journal of Physics: Condensed Matter}\ }\textbf
  {\bibinfo {volume} {19}},\ \bibinfo {pages} {233202} (\bibinfo {year}
  {2007})}\BibitemShut {NoStop}%
\bibitem [{\citenamefont {Byers}\ and\ \citenamefont
  {Flatt\'e}(1995)}]{Byers_1995}%
  \BibitemOpen
  \bibfield  {author} {\bibinfo {author} {\bibfnamefont {J.~M.}\ \bibnamefont
  {Byers}}\ and\ \bibinfo {author} {\bibfnamefont {M.~E.}\ \bibnamefont
  {Flatt\'e}},\ }\bibfield  {title} {\bibinfo {title} {Probing spatial
  correlations with nanoscale two-contact tunneling},\ }\href
  {https://doi.org/10.1103/PhysRevLett.74.306} {\bibfield  {journal} {\bibinfo
  {journal} {Phys. Rev. Lett.}\ }\textbf {\bibinfo {volume} {74}},\ \bibinfo
  {pages} {306} (\bibinfo {year} {1995})}\BibitemShut {NoStop}%
\bibitem [{\citenamefont {Deutscher}\ and\ \citenamefont
  {Feinberg}(2000)}]{Deutscher_2000}%
  \BibitemOpen
  \bibfield  {author} {\bibinfo {author} {\bibfnamefont {G.}~\bibnamefont
  {Deutscher}}\ and\ \bibinfo {author} {\bibfnamefont {D.}~\bibnamefont
  {Feinberg}},\ }\bibfield  {title} {\bibinfo {title} {Coupling
  superconducting-ferromagnetic point contacts by andreev reflections},\ }\href
  {https://doi.org/10.1063/1.125796} {\bibfield  {journal} {\bibinfo  {journal}
  {Applied Physics Letters}\ }\textbf {\bibinfo {volume} {76}},\ \bibinfo
  {pages} {487} (\bibinfo {year} {2000})},\ \Eprint
  {https://arxiv.org/abs/https://doi.org/10.1063/1.125796}
  {https://doi.org/10.1063/1.125796} \BibitemShut {NoStop}%
\bibitem [{\citenamefont {Falci}\ \emph {et~al.}(2001)\citenamefont {Falci},
  \citenamefont {Feinberg},\ and\ \citenamefont {Hekking}}]{Falci_2001}%
  \BibitemOpen
  \bibfield  {author} {\bibinfo {author} {\bibfnamefont {G.}~\bibnamefont
  {Falci}}, \bibinfo {author} {\bibfnamefont {D.}~\bibnamefont {Feinberg}},\
  and\ \bibinfo {author} {\bibfnamefont {F.~W.~J.}\ \bibnamefont {Hekking}},\
  }\bibfield  {title} {\bibinfo {title} {Correlated tunneling into a
  superconductor in a multiprobe hybrid structure},\ }\href
  {https://doi.org/10.1209/epl/i2001-00303-0} {\bibfield  {journal} {\bibinfo
  {journal} {Europhysics Letters ({EPL})}\ }\textbf {\bibinfo {volume} {54}},\
  \bibinfo {pages} {255} (\bibinfo {year} {2001})}\BibitemShut {NoStop}%
\bibitem [{\citenamefont {Lesovik}\ \emph {et~al.}(2001)\citenamefont
  {Lesovik}, \citenamefont {Martin},\ and\ \citenamefont
  {Blatter}}]{Lesovik_2001}%
  \BibitemOpen
  \bibfield  {author} {\bibinfo {author} {\bibfnamefont {G.~B.}\ \bibnamefont
  {Lesovik}}, \bibinfo {author} {\bibfnamefont {T.}~\bibnamefont {Martin}},\
  and\ \bibinfo {author} {\bibfnamefont {G.}~\bibnamefont {Blatter}},\
  }\bibfield  {title} {\bibinfo {title} {Electronic entanglement in the
  vicinity of a superconductor},\ }\href
  {https://doi.org/10.1007/s10051-001-8675-4} {\bibfield  {journal} {\bibinfo
  {journal} {Eur. Phys. J. B}\ }\textbf {\bibinfo {volume} {24}},\ \bibinfo
  {pages} {287} (\bibinfo {year} {2001})}\BibitemShut {NoStop}%
\bibitem [{\citenamefont {Chtchelkatchev}\ \emph {et~al.}(2002)\citenamefont
  {Chtchelkatchev}, \citenamefont {Blatter}, \citenamefont {Lesovik},\ and\
  \citenamefont {Martin}}]{Chtchelkatchev_2002}%
  \BibitemOpen
  \bibfield  {author} {\bibinfo {author} {\bibfnamefont {N.~M.}\ \bibnamefont
  {Chtchelkatchev}}, \bibinfo {author} {\bibfnamefont {G.}~\bibnamefont
  {Blatter}}, \bibinfo {author} {\bibfnamefont {G.~B.}\ \bibnamefont
  {Lesovik}},\ and\ \bibinfo {author} {\bibfnamefont {T.}~\bibnamefont
  {Martin}},\ }\bibfield  {title} {\bibinfo {title} {Bell inequalities and
  entanglement in solid-state devices},\ }\href
  {https://doi.org/10.1103/PhysRevB.66.161320} {\bibfield  {journal} {\bibinfo
  {journal} {Phys. Rev. B}\ }\textbf {\bibinfo {volume} {66}},\ \bibinfo
  {pages} {161320} (\bibinfo {year} {2002})}\BibitemShut {NoStop}%
\bibitem [{\citenamefont {Bena}\ \emph {et~al.}(2002)\citenamefont {Bena},
  \citenamefont {Vishveshwara}, \citenamefont {Balents},\ and\ \citenamefont
  {Fisher}}]{Bena_2002}%
  \BibitemOpen
  \bibfield  {author} {\bibinfo {author} {\bibfnamefont {C.}~\bibnamefont
  {Bena}}, \bibinfo {author} {\bibfnamefont {S.}~\bibnamefont {Vishveshwara}},
  \bibinfo {author} {\bibfnamefont {L.}~\bibnamefont {Balents}},\ and\ \bibinfo
  {author} {\bibfnamefont {M.~P.~A.}\ \bibnamefont {Fisher}},\ }\bibfield
  {title} {\bibinfo {title} {Quantum entanglement in carbon nanotubes},\ }\href
  {https://doi.org/10.1103/PhysRevLett.89.037901} {\bibfield  {journal}
  {\bibinfo  {journal} {Phys. Rev. Lett.}\ }\textbf {\bibinfo {volume} {89}},\
  \bibinfo {pages} {037901} (\bibinfo {year} {2002})}\BibitemShut {NoStop}%
\bibitem [{\citenamefont {Feinberg}(2003)}]{Feinberg_2003}%
  \BibitemOpen
  \bibfield  {author} {\bibinfo {author} {\bibfnamefont {D.}~\bibnamefont
  {Feinberg}},\ }\bibfield  {title} {\bibinfo {title} {Andreev scattering and
  cotunneling between two superconductor-normal metal interfaces: the dirty
  limit},\ }\href {https://doi.org/10.1140/epjb/e2003-00361-6} {\bibfield
  {journal} {\bibinfo  {journal} {The European Physical Journal B - Condensed
  Matter and Complex Systems}\ }\textbf {\bibinfo {volume} {36}},\ \bibinfo
  {pages} {419} (\bibinfo {year} {2003})}\BibitemShut {NoStop}%
\bibitem [{\citenamefont {M\'elin}\ and\ \citenamefont
  {Peysson}(2003)}]{Melin_2003}%
  \BibitemOpen
  \bibfield  {author} {\bibinfo {author} {\bibfnamefont {R.}~\bibnamefont
  {M\'elin}}\ and\ \bibinfo {author} {\bibfnamefont {S.}~\bibnamefont
  {Peysson}},\ }\bibfield  {title} {\bibinfo {title} {Crossed andreev
  reflection at ferromagnetic domain walls},\ }\href
  {https://doi.org/10.1103/PhysRevB.68.174515} {\bibfield  {journal} {\bibinfo
  {journal} {Phys. Rev. B}\ }\textbf {\bibinfo {volume} {68}},\ \bibinfo
  {pages} {174515} (\bibinfo {year} {2003})}\BibitemShut {NoStop}%
\bibitem [{\citenamefont {Yamashita}\ \emph {et~al.}(2003)\citenamefont
  {Yamashita}, \citenamefont {Takahashi},\ and\ \citenamefont
  {Maekawa}}]{Yamashita_2003}%
  \BibitemOpen
  \bibfield  {author} {\bibinfo {author} {\bibfnamefont {T.}~\bibnamefont
  {Yamashita}}, \bibinfo {author} {\bibfnamefont {S.}~\bibnamefont
  {Takahashi}},\ and\ \bibinfo {author} {\bibfnamefont {S.}~\bibnamefont
  {Maekawa}},\ }\bibfield  {title} {\bibinfo {title} {Crossed andreev
  reflection in structures consisting of a superconductor with ferromagnetic
  leads},\ }\href {https://doi.org/10.1103/PhysRevB.68.174504} {\bibfield
  {journal} {\bibinfo  {journal} {Phys. Rev. B}\ }\textbf {\bibinfo {volume}
  {68}},\ \bibinfo {pages} {174504} (\bibinfo {year} {2003})}\BibitemShut
  {NoStop}%
\bibitem [{\citenamefont {Stefanakis}\ and\ \citenamefont
  {lin}(2003)}]{Stefanakis_2003}%
  \BibitemOpen
  \bibfield  {author} {\bibinfo {author} {\bibfnamefont {N.}~\bibnamefont
  {Stefanakis}}\ and\ \bibinfo {author} {\bibfnamefont {R.~M.}\ \bibnamefont
  {lin}},\ }\bibfield  {title} {\bibinfo {title} {Transport properties of
  ferromagnet-d-wave superconductor ferromagnet double junctions},\ }\href
  {https://doi.org/10.1088/0953-8984/15/24/317} {\bibfield  {journal} {\bibinfo
   {journal} {Journal of Physics: Condensed Matter}\ }\textbf {\bibinfo
  {volume} {15}},\ \bibinfo {pages} {4239} (\bibinfo {year}
  {2003})}\BibitemShut {NoStop}%
\bibitem [{\citenamefont {M\'elin}\ and\ \citenamefont
  {Feinberg}(2004)}]{Melin_2004}%
  \BibitemOpen
  \bibfield  {author} {\bibinfo {author} {\bibfnamefont {R.}~\bibnamefont
  {M\'elin}}\ and\ \bibinfo {author} {\bibfnamefont {D.}~\bibnamefont
  {Feinberg}},\ }\bibfield  {title} {\bibinfo {title} {Sign of the crossed
  conductances at a ferromagnet/superconductor/ferromagnet double interface},\
  }\href {https://doi.org/10.1103/PhysRevB.70.174509} {\bibfield  {journal}
  {\bibinfo  {journal} {Phys. Rev. B}\ }\textbf {\bibinfo {volume} {70}},\
  \bibinfo {pages} {174509} (\bibinfo {year} {2004})}\BibitemShut {NoStop}%
\bibitem [{\citenamefont {Prada}\ and\ \citenamefont
  {Sols}(2004)}]{Prada_2004}%
  \BibitemOpen
  \bibfield  {author} {\bibinfo {author} {\bibfnamefont {E.}~\bibnamefont
  {Prada}}\ and\ \bibinfo {author} {\bibfnamefont {F.}~\bibnamefont {Sols}},\
  }\bibfield  {title} {\bibinfo {title} {Entangled electron current through
  finite size normal-superconductor tunneling structures},\ }\href
  {https://doi.org/10.1140/epjb/e2004-00284-8} {\bibfield  {journal} {\bibinfo
  {journal} {The European Physical Journal B - Condensed Matter and Complex
  Systems}\ }\textbf {\bibinfo {volume} {40}},\ \bibinfo {pages} {379}
  (\bibinfo {year} {2004})}\BibitemShut {NoStop}%
\bibitem [{\citenamefont {Beckmann}\ \emph {et~al.}(2004)\citenamefont
  {Beckmann}, \citenamefont {Weber},\ and\ \citenamefont
  {v.~L\"ohneysen}}]{Beckmann_2004}%
  \BibitemOpen
  \bibfield  {author} {\bibinfo {author} {\bibfnamefont {D.}~\bibnamefont
  {Beckmann}}, \bibinfo {author} {\bibfnamefont {H.~B.}\ \bibnamefont
  {Weber}},\ and\ \bibinfo {author} {\bibfnamefont {H.}~\bibnamefont
  {v.~L\"ohneysen}},\ }\bibfield  {title} {\bibinfo {title} {Evidence for
  crossed andreev reflection in superconductor-ferromagnet hybrid structures},\
  }\href {https://doi.org/10.1103/PhysRevLett.93.197003} {\bibfield  {journal}
  {\bibinfo  {journal} {Phys. Rev. Lett.}\ }\textbf {\bibinfo {volume} {93}},\
  \bibinfo {pages} {197003} (\bibinfo {year} {2004})}\BibitemShut {NoStop}%
\bibitem [{\citenamefont {Russo}\ \emph {et~al.}(2005)\citenamefont {Russo},
  \citenamefont {Kroug}, \citenamefont {Klapwijk},\ and\ \citenamefont
  {Morpurgo}}]{Russo_2005}%
  \BibitemOpen
  \bibfield  {author} {\bibinfo {author} {\bibfnamefont {S.}~\bibnamefont
  {Russo}}, \bibinfo {author} {\bibfnamefont {M.}~\bibnamefont {Kroug}},
  \bibinfo {author} {\bibfnamefont {T.~M.}\ \bibnamefont {Klapwijk}},\ and\
  \bibinfo {author} {\bibfnamefont {A.~F.}\ \bibnamefont {Morpurgo}},\
  }\bibfield  {title} {\bibinfo {title} {Experimental observation of
  bias-dependent nonlocal andreev reflection},\ }\href
  {https://doi.org/10.1103/PhysRevLett.95.027002} {\bibfield  {journal}
  {\bibinfo  {journal} {Phys. Rev. Lett.}\ }\textbf {\bibinfo {volume} {95}},\
  \bibinfo {pages} {027002} (\bibinfo {year} {2005})}\BibitemShut {NoStop}%
\bibitem [{\citenamefont {Brinkman}\ and\ \citenamefont
  {Golubov}(2006)}]{Brinkman_2006}%
  \BibitemOpen
  \bibfield  {author} {\bibinfo {author} {\bibfnamefont {A.}~\bibnamefont
  {Brinkman}}\ and\ \bibinfo {author} {\bibfnamefont {A.~A.}\ \bibnamefont
  {Golubov}},\ }\bibfield  {title} {\bibinfo {title} {Crossed andreev
  reflection in diffusive contacts: Quasiclassical keldysh-usadel formalism},\
  }\href {https://doi.org/10.1103/PhysRevB.74.214512} {\bibfield  {journal}
  {\bibinfo  {journal} {Phys. Rev. B}\ }\textbf {\bibinfo {volume} {74}},\
  \bibinfo {pages} {214512} (\bibinfo {year} {2006})}\BibitemShut {NoStop}%
\bibitem [{\citenamefont {Cadden-Zimansky}\ and\ \citenamefont
  {Chandrasekhar}(2006)}]{Cadden_2006}%
  \BibitemOpen
  \bibfield  {author} {\bibinfo {author} {\bibfnamefont {P.}~\bibnamefont
  {Cadden-Zimansky}}\ and\ \bibinfo {author} {\bibfnamefont {V.}~\bibnamefont
  {Chandrasekhar}},\ }\bibfield  {title} {\bibinfo {title} {Nonlocal
  correlations in normal-metal superconducting systems},\ }\href
  {https://doi.org/10.1103/PhysRevLett.97.237003} {\bibfield  {journal}
  {\bibinfo  {journal} {Phys. Rev. Lett.}\ }\textbf {\bibinfo {volume} {97}},\
  \bibinfo {pages} {237003} (\bibinfo {year} {2006})}\BibitemShut {NoStop}%
\bibitem [{\citenamefont {Yeyati}\ \emph {et~al.}(2007)\citenamefont {Yeyati},
  \citenamefont {Bergeret}, \citenamefont {Mart{\'i}n-Rodero},\ and\
  \citenamefont {Klapwijk}}]{Yeyati_2007}%
  \BibitemOpen
  \bibfield  {author} {\bibinfo {author} {\bibfnamefont {A.~L.}\ \bibnamefont
  {Yeyati}}, \bibinfo {author} {\bibfnamefont {F.~S.}\ \bibnamefont
  {Bergeret}}, \bibinfo {author} {\bibfnamefont {A.}~\bibnamefont
  {Mart{\'i}n-Rodero}},\ and\ \bibinfo {author} {\bibfnamefont {T.~M.}\
  \bibnamefont {Klapwijk}},\ }\bibfield  {title} {\bibinfo {title} {Entangled
  andreev pairs and collective excitations in nanoscale superconductors},\
  }\href {https://doi.org/10.1038/nphys621} {\bibfield  {journal} {\bibinfo
  {journal} {Nature Physics}\ }\textbf {\bibinfo {volume} {3}},\ \bibinfo
  {pages} {455} (\bibinfo {year} {2007})}\BibitemShut {NoStop}%
\bibitem [{\citenamefont {Beckmann}\ and\ \citenamefont
  {v.~L{\"o}hneysen}(2007)}]{Beckmann_2007}%
  \BibitemOpen
  \bibfield  {author} {\bibinfo {author} {\bibfnamefont {D.}~\bibnamefont
  {Beckmann}}\ and\ \bibinfo {author} {\bibfnamefont {H.}~\bibnamefont
  {v.~L{\"o}hneysen}},\ }\bibfield  {title} {\bibinfo {title} {Negative
  four-terminal resistance as a probe of crossed andreev reflection},\ }\href
  {https://doi.org/10.1007/s00339-007-4193-4} {\bibfield  {journal} {\bibinfo
  {journal} {Applied Physics A}\ }\textbf {\bibinfo {volume} {89}},\ \bibinfo
  {pages} {603} (\bibinfo {year} {2007})}\BibitemShut {NoStop}%
\bibitem [{\citenamefont {Kalenkov}\ and\ \citenamefont
  {Zaikin}(2007{\natexlab{a}})}]{Kalenkov_2007}%
  \BibitemOpen
  \bibfield  {author} {\bibinfo {author} {\bibfnamefont {M.~S.}\ \bibnamefont
  {Kalenkov}}\ and\ \bibinfo {author} {\bibfnamefont {A.~D.}\ \bibnamefont
  {Zaikin}},\ }\bibfield  {title} {\bibinfo {title} {Nonlocal andreev
  reflection at high transmissions},\ }\href
  {https://doi.org/10.1103/PhysRevB.75.172503} {\bibfield  {journal} {\bibinfo
  {journal} {Phys. Rev. B}\ }\textbf {\bibinfo {volume} {75}},\ \bibinfo
  {pages} {172503} (\bibinfo {year} {2007}{\natexlab{a}})}\BibitemShut
  {NoStop}%
\bibitem [{\citenamefont {Kalenkov}\ and\ \citenamefont
  {Zaikin}(2007{\natexlab{b}})}]{Kalenkov_2007b}%
  \BibitemOpen
  \bibfield  {author} {\bibinfo {author} {\bibfnamefont {M.~S.}\ \bibnamefont
  {Kalenkov}}\ and\ \bibinfo {author} {\bibfnamefont {A.~D.}\ \bibnamefont
  {Zaikin}},\ }\bibfield  {title} {\bibinfo {title} {Crossed andreev reflection
  at spin-active interfaces},\ }\href
  {https://doi.org/10.1103/PhysRevB.76.224506} {\bibfield  {journal} {\bibinfo
  {journal} {Phys. Rev. B}\ }\textbf {\bibinfo {volume} {76}},\ \bibinfo
  {pages} {224506} (\bibinfo {year} {2007}{\natexlab{b}})}\BibitemShut
  {NoStop}%
\bibitem [{\citenamefont {Benjamin}\ and\ \citenamefont
  {Pachos}(2008)}]{Benjamin_2008}%
  \BibitemOpen
  \bibfield  {author} {\bibinfo {author} {\bibfnamefont {C.}~\bibnamefont
  {Benjamin}}\ and\ \bibinfo {author} {\bibfnamefont {J.~K.}\ \bibnamefont
  {Pachos}},\ }\bibfield  {title} {\bibinfo {title} {Detecting entangled states
  in graphene via crossed andreev reflection},\ }\href
  {https://doi.org/10.1103/PhysRevB.78.235403} {\bibfield  {journal} {\bibinfo
  {journal} {Phys. Rev. B}\ }\textbf {\bibinfo {volume} {78}},\ \bibinfo
  {pages} {235403} (\bibinfo {year} {2008})}\BibitemShut {NoStop}%
\bibitem [{\citenamefont {Golubev}\ \emph {et~al.}(2009)\citenamefont
  {Golubev}, \citenamefont {Kalenkov},\ and\ \citenamefont
  {Zaikin}}]{Golubev_2009}%
  \BibitemOpen
  \bibfield  {author} {\bibinfo {author} {\bibfnamefont {D.~S.}\ \bibnamefont
  {Golubev}}, \bibinfo {author} {\bibfnamefont {M.~S.}\ \bibnamefont
  {Kalenkov}},\ and\ \bibinfo {author} {\bibfnamefont {A.~D.}\ \bibnamefont
  {Zaikin}},\ }\bibfield  {title} {\bibinfo {title} {Crossed andreev reflection
  and charge imbalance in diffusive normal-superconducting-normal structures},\
  }\href {https://doi.org/10.1103/PhysRevLett.103.067006} {\bibfield  {journal}
  {\bibinfo  {journal} {Phys. Rev. Lett.}\ }\textbf {\bibinfo {volume} {103}},\
  \bibinfo {pages} {067006} (\bibinfo {year} {2009})}\BibitemShut {NoStop}%
\bibitem [{\citenamefont {Cadden-Zimansky}\ \emph {et~al.}(2009)\citenamefont
  {Cadden-Zimansky}, \citenamefont {Wei},\ and\ \citenamefont
  {Chandrasekhar}}]{Cadden_2009}%
  \BibitemOpen
  \bibfield  {author} {\bibinfo {author} {\bibfnamefont {P.}~\bibnamefont
  {Cadden-Zimansky}}, \bibinfo {author} {\bibfnamefont {J.}~\bibnamefont
  {Wei}},\ and\ \bibinfo {author} {\bibfnamefont {V.}~\bibnamefont
  {Chandrasekhar}},\ }\bibfield  {title} {\bibinfo {title}
  {Cooper-pair-mediated coherence between two normal metals},\ }\href
  {http://dx.doi.org/10.1038/nphys1252} {\bibfield  {journal} {\bibinfo
  {journal} {Nature Physics}\ }\textbf {\bibinfo {volume} {5}},\ \bibinfo
  {pages} {393 EP } (\bibinfo {year} {2009})}\BibitemShut {NoStop}%
\bibitem [{\citenamefont {Haugen}\ \emph {et~al.}(2010)\citenamefont {Haugen},
  \citenamefont {Huertas-Hernando}, \citenamefont {Brataas},\ and\
  \citenamefont {Waintal}}]{Haugen_2010}%
  \BibitemOpen
  \bibfield  {author} {\bibinfo {author} {\bibfnamefont {H.}~\bibnamefont
  {Haugen}}, \bibinfo {author} {\bibfnamefont {D.}~\bibnamefont
  {Huertas-Hernando}}, \bibinfo {author} {\bibfnamefont {A.}~\bibnamefont
  {Brataas}},\ and\ \bibinfo {author} {\bibfnamefont {X.}~\bibnamefont
  {Waintal}},\ }\bibfield  {title} {\bibinfo {title} {Crossed andreev
  reflection versus electron transfer in three-terminal graphene devices},\
  }\href {https://doi.org/10.1103/PhysRevB.81.174523} {\bibfield  {journal}
  {\bibinfo  {journal} {Phys. Rev. B}\ }\textbf {\bibinfo {volume} {81}},\
  \bibinfo {pages} {174523} (\bibinfo {year} {2010})}\BibitemShut {NoStop}%
\bibitem [{\citenamefont {Burset}\ \emph {et~al.}(2011)\citenamefont {Burset},
  \citenamefont {Herrera},\ and\ \citenamefont {Yeyati}}]{Burset_2011}%
  \BibitemOpen
  \bibfield  {author} {\bibinfo {author} {\bibfnamefont {P.}~\bibnamefont
  {Burset}}, \bibinfo {author} {\bibfnamefont {W.~J.}\ \bibnamefont
  {Herrera}},\ and\ \bibinfo {author} {\bibfnamefont {A.~L.}\ \bibnamefont
  {Yeyati}},\ }\bibfield  {title} {\bibinfo {title} {Microscopic theory of
  cooper pair beam splitters based on carbon nanotubes},\ }\href
  {https://doi.org/10.1103/PhysRevB.84.115448} {\bibfield  {journal} {\bibinfo
  {journal} {Phys. Rev. B}\ }\textbf {\bibinfo {volume} {84}},\ \bibinfo
  {pages} {115448} (\bibinfo {year} {2011})}\BibitemShut {NoStop}%
\bibitem [{\citenamefont {G\'omez}\ \emph {et~al.}(2012)\citenamefont
  {G\'omez}, \citenamefont {Burset}, \citenamefont {Herrera},\ and\
  \citenamefont {Yeyati}}]{Gomez_2012}%
  \BibitemOpen
  \bibfield  {author} {\bibinfo {author} {\bibfnamefont {S.}~\bibnamefont
  {G\'omez}}, \bibinfo {author} {\bibfnamefont {P.}~\bibnamefont {Burset}},
  \bibinfo {author} {\bibfnamefont {W.~J.}\ \bibnamefont {Herrera}},\ and\
  \bibinfo {author} {\bibfnamefont {A.~L.}\ \bibnamefont {Yeyati}},\ }\bibfield
   {title} {\bibinfo {title} {Selective focusing of electrons and holes in a
  graphene-based superconducting lens},\ }\href
  {https://doi.org/10.1103/PhysRevB.85.115411} {\bibfield  {journal} {\bibinfo
  {journal} {Phys. Rev. B}\ }\textbf {\bibinfo {volume} {85}},\ \bibinfo
  {pages} {115411} (\bibinfo {year} {2012})}\BibitemShut {NoStop}%
\bibitem [{\citenamefont {Beiranvand}\ \emph {et~al.}(2017)\citenamefont
  {Beiranvand}, \citenamefont {Hamzehpour},\ and\ \citenamefont
  {Alidoust}}]{Beiranvand_2017}%
  \BibitemOpen
  \bibfield  {author} {\bibinfo {author} {\bibfnamefont {R.}~\bibnamefont
  {Beiranvand}}, \bibinfo {author} {\bibfnamefont {H.}~\bibnamefont
  {Hamzehpour}},\ and\ \bibinfo {author} {\bibfnamefont {M.}~\bibnamefont
  {Alidoust}},\ }\bibfield  {title} {\bibinfo {title} {Nonlocal andreev
  entanglements and triplet correlations in graphene with spin-orbit
  coupling},\ }\href {https://doi.org/10.1103/PhysRevB.96.161403} {\bibfield
  {journal} {\bibinfo  {journal} {Phys. Rev. B}\ }\textbf {\bibinfo {volume}
  {96}},\ \bibinfo {pages} {161403} (\bibinfo {year} {2017})}\BibitemShut
  {NoStop}%
\bibitem [{\citenamefont {Beconcini}\ \emph {et~al.}(2018)\citenamefont
  {Beconcini}, \citenamefont {Polini},\ and\ \citenamefont
  {Taddei}}]{Beconcini_2018}%
  \BibitemOpen
  \bibfield  {author} {\bibinfo {author} {\bibfnamefont {M.}~\bibnamefont
  {Beconcini}}, \bibinfo {author} {\bibfnamefont {M.}~\bibnamefont {Polini}},\
  and\ \bibinfo {author} {\bibfnamefont {F.}~\bibnamefont {Taddei}},\
  }\bibfield  {title} {\bibinfo {title} {Nonlocal superconducting correlations
  in graphene in the quantum hall regime},\ }\href
  {https://doi.org/10.1103/PhysRevB.97.201403} {\bibfield  {journal} {\bibinfo
  {journal} {Phys. Rev. B}\ }\textbf {\bibinfo {volume} {97}},\ \bibinfo
  {pages} {201403} (\bibinfo {year} {2018})}\BibitemShut {NoStop}%
\bibitem [{\citenamefont {Golubev}\ and\ \citenamefont
  {Zaikin}(2019)}]{Golubev_2019}%
  \BibitemOpen
  \bibfield  {author} {\bibinfo {author} {\bibfnamefont {D.~S.}\ \bibnamefont
  {Golubev}}\ and\ \bibinfo {author} {\bibfnamefont {A.~D.}\ \bibnamefont
  {Zaikin}},\ }\bibfield  {title} {\bibinfo {title} {Cross-correlated shot
  noise in three-terminal superconducting hybrid nanostructures},\ }\href
  {https://doi.org/10.1103/PhysRevB.99.144504} {\bibfield  {journal} {\bibinfo
  {journal} {Phys. Rev. B}\ }\textbf {\bibinfo {volume} {99}},\ \bibinfo
  {pages} {144504} (\bibinfo {year} {2019})}\BibitemShut {NoStop}%
\bibitem [{\citenamefont {Wu}\ \emph {et~al.}(2020)\citenamefont {Wu},
  \citenamefont {Meng}, \citenamefont {Kong}, \citenamefont {Zhang},
  \citenamefont {Bai},\ and\ \citenamefont {Xu}}]{Wu_2020}%
  \BibitemOpen
  \bibfield  {author} {\bibinfo {author} {\bibfnamefont {X.}~\bibnamefont
  {Wu}}, \bibinfo {author} {\bibfnamefont {H.}~\bibnamefont {Meng}}, \bibinfo
  {author} {\bibfnamefont {F.}~\bibnamefont {Kong}}, \bibinfo {author}
  {\bibfnamefont {H.}~\bibnamefont {Zhang}}, \bibinfo {author} {\bibfnamefont
  {Y.}~\bibnamefont {Bai}},\ and\ \bibinfo {author} {\bibfnamefont
  {N.}~\bibnamefont {Xu}},\ }\bibfield  {title} {\bibinfo {title} {Tunable
  nonlocal valley-entangled cooper pair splitter realized in bilayer-graphene
  van der waals spin valves},\ }\href
  {https://doi.org/10.1103/PhysRevB.101.125406} {\bibfield  {journal} {\bibinfo
   {journal} {Phys. Rev. B}\ }\textbf {\bibinfo {volume} {101}},\ \bibinfo
  {pages} {125406} (\bibinfo {year} {2020})}\BibitemShut {NoStop}%
\bibitem [{\citenamefont {Jakobsen}\ \emph {et~al.}(2021)\citenamefont
  {Jakobsen}, \citenamefont {Brataas},\ and\ \citenamefont
  {Qaiumzadeh}}]{Jakobsen_2021}%
  \BibitemOpen
  \bibfield  {author} {\bibinfo {author} {\bibfnamefont {M.~F.}\ \bibnamefont
  {Jakobsen}}, \bibinfo {author} {\bibfnamefont {A.}~\bibnamefont {Brataas}},\
  and\ \bibinfo {author} {\bibfnamefont {A.}~\bibnamefont {Qaiumzadeh}},\
  }\bibfield  {title} {\bibinfo {title} {Electrically controlled crossed
  andreev reflection in two-dimensional antiferromagnets},\ }\href
  {https://doi.org/10.1103/PhysRevLett.127.017701} {\bibfield  {journal}
  {\bibinfo  {journal} {Phys. Rev. Lett.}\ }\textbf {\bibinfo {volume} {127}},\
  \bibinfo {pages} {017701} (\bibinfo {year} {2021})}\BibitemShut {NoStop}%
\bibitem [{\citenamefont {Recher}\ \emph {et~al.}(2001)\citenamefont {Recher},
  \citenamefont {Sukhorukov},\ and\ \citenamefont {Loss}}]{Recher_2001}%
  \BibitemOpen
  \bibfield  {author} {\bibinfo {author} {\bibfnamefont {P.}~\bibnamefont
  {Recher}}, \bibinfo {author} {\bibfnamefont {E.~V.}\ \bibnamefont
  {Sukhorukov}},\ and\ \bibinfo {author} {\bibfnamefont {D.}~\bibnamefont
  {Loss}},\ }\bibfield  {title} {\bibinfo {title} {Andreev tunneling, coulomb
  blockade, and resonant transport of nonlocal spin-entangled electrons},\
  }\href {https://doi.org/10.1103/PhysRevB.63.165314} {\bibfield  {journal}
  {\bibinfo  {journal} {Phys. Rev. B}\ }\textbf {\bibinfo {volume} {63}},\
  \bibinfo {pages} {165314} (\bibinfo {year} {2001})}\BibitemShut {NoStop}%
\bibitem [{\citenamefont {Recher}\ and\ \citenamefont
  {Loss}(2003)}]{Recher_2003}%
  \BibitemOpen
  \bibfield  {author} {\bibinfo {author} {\bibfnamefont {P.}~\bibnamefont
  {Recher}}\ and\ \bibinfo {author} {\bibfnamefont {D.}~\bibnamefont {Loss}},\
  }\bibfield  {title} {\bibinfo {title} {Dynamical coulomb blockade and
  spin-entangled electrons},\ }\href
  {https://doi.org/10.1103/PhysRevLett.91.267003} {\bibfield  {journal}
  {\bibinfo  {journal} {Phys. Rev. Lett.}\ }\textbf {\bibinfo {volume} {91}},\
  \bibinfo {pages} {267003} (\bibinfo {year} {2003})}\BibitemShut {NoStop}%
\bibitem [{\citenamefont {Hofstetter}\ \emph {et~al.}(2009)\citenamefont
  {Hofstetter}, \citenamefont {Csonka}, \citenamefont {Nyg{\aa}rd},\ and\
  \citenamefont {Sch{\"o}nenberger}}]{Hofstetter_2009}%
  \BibitemOpen
  \bibfield  {author} {\bibinfo {author} {\bibfnamefont {L.}~\bibnamefont
  {Hofstetter}}, \bibinfo {author} {\bibfnamefont {S.}~\bibnamefont {Csonka}},
  \bibinfo {author} {\bibfnamefont {J.}~\bibnamefont {Nyg{\aa}rd}},\ and\
  \bibinfo {author} {\bibfnamefont {C.}~\bibnamefont {Sch{\"o}nenberger}},\
  }\bibfield  {title} {\bibinfo {title} {Cooper pair splitter realized in a
  two-quantum-dot y-junction},\ }\href {https://doi.org/10.1038/nature08432}
  {\bibfield  {journal} {\bibinfo  {journal} {Nature}\ }\textbf {\bibinfo
  {volume} {461}},\ \bibinfo {pages} {960} (\bibinfo {year}
  {2009})}\BibitemShut {NoStop}%
\bibitem [{\citenamefont {Herrmann}\ \emph
  {et~al.}(2010{\natexlab{a}})\citenamefont {Herrmann}, \citenamefont
  {Portier}, \citenamefont {Roche}, \citenamefont {Yeyati}, \citenamefont
  {Kontos},\ and\ \citenamefont {Strunk}}]{Herrmann_2010}%
  \BibitemOpen
  \bibfield  {author} {\bibinfo {author} {\bibfnamefont {L.~G.}\ \bibnamefont
  {Herrmann}}, \bibinfo {author} {\bibfnamefont {F.}~\bibnamefont {Portier}},
  \bibinfo {author} {\bibfnamefont {P.}~\bibnamefont {Roche}}, \bibinfo
  {author} {\bibfnamefont {A.~L.}\ \bibnamefont {Yeyati}}, \bibinfo {author}
  {\bibfnamefont {T.}~\bibnamefont {Kontos}},\ and\ \bibinfo {author}
  {\bibfnamefont {C.}~\bibnamefont {Strunk}},\ }\bibfield  {title} {\bibinfo
  {title} {Carbon nanotubes as cooper-pair beam splitters},\ }\href
  {https://doi.org/10.1103/PhysRevLett.104.026801} {\bibfield  {journal}
  {\bibinfo  {journal} {Phys. Rev. Lett.}\ }\textbf {\bibinfo {volume} {104}},\
  \bibinfo {pages} {026801} (\bibinfo {year} {2010}{\natexlab{a}})}\BibitemShut
  {NoStop}%
\bibitem [{\citenamefont {Schindele}\ \emph {et~al.}(2012)\citenamefont
  {Schindele}, \citenamefont {Baumgartner},\ and\ \citenamefont
  {Sch\"onenberger}}]{Schindele_2012}%
  \BibitemOpen
  \bibfield  {author} {\bibinfo {author} {\bibfnamefont {J.}~\bibnamefont
  {Schindele}}, \bibinfo {author} {\bibfnamefont {A.}~\bibnamefont
  {Baumgartner}},\ and\ \bibinfo {author} {\bibfnamefont {C.}~\bibnamefont
  {Sch\"onenberger}},\ }\bibfield  {title} {\bibinfo {title} {Near-unity cooper
  pair splitting efficiency},\ }\href
  {https://doi.org/10.1103/PhysRevLett.109.157002} {\bibfield  {journal}
  {\bibinfo  {journal} {Phys. Rev. Lett.}\ }\textbf {\bibinfo {volume} {109}},\
  \bibinfo {pages} {157002} (\bibinfo {year} {2012})}\BibitemShut {NoStop}%
\bibitem [{\citenamefont {Cottet}\ \emph {et~al.}(2012)\citenamefont {Cottet},
  \citenamefont {Kontos},\ and\ \citenamefont {Yeyati}}]{Cottet_2012}%
  \BibitemOpen
  \bibfield  {author} {\bibinfo {author} {\bibfnamefont {A.}~\bibnamefont
  {Cottet}}, \bibinfo {author} {\bibfnamefont {T.}~\bibnamefont {Kontos}},\
  and\ \bibinfo {author} {\bibfnamefont {A.~L.}\ \bibnamefont {Yeyati}},\
  }\bibfield  {title} {\bibinfo {title} {Subradiant split cooper pairs},\
  }\href {https://doi.org/10.1103/PhysRevLett.108.166803} {\bibfield  {journal}
  {\bibinfo  {journal} {Phys. Rev. Lett.}\ }\textbf {\bibinfo {volume} {108}},\
  \bibinfo {pages} {166803} (\bibinfo {year} {2012})}\BibitemShut {NoStop}%
\bibitem [{\citenamefont {Micha\l{}ek}\ \emph {et~al.}(2013)\citenamefont
  {Micha\l{}ek}, \citenamefont {Bu\l{}ka}, \citenamefont
  {Doma\ifmmode~\acute{n}\else \'{n}\fi{}ski},\ and\ \citenamefont
  {Wysoki\ifmmode~\acute{n}\else \'{n}\fi{}ski}}]{Michalek_2013}%
  \BibitemOpen
  \bibfield  {author} {\bibinfo {author} {\bibfnamefont {G.}~\bibnamefont
  {Micha\l{}ek}}, \bibinfo {author} {\bibfnamefont {B.~R.}\ \bibnamefont
  {Bu\l{}ka}}, \bibinfo {author} {\bibfnamefont {T.}~\bibnamefont
  {Doma\ifmmode~\acute{n}\else \'{n}\fi{}ski}},\ and\ \bibinfo {author}
  {\bibfnamefont {K.~I.}\ \bibnamefont {Wysoki\ifmmode~\acute{n}\else
  \'{n}\fi{}ski}},\ }\bibfield  {title} {\bibinfo {title} {Interplay between
  direct and crossed andreev reflections in hybrid nanostructures},\ }\href
  {https://doi.org/10.1103/PhysRevB.88.155425} {\bibfield  {journal} {\bibinfo
  {journal} {Phys. Rev. B}\ }\textbf {\bibinfo {volume} {88}},\ \bibinfo
  {pages} {155425} (\bibinfo {year} {2013})}\BibitemShut {NoStop}%
\bibitem [{\citenamefont {Micha{\l}ek}\ \emph {et~al.}(2015)\citenamefont
  {Micha{\l}ek}, \citenamefont {Doma{\'{n}}ski}, \citenamefont {Bu{\l}ka},\
  and\ \citenamefont {Wysoki{\'{n}}ski}}]{Michalek_2015}%
  \BibitemOpen
  \bibfield  {author} {\bibinfo {author} {\bibfnamefont {G.}~\bibnamefont
  {Micha{\l}ek}}, \bibinfo {author} {\bibfnamefont {T.}~\bibnamefont
  {Doma{\'{n}}ski}}, \bibinfo {author} {\bibfnamefont {B.~R.}\ \bibnamefont
  {Bu{\l}ka}},\ and\ \bibinfo {author} {\bibfnamefont {K.~I.}\ \bibnamefont
  {Wysoki{\'{n}}ski}},\ }\bibfield  {title} {\bibinfo {title} {Novel non-local
  effects in three-terminal hybrid devices with quantum dot},\ }\href
  {https://doi.org/10.1038/srep14572} {\bibfield  {journal} {\bibinfo
  {journal} {Scientific Reports}\ }\textbf {\bibinfo {volume} {5}},\ \bibinfo
  {pages} {14572} (\bibinfo {year} {2015})}\BibitemShut {NoStop}%
\bibitem [{\citenamefont {Michałek}\ \emph {et~al.}(2017)\citenamefont
  {Michałek}, \citenamefont {Domanski},\ and\ \citenamefont
  {Wysokinski}}]{Michalek_2017}%
  \BibitemOpen
  \bibfield  {author} {\bibinfo {author} {\bibfnamefont {G.}~\bibnamefont
  {Michałek}}, \bibinfo {author} {\bibfnamefont {T.}~\bibnamefont
  {Domanski}},\ and\ \bibinfo {author} {\bibfnamefont {K.}~\bibnamefont
  {Wysokinski}},\ }\bibfield  {title} {\bibinfo {title} {Cooper pair splitting
  efficiency in the hybrid three-terminal quantum dot},\ }\href
  {https://doi.org/10.1007/s10948-016-3757-y} {\bibfield  {journal} {\bibinfo
  {journal} {Journal of Superconductivity and Novel Magnetism}\ }\textbf
  {\bibinfo {volume} {30}} (\bibinfo {year} {2017})}\BibitemShut {NoStop}%
\bibitem [{\citenamefont {Bocian}\ \emph {et~al.}(2018)\citenamefont {Bocian},
  \citenamefont {Rudzi\ifmmode~\acute{n}\else \'{n}\fi{}ski},\ and\
  \citenamefont {Weymann}}]{Bocian_2018}%
  \BibitemOpen
  \bibfield  {author} {\bibinfo {author} {\bibfnamefont {K.}~\bibnamefont
  {Bocian}}, \bibinfo {author} {\bibfnamefont {W.}~\bibnamefont
  {Rudzi\ifmmode~\acute{n}\else \'{n}\fi{}ski}},\ and\ \bibinfo {author}
  {\bibfnamefont {I.}~\bibnamefont {Weymann}},\ }\bibfield  {title} {\bibinfo
  {title} {Splitting efficiency and interference effects in a cooper pair
  splitter based on a triple quantum dot with ferromagnetic contacts},\ }\href
  {https://doi.org/10.1103/PhysRevB.97.195441} {\bibfield  {journal} {\bibinfo
  {journal} {Phys. Rev. B}\ }\textbf {\bibinfo {volume} {97}},\ \bibinfo
  {pages} {195441} (\bibinfo {year} {2018})}\BibitemShut {NoStop}%
\bibitem [{\citenamefont {Takahashi}\ \emph {et~al.}(2006)\citenamefont
  {Takahashi}, \citenamefont {Yamashita},\ and\ \citenamefont
  {Maekawa}}]{Takahashi_2006}%
  \BibitemOpen
  \bibfield  {author} {\bibinfo {author} {\bibfnamefont {S.}~\bibnamefont
  {Takahashi}}, \bibinfo {author} {\bibfnamefont {T.}~\bibnamefont
  {Yamashita}},\ and\ \bibinfo {author} {\bibfnamefont {S.}~\bibnamefont
  {Maekawa}},\ }\bibfield  {title} {\bibinfo {title} {Quantum interference due
  to crossed andreev reflection in a d-wave superconductor with two
  nano-contacts},\ }\href
  {https://doi.org/https://doi.org/10.1016/j.jpcs.2005.10.163} {\bibfield
  {journal} {\bibinfo  {journal} {Journal of Physics and Chemistry of Solids}\
  }\textbf {\bibinfo {volume} {67}},\ \bibinfo {pages} {325} (\bibinfo {year}
  {2006})},\ \bibinfo {note} {spectroscopies in Novel Superconductors
  2004}\BibitemShut {NoStop}%
\bibitem [{\citenamefont {Burset}\ \emph {et~al.}(2009)\citenamefont {Burset},
  \citenamefont {Herrera},\ and\ \citenamefont {Levy~Yeyati}}]{Herrera_2009}%
  \BibitemOpen
  \bibfield  {author} {\bibinfo {author} {\bibfnamefont {P.}~\bibnamefont
  {Burset}}, \bibinfo {author} {\bibfnamefont {W.}~\bibnamefont {Herrera}},\
  and\ \bibinfo {author} {\bibfnamefont {A.}~\bibnamefont {Levy~Yeyati}},\
  }\bibfield  {title} {\bibinfo {title} {Proximity-induced interface bound
  states in superconductor-graphene junctions},\ }\href
  {https://doi.org/10.1103/PhysRevB.80.041402} {\bibfield  {journal} {\bibinfo
  {journal} {Phys. Rev. B}\ }\textbf {\bibinfo {volume} {80}},\ \bibinfo
  {pages} {041402} (\bibinfo {year} {2009})}\BibitemShut {NoStop}%
\bibitem [{\citenamefont {{Celis Gil}}\ \emph {et~al.}(2017)\citenamefont
  {{Celis Gil}}, \citenamefont {{Gomez P.}},\ and\ \citenamefont
  {Herrera}}]{Celis_2017}%
  \BibitemOpen
  \bibfield  {author} {\bibinfo {author} {\bibfnamefont {J.}~\bibnamefont
  {{Celis Gil}}}, \bibinfo {author} {\bibfnamefont {S.}~\bibnamefont {{Gomez
  P.}}},\ and\ \bibinfo {author} {\bibfnamefont {W.~J.}\ \bibnamefont
  {Herrera}},\ }\bibfield  {title} {\bibinfo {title} {Noise cross-correlation
  and cooper pair splitting efficiency in multi-teminal superconductor
  junctions},\ }\href
  {https://doi.org/https://doi.org/10.1016/j.ssc.2017.04.009} {\bibfield
  {journal} {\bibinfo  {journal} {Solid State Communications}\ }\textbf
  {\bibinfo {volume} {258}},\ \bibinfo {pages} {25} (\bibinfo {year}
  {2017})}\BibitemShut {NoStop}%
\bibitem [{\citenamefont {Zhang}\ \emph {et~al.}(2015)\citenamefont {Zhang},
  \citenamefont {Deng}, \citenamefont {Sun},\ and\ \citenamefont
  {Qiao}}]{Zhang_2015}%
  \BibitemOpen
  \bibfield  {author} {\bibinfo {author} {\bibfnamefont {Y.-T.}\ \bibnamefont
  {Zhang}}, \bibinfo {author} {\bibfnamefont {X.}~\bibnamefont {Deng}},
  \bibinfo {author} {\bibfnamefont {Q.-F.}\ \bibnamefont {Sun}},\ and\ \bibinfo
  {author} {\bibfnamefont {Z.}~\bibnamefont {Qiao}},\ }\bibfield  {title}
  {\bibinfo {title} {High-efficiency cooper-pair splitter in quantum anomalous
  hall insulator proximity-coupled with superconductor},\ }\href
  {https://doi.org/10.1038/srep14892} {\bibfield  {journal} {\bibinfo
  {journal} {Scientific Reports}\ }\textbf {\bibinfo {volume} {5}},\ \bibinfo
  {pages} {14892} (\bibinfo {year} {2015})}\BibitemShut {NoStop}%
\bibitem [{\citenamefont {Hou}\ \emph {et~al.}(2016)\citenamefont {Hou},
  \citenamefont {Xing}, \citenamefont {Guo},\ and\ \citenamefont
  {Sun}}]{Hou_2016}%
  \BibitemOpen
  \bibfield  {author} {\bibinfo {author} {\bibfnamefont {Z.}~\bibnamefont
  {Hou}}, \bibinfo {author} {\bibfnamefont {Y.}~\bibnamefont {Xing}}, \bibinfo
  {author} {\bibfnamefont {A.-M.}\ \bibnamefont {Guo}},\ and\ \bibinfo {author}
  {\bibfnamefont {Q.-F.}\ \bibnamefont {Sun}},\ }\bibfield  {title} {\bibinfo
  {title} {Crossed andreev effects in two-dimensional quantum hall systems},\
  }\href {https://doi.org/10.1103/PhysRevB.94.064516} {\bibfield  {journal}
  {\bibinfo  {journal} {Phys. Rev. B}\ }\textbf {\bibinfo {volume} {94}},\
  \bibinfo {pages} {064516} (\bibinfo {year} {2016})}\BibitemShut {NoStop}%
\bibitem [{\citenamefont {Volovik}(2003)}]{Volovik}%
  \BibitemOpen
  \bibfield  {author} {\bibinfo {author} {\bibfnamefont {G.~E.}\ \bibnamefont
  {Volovik}},\ }\href@noop {} {\emph {\bibinfo {title} {{The Universe in a
  Helium Droplet}}}},\ OUP International series of monographs on physics\
  (\bibinfo  {publisher} {Clarendon Press},\ \bibinfo {address} {Oxford},\
  \bibinfo {year} {2003})\BibitemShut {NoStop}%
\bibitem [{\citenamefont {Volovik}(1997)}]{Volovik_1997}%
  \BibitemOpen
  \bibfield  {author} {\bibinfo {author} {\bibfnamefont {G.~E.}\ \bibnamefont
  {Volovik}},\ }\bibfield  {title} {\bibinfo {title} {On edge states in
  superconductors with time inversion symmetry breaking},\ }\href
  {https://doi.org/10.1134/1.567563} {\bibfield  {journal} {\bibinfo  {journal}
  {Journal of Experimental and Theoretical Physics Letters}\ }\textbf {\bibinfo
  {volume} {66}},\ \bibinfo {pages} {522} (\bibinfo {year} {1997})}\BibitemShut
  {NoStop}%
\bibitem [{\citenamefont {Hasan}\ and\ \citenamefont {Kane}(2010)}]{Kane_2010}%
  \BibitemOpen
  \bibfield  {author} {\bibinfo {author} {\bibfnamefont {M.~Z.}\ \bibnamefont
  {Hasan}}\ and\ \bibinfo {author} {\bibfnamefont {C.~L.}\ \bibnamefont
  {Kane}},\ }\bibfield  {title} {\bibinfo {title} {Colloquium: Topological
  insulators},\ }\href {https://doi.org/10.1103/RevModPhys.82.3045} {\bibfield
  {journal} {\bibinfo  {journal} {Rev. Mod. Phys.}\ }\textbf {\bibinfo {volume}
  {82}},\ \bibinfo {pages} {3045} (\bibinfo {year} {2010})}\BibitemShut
  {NoStop}%
\bibitem [{\citenamefont {Qi}\ and\ \citenamefont {Zhang}(2011)}]{Zhang_2011}%
  \BibitemOpen
  \bibfield  {author} {\bibinfo {author} {\bibfnamefont {X.-L.}\ \bibnamefont
  {Qi}}\ and\ \bibinfo {author} {\bibfnamefont {S.-C.}\ \bibnamefont {Zhang}},\
  }\bibfield  {title} {\bibinfo {title} {Topological insulators and
  superconductors},\ }\href {https://doi.org/10.1103/RevModPhys.83.1057}
  {\bibfield  {journal} {\bibinfo  {journal} {Rev. Mod. Phys.}\ }\textbf
  {\bibinfo {volume} {83}},\ \bibinfo {pages} {1057} (\bibinfo {year}
  {2011})}\BibitemShut {NoStop}%
\bibitem [{\citenamefont {Huang}\ \emph {et~al.}(2014)\citenamefont {Huang},
  \citenamefont {Taylor},\ and\ \citenamefont {Kallin}}]{Kallin_2014}%
  \BibitemOpen
  \bibfield  {author} {\bibinfo {author} {\bibfnamefont {W.}~\bibnamefont
  {Huang}}, \bibinfo {author} {\bibfnamefont {E.}~\bibnamefont {Taylor}},\ and\
  \bibinfo {author} {\bibfnamefont {C.}~\bibnamefont {Kallin}},\ }\bibfield
  {title} {\bibinfo {title} {Vanishing edge currents in non-$p$-wave
  topological chiral superconductors},\ }\href
  {https://doi.org/10.1103/PhysRevB.90.224519} {\bibfield  {journal} {\bibinfo
  {journal} {Phys. Rev. B}\ }\textbf {\bibinfo {volume} {90}},\ \bibinfo
  {pages} {224519} (\bibinfo {year} {2014})}\BibitemShut {NoStop}%
\bibitem [{\citenamefont {Schnyder}\ and\ \citenamefont
  {Brydon}(2015)}]{Schnyder_2015}%
  \BibitemOpen
  \bibfield  {author} {\bibinfo {author} {\bibfnamefont {A.~P.}\ \bibnamefont
  {Schnyder}}\ and\ \bibinfo {author} {\bibfnamefont {P.~M.~R.}\ \bibnamefont
  {Brydon}},\ }\bibfield  {title} {\bibinfo {title} {Topological surface states
  in nodal superconductors},\ }\href
  {http://stacks.iop.org/0953-8984/27/i=24/a=243201} {\bibfield  {journal}
  {\bibinfo  {journal} {Journal of Physics: Condensed Matter}\ }\textbf
  {\bibinfo {volume} {27}},\ \bibinfo {pages} {243201} (\bibinfo {year}
  {2015})}\BibitemShut {NoStop}%
\bibitem [{\citenamefont {Sato}\ and\ \citenamefont {Ando}(2017)}]{Sato_2017}%
  \BibitemOpen
  \bibfield  {author} {\bibinfo {author} {\bibfnamefont {M.}~\bibnamefont
  {Sato}}\ and\ \bibinfo {author} {\bibfnamefont {Y.}~\bibnamefont {Ando}},\
  }\bibfield  {title} {\bibinfo {title} {Topological superconductors: a
  review},\ }\href {http://stacks.iop.org/0034-4885/80/i=7/a=076501} {\bibfield
   {journal} {\bibinfo  {journal} {Reports on Progress in Physics}\ }\textbf
  {\bibinfo {volume} {80}},\ \bibinfo {pages} {076501} (\bibinfo {year}
  {2017})}\BibitemShut {NoStop}%
\bibitem [{\citenamefont {Haim}\ and\ \citenamefont {Oreg}(2019)}]{Oreg_2018}%
  \BibitemOpen
  \bibfield  {author} {\bibinfo {author} {\bibfnamefont {A.}~\bibnamefont
  {Haim}}\ and\ \bibinfo {author} {\bibfnamefont {Y.}~\bibnamefont {Oreg}},\
  }\bibfield  {title} {\bibinfo {title} {Time-reversal-invariant topological
  superconductivity in one and two dimensions},\ }\bibfield  {journal}
  {\bibinfo  {journal} {Physics Reports}\ }\href
  {https://doi.org/10.1016/j.physrep.2019.08.002}
  {10.1016/j.physrep.2019.08.002} (\bibinfo {year} {2019})\BibitemShut
  {NoStop}%
\bibitem [{\citenamefont {Wilczek}(2009)}]{Wilczek_2009}%
  \BibitemOpen
  \bibfield  {author} {\bibinfo {author} {\bibfnamefont {F.}~\bibnamefont
  {Wilczek}},\ }\bibfield  {title} {\bibinfo {title} {Majorana returns},\
  }\href {https://doi.org/10.1038/nphys1380} {\bibfield  {journal} {\bibinfo
  {journal} {Nature Physics}\ }\textbf {\bibinfo {volume} {5}},\ \bibinfo
  {pages} {614 EP } (\bibinfo {year} {2009})}\BibitemShut {NoStop}%
\bibitem [{\citenamefont {Alicea}\ \emph {et~al.}(2011)\citenamefont {Alicea},
  \citenamefont {Oreg}, \citenamefont {Refael}, \citenamefont {von Oppen},\
  and\ \citenamefont {Fisher}}]{Alicea_2011}%
  \BibitemOpen
  \bibfield  {author} {\bibinfo {author} {\bibfnamefont {J.}~\bibnamefont
  {Alicea}}, \bibinfo {author} {\bibfnamefont {Y.}~\bibnamefont {Oreg}},
  \bibinfo {author} {\bibfnamefont {G.}~\bibnamefont {Refael}}, \bibinfo
  {author} {\bibfnamefont {F.}~\bibnamefont {von Oppen}},\ and\ \bibinfo
  {author} {\bibfnamefont {M.~P.~A.}\ \bibnamefont {Fisher}},\ }\bibfield
  {title} {\bibinfo {title} {Non-abelian statistics and topological quantum
  information processing in 1d wire networks},\ }\href
  {http://dx.doi.org/10.1038/nphys1915} {\bibfield  {journal} {\bibinfo
  {journal} {Nature Physics}\ }\textbf {\bibinfo {volume} {7}},\ \bibinfo
  {pages} {412 EP } (\bibinfo {year} {2011})}\BibitemShut {NoStop}%
\bibitem [{\citenamefont {Pachos}(2012)}]{Pachos}%
  \BibitemOpen
  \bibfield  {author} {\bibinfo {author} {\bibfnamefont {J.~K.}\ \bibnamefont
  {Pachos}},\ }\href {https://doi.org/10.1017/CBO9780511792908} {\emph
  {\bibinfo {title} {Introduction to Topological Quantum Computation}}}\
  (\bibinfo  {publisher} {Cambridge University Press},\ \bibinfo {year}
  {2012})\BibitemShut {NoStop}%
\bibitem [{\citenamefont {Elliott}\ and\ \citenamefont
  {Franz}(2015)}]{Elliott}%
  \BibitemOpen
  \bibfield  {author} {\bibinfo {author} {\bibfnamefont {S.~R.}\ \bibnamefont
  {Elliott}}\ and\ \bibinfo {author} {\bibfnamefont {M.}~\bibnamefont
  {Franz}},\ }\bibfield  {title} {\bibinfo {title} {Colloquium: Majorana
  fermions in nuclear, particle, and solid-state physics},\ }\href
  {https://doi.org/10.1103/RevModPhys.87.137} {\bibfield  {journal} {\bibinfo
  {journal} {Rev. Mod. Phys.}\ }\textbf {\bibinfo {volume} {87}},\ \bibinfo
  {pages} {137} (\bibinfo {year} {2015})}\BibitemShut {NoStop}%
\bibitem [{\citenamefont {Beenakker}(2013)}]{Beenakker_2013}%
  \BibitemOpen
  \bibfield  {author} {\bibinfo {author} {\bibfnamefont {C.}~\bibnamefont
  {Beenakker}},\ }\bibfield  {title} {\bibinfo {title} {Search for majorana
  fermions in superconductors},\ }\href
  {https://doi.org/10.1146/annurev-conmatphys-030212-184337} {\bibfield
  {journal} {\bibinfo  {journal} {Annual Review of Condensed Matter Physics}\
  }\textbf {\bibinfo {volume} {4}},\ \bibinfo {pages} {113} (\bibinfo {year}
  {2013})},\ \Eprint
  {https://arxiv.org/abs/https://doi.org/10.1146/annurev-conmatphys-030212-184337}
  {https://doi.org/10.1146/annurev-conmatphys-030212-184337} \BibitemShut
  {NoStop}%
\bibitem [{\citenamefont {Sarma}\ \emph {et~al.}(2015)\citenamefont {Sarma},
  \citenamefont {Freedman},\ and\ \citenamefont {Nayak}}]{Sarma_2015}%
  \BibitemOpen
  \bibfield  {author} {\bibinfo {author} {\bibfnamefont {S.~D.}\ \bibnamefont
  {Sarma}}, \bibinfo {author} {\bibfnamefont {M.}~\bibnamefont {Freedman}},\
  and\ \bibinfo {author} {\bibfnamefont {C.}~\bibnamefont {Nayak}},\ }\bibfield
   {title} {\bibinfo {title} {Majorana zero modes and topological quantum
  computation},\ }\href {https://doi.org/10.1038/npjqi.2015.1} {\bibfield
  {journal} {\bibinfo  {journal} {Npj Quantum Information}\ }\textbf {\bibinfo
  {volume} {1}},\ \bibinfo {pages} {15001 EP } (\bibinfo {year}
  {2015})}\BibitemShut {NoStop}%
\bibitem [{\citenamefont {Fu}\ and\ \citenamefont {Kane}(2008)}]{Kane_2008}%
  \BibitemOpen
  \bibfield  {author} {\bibinfo {author} {\bibfnamefont {L.}~\bibnamefont
  {Fu}}\ and\ \bibinfo {author} {\bibfnamefont {C.~L.}\ \bibnamefont {Kane}},\
  }\bibfield  {title} {\bibinfo {title} {Superconducting proximity effect and
  majorana fermions at the surface of a topological insulator},\ }\href
  {https://doi.org/10.1103/PhysRevLett.100.096407} {\bibfield  {journal}
  {\bibinfo  {journal} {Phys. Rev. Lett.}\ }\textbf {\bibinfo {volume} {100}},\
  \bibinfo {pages} {096407} (\bibinfo {year} {2008})}\BibitemShut {NoStop}%
\bibitem [{\citenamefont {Tanaka}\ \emph {et~al.}(2009)\citenamefont {Tanaka},
  \citenamefont {Yokoyama},\ and\ \citenamefont {Nagaosa}}]{Tanaka_2009}%
  \BibitemOpen
  \bibfield  {author} {\bibinfo {author} {\bibfnamefont {Y.}~\bibnamefont
  {Tanaka}}, \bibinfo {author} {\bibfnamefont {T.}~\bibnamefont {Yokoyama}},\
  and\ \bibinfo {author} {\bibfnamefont {N.}~\bibnamefont {Nagaosa}},\
  }\bibfield  {title} {\bibinfo {title} {Manipulation of the majorana fermion,
  andreev reflection, and josephson current on topological insulators},\ }\href
  {https://doi.org/10.1103/PhysRevLett.103.107002} {\bibfield  {journal}
  {\bibinfo  {journal} {Phys. Rev. Lett.}\ }\textbf {\bibinfo {volume} {103}},\
  \bibinfo {pages} {107002} (\bibinfo {year} {2009})}\BibitemShut {NoStop}%
\bibitem [{\citenamefont {Stanescu}\ \emph {et~al.}(2010)\citenamefont
  {Stanescu}, \citenamefont {Sau}, \citenamefont {Lutchyn},\ and\ \citenamefont
  {Das~Sarma}}]{Stanescu_2010}%
  \BibitemOpen
  \bibfield  {author} {\bibinfo {author} {\bibfnamefont {T.~D.}\ \bibnamefont
  {Stanescu}}, \bibinfo {author} {\bibfnamefont {J.~D.}\ \bibnamefont {Sau}},
  \bibinfo {author} {\bibfnamefont {R.~M.}\ \bibnamefont {Lutchyn}},\ and\
  \bibinfo {author} {\bibfnamefont {S.}~\bibnamefont {Das~Sarma}},\ }\bibfield
  {title} {\bibinfo {title} {Proximity effect at the
  superconductor--topological insulator interface},\ }\href
  {https://doi.org/10.1103/PhysRevB.81.241310} {\bibfield  {journal} {\bibinfo
  {journal} {Phys. Rev. B}\ }\textbf {\bibinfo {volume} {81}},\ \bibinfo
  {pages} {241310} (\bibinfo {year} {2010})}\BibitemShut {NoStop}%
\bibitem [{\citenamefont {Linder}\ \emph {et~al.}(2010)\citenamefont {Linder},
  \citenamefont {Tanaka}, \citenamefont {Yokoyama}, \citenamefont {Sudb\o{}},\
  and\ \citenamefont {Nagaosa}}]{Nagaosa_2010}%
  \BibitemOpen
  \bibfield  {author} {\bibinfo {author} {\bibfnamefont {J.}~\bibnamefont
  {Linder}}, \bibinfo {author} {\bibfnamefont {Y.}~\bibnamefont {Tanaka}},
  \bibinfo {author} {\bibfnamefont {T.}~\bibnamefont {Yokoyama}}, \bibinfo
  {author} {\bibfnamefont {A.}~\bibnamefont {Sudb\o{}}},\ and\ \bibinfo
  {author} {\bibfnamefont {N.}~\bibnamefont {Nagaosa}},\ }\bibfield  {title}
  {\bibinfo {title} {Interplay between superconductivity and ferromagnetism on
  a topological insulator},\ }\href
  {https://doi.org/10.1103/PhysRevB.81.184525} {\bibfield  {journal} {\bibinfo
  {journal} {Phys. Rev. B}\ }\textbf {\bibinfo {volume} {81}},\ \bibinfo
  {pages} {184525} (\bibinfo {year} {2010})}\BibitemShut {NoStop}%
\bibitem [{\citenamefont {Zareapour}\ \emph {et~al.}(2012)\citenamefont
  {Zareapour}, \citenamefont {Hayat}, \citenamefont {Zhao}, \citenamefont
  {Kreshchuk}, \citenamefont {Jain}, \citenamefont {Kwok}, \citenamefont {Lee},
  \citenamefont {Cheong}, \citenamefont {Xu}, \citenamefont {Yang},
  \citenamefont {Gu}, \citenamefont {Jia}, \citenamefont {Cava},\ and\
  \citenamefont {Burch}}]{Zareapour2012}%
  \BibitemOpen
  \bibfield  {author} {\bibinfo {author} {\bibfnamefont {P.}~\bibnamefont
  {Zareapour}}, \bibinfo {author} {\bibfnamefont {A.}~\bibnamefont {Hayat}},
  \bibinfo {author} {\bibfnamefont {S.~Y.~F.}\ \bibnamefont {Zhao}}, \bibinfo
  {author} {\bibfnamefont {M.}~\bibnamefont {Kreshchuk}}, \bibinfo {author}
  {\bibfnamefont {A.}~\bibnamefont {Jain}}, \bibinfo {author} {\bibfnamefont
  {D.~C.}\ \bibnamefont {Kwok}}, \bibinfo {author} {\bibfnamefont
  {N.}~\bibnamefont {Lee}}, \bibinfo {author} {\bibfnamefont {S.-W.}\
  \bibnamefont {Cheong}}, \bibinfo {author} {\bibfnamefont {Z.}~\bibnamefont
  {Xu}}, \bibinfo {author} {\bibfnamefont {A.}~\bibnamefont {Yang}}, \bibinfo
  {author} {\bibfnamefont {G.~D.}\ \bibnamefont {Gu}}, \bibinfo {author}
  {\bibfnamefont {S.}~\bibnamefont {Jia}}, \bibinfo {author} {\bibfnamefont
  {R.~J.}\ \bibnamefont {Cava}},\ and\ \bibinfo {author} {\bibfnamefont
  {K.~S.}\ \bibnamefont {Burch}},\ }\bibfield  {title} {\bibinfo {title}
  {Proximity-induced high-temperature superconductivity in the topological
  insulators bi2se3 and bi2te3},\ }\href {https://doi.org/10.1038/ncomms2042}
  {\bibfield  {journal} {\bibinfo  {journal} {Nature Communications}\ }\textbf
  {\bibinfo {volume} {3}},\ \bibinfo {pages} {1056} (\bibinfo {year}
  {2012})}\BibitemShut {NoStop}%
\bibitem [{\citenamefont {Koren}\ \emph {et~al.}(2013)\citenamefont {Koren},
  \citenamefont {Kirzhner}, \citenamefont {Kalcheim},\ and\ \citenamefont
  {Millo}}]{Koren_2013}%
  \BibitemOpen
  \bibfield  {author} {\bibinfo {author} {\bibfnamefont {G.}~\bibnamefont
  {Koren}}, \bibinfo {author} {\bibfnamefont {T.}~\bibnamefont {Kirzhner}},
  \bibinfo {author} {\bibfnamefont {Y.}~\bibnamefont {Kalcheim}},\ and\
  \bibinfo {author} {\bibfnamefont {O.}~\bibnamefont {Millo}},\ }\bibfield
  {title} {\bibinfo {title} {Signature of proximity-induced px + ipy triplet
  pairing in the doped topological insulator bi2se3 by the s-wave
  superconductor nbn},\ }\href {https://doi.org/10.1209/0295-5075/103/67010}
  {\bibfield  {journal} {\bibinfo  {journal} {Europhysics Letters}\ }\textbf
  {\bibinfo {volume} {103}},\ \bibinfo {pages} {67010} (\bibinfo {year}
  {2013})}\BibitemShut {NoStop}%
\bibitem [{\citenamefont {Xu}\ \emph {et~al.}(2015)\citenamefont {Xu},
  \citenamefont {Wang}, \citenamefont {Liu}, \citenamefont {Ge}, \citenamefont
  {Yang}, \citenamefont {Liu}, \citenamefont {Xu}, \citenamefont {Guan},
  \citenamefont {Gao}, \citenamefont {Qian}, \citenamefont {Liu}, \citenamefont
  {Wang}, \citenamefont {Zhang}, \citenamefont {Xue},\ and\ \citenamefont
  {Jia}}]{Xu_2015}%
  \BibitemOpen
  \bibfield  {author} {\bibinfo {author} {\bibfnamefont {J.-P.}\ \bibnamefont
  {Xu}}, \bibinfo {author} {\bibfnamefont {M.-X.}\ \bibnamefont {Wang}},
  \bibinfo {author} {\bibfnamefont {Z.~L.}\ \bibnamefont {Liu}}, \bibinfo
  {author} {\bibfnamefont {J.-F.}\ \bibnamefont {Ge}}, \bibinfo {author}
  {\bibfnamefont {X.}~\bibnamefont {Yang}}, \bibinfo {author} {\bibfnamefont
  {C.}~\bibnamefont {Liu}}, \bibinfo {author} {\bibfnamefont {Z.~A.}\
  \bibnamefont {Xu}}, \bibinfo {author} {\bibfnamefont {D.}~\bibnamefont
  {Guan}}, \bibinfo {author} {\bibfnamefont {C.~L.}\ \bibnamefont {Gao}},
  \bibinfo {author} {\bibfnamefont {D.}~\bibnamefont {Qian}}, \bibinfo {author}
  {\bibfnamefont {Y.}~\bibnamefont {Liu}}, \bibinfo {author} {\bibfnamefont
  {Q.-H.}\ \bibnamefont {Wang}}, \bibinfo {author} {\bibfnamefont {F.-C.}\
  \bibnamefont {Zhang}}, \bibinfo {author} {\bibfnamefont {Q.-K.}\ \bibnamefont
  {Xue}},\ and\ \bibinfo {author} {\bibfnamefont {J.-F.}\ \bibnamefont {Jia}},\
  }\bibfield  {title} {\bibinfo {title} {Experimental detection of a majorana
  mode in the core of a magnetic vortex inside a topological
  insulator-superconductor
  ${\mathrm{bi}}_{2}{\mathrm{te}}_{3}/{\mathrm{nbse}}_{2}$ heterostructure},\
  }\href {https://doi.org/10.1103/PhysRevLett.114.017001} {\bibfield  {journal}
  {\bibinfo  {journal} {Phys. Rev. Lett.}\ }\textbf {\bibinfo {volume} {114}},\
  \bibinfo {pages} {017001} (\bibinfo {year} {2015})}\BibitemShut {NoStop}%
\bibitem [{\citenamefont {Sun}\ \emph {et~al.}(2016)\citenamefont {Sun},
  \citenamefont {Zhang}, \citenamefont {Hu}, \citenamefont {Li}, \citenamefont
  {Wang}, \citenamefont {Ma}, \citenamefont {Xu}, \citenamefont {Gao},
  \citenamefont {Guan}, \citenamefont {Li}, \citenamefont {Liu}, \citenamefont
  {Qian}, \citenamefont {Zhou}, \citenamefont {Fu}, \citenamefont {Li},
  \citenamefont {Zhang},\ and\ \citenamefont {Jia}}]{Sun_2016_TI}%
  \BibitemOpen
  \bibfield  {author} {\bibinfo {author} {\bibfnamefont {H.-H.}\ \bibnamefont
  {Sun}}, \bibinfo {author} {\bibfnamefont {K.-W.}\ \bibnamefont {Zhang}},
  \bibinfo {author} {\bibfnamefont {L.-H.}\ \bibnamefont {Hu}}, \bibinfo
  {author} {\bibfnamefont {C.}~\bibnamefont {Li}}, \bibinfo {author}
  {\bibfnamefont {G.-Y.}\ \bibnamefont {Wang}}, \bibinfo {author}
  {\bibfnamefont {H.-Y.}\ \bibnamefont {Ma}}, \bibinfo {author} {\bibfnamefont
  {Z.-A.}\ \bibnamefont {Xu}}, \bibinfo {author} {\bibfnamefont {C.-L.}\
  \bibnamefont {Gao}}, \bibinfo {author} {\bibfnamefont {D.-D.}\ \bibnamefont
  {Guan}}, \bibinfo {author} {\bibfnamefont {Y.-Y.}\ \bibnamefont {Li}},
  \bibinfo {author} {\bibfnamefont {C.}~\bibnamefont {Liu}}, \bibinfo {author}
  {\bibfnamefont {D.}~\bibnamefont {Qian}}, \bibinfo {author} {\bibfnamefont
  {Y.}~\bibnamefont {Zhou}}, \bibinfo {author} {\bibfnamefont {L.}~\bibnamefont
  {Fu}}, \bibinfo {author} {\bibfnamefont {S.-C.}\ \bibnamefont {Li}}, \bibinfo
  {author} {\bibfnamefont {F.-C.}\ \bibnamefont {Zhang}},\ and\ \bibinfo
  {author} {\bibfnamefont {J.-F.}\ \bibnamefont {Jia}},\ }\bibfield  {title}
  {\bibinfo {title} {Majorana zero mode detected with spin selective andreev
  reflection in the vortex of a topological superconductor},\ }\href
  {https://doi.org/10.1103/PhysRevLett.116.257003} {\bibfield  {journal}
  {\bibinfo  {journal} {Phys. Rev. Lett.}\ }\textbf {\bibinfo {volume} {116}},\
  \bibinfo {pages} {257003} (\bibinfo {year} {2016})}\BibitemShut {NoStop}%
\bibitem [{\citenamefont {Sun}\ and\ \citenamefont {Jia}(2017)}]{Sun_2017}%
  \BibitemOpen
  \bibfield  {author} {\bibinfo {author} {\bibfnamefont {H.-H.}\ \bibnamefont
  {Sun}}\ and\ \bibinfo {author} {\bibfnamefont {J.-F.}\ \bibnamefont {Jia}},\
  }\bibfield  {title} {\bibinfo {title} {Detection of majorana zero mode in the
  vortex},\ }\href {https://doi.org/10.1038/s41535-017-0037-4} {\bibfield
  {journal} {\bibinfo  {journal} {npj Quantum Materials}\ }\textbf {\bibinfo
  {volume} {2}},\ \bibinfo {pages} {34} (\bibinfo {year} {2017})}\BibitemShut
  {NoStop}%
\bibitem [{\citenamefont {Dai}\ \emph {et~al.}(2017)\citenamefont {Dai},
  \citenamefont {Richardella}, \citenamefont {Du}, \citenamefont {Zhao},
  \citenamefont {Liu}, \citenamefont {Liu}, \citenamefont {Huang},
  \citenamefont {Sankar}, \citenamefont {Chou}, \citenamefont {Samarth},\ and\
  \citenamefont {Li}}]{Dai_2017}%
  \BibitemOpen
  \bibfield  {author} {\bibinfo {author} {\bibfnamefont {W.}~\bibnamefont
  {Dai}}, \bibinfo {author} {\bibfnamefont {A.}~\bibnamefont {Richardella}},
  \bibinfo {author} {\bibfnamefont {R.}~\bibnamefont {Du}}, \bibinfo {author}
  {\bibfnamefont {W.}~\bibnamefont {Zhao}}, \bibinfo {author} {\bibfnamefont
  {X.}~\bibnamefont {Liu}}, \bibinfo {author} {\bibfnamefont {C.~X.}\
  \bibnamefont {Liu}}, \bibinfo {author} {\bibfnamefont {S.-H.}\ \bibnamefont
  {Huang}}, \bibinfo {author} {\bibfnamefont {R.}~\bibnamefont {Sankar}},
  \bibinfo {author} {\bibfnamefont {F.}~\bibnamefont {Chou}}, \bibinfo {author}
  {\bibfnamefont {N.}~\bibnamefont {Samarth}},\ and\ \bibinfo {author}
  {\bibfnamefont {Q.}~\bibnamefont {Li}},\ }\bibfield  {title} {\bibinfo
  {title} {Proximity-effect-induced superconducting gap in topological surface
  states -- a point contact spectroscopy study of nbse2/bi2se3
  superconductor-topological insulator heterostructures},\ }\href
  {https://doi.org/10.1038/s41598-017-07990-3} {\bibfield  {journal} {\bibinfo
  {journal} {Scientific Reports}\ }\textbf {\bibinfo {volume} {7}},\ \bibinfo
  {pages} {7631} (\bibinfo {year} {2017})}\BibitemShut {NoStop}%
\bibitem [{\citenamefont {Casas}\ \emph
  {et~al.}(2019{\natexlab{a}})\citenamefont {Casas}, \citenamefont {Arrachea},
  \citenamefont {Herrera},\ and\ \citenamefont {Yeyati}}]{Casas_2019_2}%
  \BibitemOpen
  \bibfield  {author} {\bibinfo {author} {\bibfnamefont {O.~E.}\ \bibnamefont
  {Casas}}, \bibinfo {author} {\bibfnamefont {L.}~\bibnamefont {Arrachea}},
  \bibinfo {author} {\bibfnamefont {W.~J.}\ \bibnamefont {Herrera}},\ and\
  \bibinfo {author} {\bibfnamefont {A.~L.}\ \bibnamefont {Yeyati}},\ }\bibfield
   {title} {\bibinfo {title} {Proximity induced time-reversal topological
  superconductivity in ${\mathrm{bi}}_{2}{\mathrm{se}}_{3}$ films without phase
  tuning},\ }\href {https://doi.org/10.1103/PhysRevB.99.161301} {\bibfield
  {journal} {\bibinfo  {journal} {Phys. Rev. B}\ }\textbf {\bibinfo {volume}
  {99}},\ \bibinfo {pages} {161301} (\bibinfo {year}
  {2019}{\natexlab{a}})}\BibitemShut {NoStop}%
\bibitem [{\citenamefont {Casas}\ \emph {et~al.}(2020)\citenamefont {Casas},
  \citenamefont {P{\'{a}}ez},\ and\ \citenamefont {Herrera}}]{Casas_2020}%
  \BibitemOpen
  \bibfield  {author} {\bibinfo {author} {\bibfnamefont {O.~E.}\ \bibnamefont
  {Casas}}, \bibinfo {author} {\bibfnamefont {S.~G.}\ \bibnamefont
  {P{\'{a}}ez}},\ and\ \bibinfo {author} {\bibfnamefont {W.~J.}\ \bibnamefont
  {Herrera}},\ }\bibfield  {title} {\bibinfo {title} {A green's function
  approach to topological insulator junctions with magnetic and superconducting
  regions},\ }\href {https://doi.org/10.1088/1361-648x/abafc9} {\bibfield
  {journal} {\bibinfo  {journal} {Journal of Physics: Condensed Matter}\
  }\textbf {\bibinfo {volume} {32}},\ \bibinfo {pages} {485302} (\bibinfo
  {year} {2020})}\BibitemShut {NoStop}%
\bibitem [{\citenamefont {Zhu}\ \emph {et~al.}(2021)\citenamefont {Zhu},
  \citenamefont {Papaj}, \citenamefont {Nie}, \citenamefont {Xu}, \citenamefont
  {Gu}, \citenamefont {Yang}, \citenamefont {Guan}, \citenamefont {Wang},
  \citenamefont {Li}, \citenamefont {Liu}, \citenamefont {Luo}, \citenamefont
  {Xu}, \citenamefont {Zheng}, \citenamefont {Fu},\ and\ \citenamefont
  {Jia}}]{science_2021}%
  \BibitemOpen
  \bibfield  {author} {\bibinfo {author} {\bibfnamefont {Z.}~\bibnamefont
  {Zhu}}, \bibinfo {author} {\bibfnamefont {M.}~\bibnamefont {Papaj}}, \bibinfo
  {author} {\bibfnamefont {X.-A.}\ \bibnamefont {Nie}}, \bibinfo {author}
  {\bibfnamefont {H.-K.}\ \bibnamefont {Xu}}, \bibinfo {author} {\bibfnamefont
  {Y.-S.}\ \bibnamefont {Gu}}, \bibinfo {author} {\bibfnamefont
  {X.}~\bibnamefont {Yang}}, \bibinfo {author} {\bibfnamefont {D.}~\bibnamefont
  {Guan}}, \bibinfo {author} {\bibfnamefont {S.}~\bibnamefont {Wang}}, \bibinfo
  {author} {\bibfnamefont {Y.}~\bibnamefont {Li}}, \bibinfo {author}
  {\bibfnamefont {C.}~\bibnamefont {Liu}}, \bibinfo {author} {\bibfnamefont
  {J.}~\bibnamefont {Luo}}, \bibinfo {author} {\bibfnamefont {Z.-A.}\
  \bibnamefont {Xu}}, \bibinfo {author} {\bibfnamefont {H.}~\bibnamefont
  {Zheng}}, \bibinfo {author} {\bibfnamefont {L.}~\bibnamefont {Fu}},\ and\
  \bibinfo {author} {\bibfnamefont {J.-F.}\ \bibnamefont {Jia}},\ }\bibfield
  {title} {\bibinfo {title} {Discovery of segmented fermi surface induced by
  cooper pair momentum},\ }\href {https://doi.org/10.1126/science.abf1077}
  {\bibfield  {journal} {\bibinfo  {journal} {Science}\ }\textbf {\bibinfo
  {volume} {374}},\ \bibinfo {pages} {1381} (\bibinfo {year} {2021})},\ \Eprint
  {https://arxiv.org/abs/https://www.science.org/doi/pdf/10.1126/science.abf1077}
  {https://www.science.org/doi/pdf/10.1126/science.abf1077} \BibitemShut
  {NoStop}%
\bibitem [{\citenamefont {Kiphart}\ \emph {et~al.}(2021)\citenamefont
  {Kiphart}, \citenamefont {Harkavyi}, \citenamefont {Balin}, \citenamefont
  {Szade}, \citenamefont {Mr{\'o}z}, \citenamefont {Ku{\'{s}}wik},
  \citenamefont {Jurga},\ and\ \citenamefont {Wiesner}}]{Kiphart2021}%
  \BibitemOpen
  \bibfield  {author} {\bibinfo {author} {\bibfnamefont {D.}~\bibnamefont
  {Kiphart}}, \bibinfo {author} {\bibfnamefont {Y.}~\bibnamefont {Harkavyi}},
  \bibinfo {author} {\bibfnamefont {K.}~\bibnamefont {Balin}}, \bibinfo
  {author} {\bibfnamefont {J.}~\bibnamefont {Szade}}, \bibinfo {author}
  {\bibfnamefont {B.}~\bibnamefont {Mr{\'o}z}}, \bibinfo {author}
  {\bibfnamefont {P.}~\bibnamefont {Ku{\'{s}}wik}}, \bibinfo {author}
  {\bibfnamefont {S.}~\bibnamefont {Jurga}},\ and\ \bibinfo {author}
  {\bibfnamefont {M.}~\bibnamefont {Wiesner}},\ }\bibfield  {title} {\bibinfo
  {title} {Investigations of proximity-induced superconductivity in the
  topological insulator bi2te3 by microraman spectroscopy},\ }\href
  {https://doi.org/10.1038/s41598-021-02475-w} {\bibfield  {journal} {\bibinfo
  {journal} {Scientific Reports}\ }\textbf {\bibinfo {volume} {11}},\ \bibinfo
  {pages} {22980} (\bibinfo {year} {2021})}\BibitemShut {NoStop}%
\bibitem [{\citenamefont {Bai}\ \emph {et~al.}(2022)\citenamefont {Bai},
  \citenamefont {Wei}, \citenamefont {Feng}, \citenamefont {Luysberg},
  \citenamefont {Bliesener}, \citenamefont {Lippertz}, \citenamefont {Uday},
  \citenamefont {Taskin}, \citenamefont {Mayer},\ and\ \citenamefont
  {Ando}}]{Bai2022}%
  \BibitemOpen
  \bibfield  {author} {\bibinfo {author} {\bibfnamefont {M.}~\bibnamefont
  {Bai}}, \bibinfo {author} {\bibfnamefont {X.-K.}\ \bibnamefont {Wei}},
  \bibinfo {author} {\bibfnamefont {J.}~\bibnamefont {Feng}}, \bibinfo {author}
  {\bibfnamefont {M.}~\bibnamefont {Luysberg}}, \bibinfo {author}
  {\bibfnamefont {A.}~\bibnamefont {Bliesener}}, \bibinfo {author}
  {\bibfnamefont {G.}~\bibnamefont {Lippertz}}, \bibinfo {author}
  {\bibfnamefont {A.}~\bibnamefont {Uday}}, \bibinfo {author} {\bibfnamefont
  {A.~A.}\ \bibnamefont {Taskin}}, \bibinfo {author} {\bibfnamefont
  {J.}~\bibnamefont {Mayer}},\ and\ \bibinfo {author} {\bibfnamefont
  {Y.}~\bibnamefont {Ando}},\ }\bibfield  {title} {\bibinfo {title}
  {Proximity-induced superconductivity in (bi1-xsbx)2te3 topological-insulator
  nanowires},\ }\href {https://doi.org/10.1038/s43246-022-00242-6} {\bibfield
  {journal} {\bibinfo  {journal} {Communications Materials}\ }\textbf {\bibinfo
  {volume} {3}},\ \bibinfo {pages} {20} (\bibinfo {year} {2022})}\BibitemShut
  {NoStop}%
\bibitem [{\citenamefont {Li}\ \emph {et~al.}(2023)\citenamefont {Li},
  \citenamefont {Zhao}, \citenamefont {Vera}, \citenamefont {Lesser},
  \citenamefont {Yi}, \citenamefont {Kumari}, \citenamefont {Yan},
  \citenamefont {Dong}, \citenamefont {Bowen}, \citenamefont {Wang},
  \citenamefont {Wang}, \citenamefont {Thompson}, \citenamefont {Watanabe},
  \citenamefont {Taniguchi}, \citenamefont {Reifsnyder~Hickey}, \citenamefont
  {Oreg}, \citenamefont {Robinson}, \citenamefont {Chang},\ and\ \citenamefont
  {Zhu}}]{Li2023}%
  \BibitemOpen
  \bibfield  {author} {\bibinfo {author} {\bibfnamefont {C.}~\bibnamefont
  {Li}}, \bibinfo {author} {\bibfnamefont {Y.-F.}\ \bibnamefont {Zhao}},
  \bibinfo {author} {\bibfnamefont {A.}~\bibnamefont {Vera}}, \bibinfo {author}
  {\bibfnamefont {O.}~\bibnamefont {Lesser}}, \bibinfo {author} {\bibfnamefont
  {H.}~\bibnamefont {Yi}}, \bibinfo {author} {\bibfnamefont {S.}~\bibnamefont
  {Kumari}}, \bibinfo {author} {\bibfnamefont {Z.}~\bibnamefont {Yan}},
  \bibinfo {author} {\bibfnamefont {C.}~\bibnamefont {Dong}}, \bibinfo {author}
  {\bibfnamefont {T.}~\bibnamefont {Bowen}}, \bibinfo {author} {\bibfnamefont
  {K.}~\bibnamefont {Wang}}, \bibinfo {author} {\bibfnamefont {H.}~\bibnamefont
  {Wang}}, \bibinfo {author} {\bibfnamefont {J.~L.}\ \bibnamefont {Thompson}},
  \bibinfo {author} {\bibfnamefont {K.}~\bibnamefont {Watanabe}}, \bibinfo
  {author} {\bibfnamefont {T.}~\bibnamefont {Taniguchi}}, \bibinfo {author}
  {\bibfnamefont {D.}~\bibnamefont {Reifsnyder~Hickey}}, \bibinfo {author}
  {\bibfnamefont {Y.}~\bibnamefont {Oreg}}, \bibinfo {author} {\bibfnamefont
  {J.~A.}\ \bibnamefont {Robinson}}, \bibinfo {author} {\bibfnamefont {C.-Z.}\
  \bibnamefont {Chang}},\ and\ \bibinfo {author} {\bibfnamefont
  {J.}~\bibnamefont {Zhu}},\ }\bibfield  {title} {\bibinfo {title}
  {Proximity-induced superconductivity in epitaxial topological
  insulator/graphene/gallium heterostructures},\ }\href
  {https://doi.org/10.1038/s41563-023-01478-4} {\bibfield  {journal} {\bibinfo
  {journal} {Nature Materials}\ }\textbf {\bibinfo {volume} {22}},\ \bibinfo
  {pages} {570} (\bibinfo {year} {2023})}\BibitemShut {NoStop}%
\bibitem [{\citenamefont {Niu}(2010)}]{Niu_2010}%
  \BibitemOpen
  \bibfield  {author} {\bibinfo {author} {\bibfnamefont {Z.~P.}\ \bibnamefont
  {Niu}},\ }\bibfield  {title} {\bibinfo {title} {Crossed andreev reflection on
  a topological insulator},\ }\href {https://doi.org/10.1063/1.3499295}
  {\bibfield  {journal} {\bibinfo  {journal} {Journal of Applied Physics}\
  }\textbf {\bibinfo {volume} {108}},\ \bibinfo {pages} {103904} (\bibinfo
  {year} {2010})}\BibitemShut {NoStop}%
\bibitem [{\citenamefont {Vali}\ and\ \citenamefont
  {Khouzestani}(2014)}]{Vali_2014}%
  \BibitemOpen
  \bibfield  {author} {\bibinfo {author} {\bibfnamefont {R.}~\bibnamefont
  {Vali}}\ and\ \bibinfo {author} {\bibfnamefont {H.~F.}\ \bibnamefont
  {Khouzestani}},\ }\bibfield  {title} {\bibinfo {title} {Nonlocal transport
  properties of topological insulator f/i/sc/i/f junction with perpendicular
  magnetization},\ }\href {https://doi.org/10.1140/epjb/e2014-40872-3}
  {\bibfield  {journal} {\bibinfo  {journal} {The European Physical Journal B}\
  }\textbf {\bibinfo {volume} {87}},\ \bibinfo {pages} {25} (\bibinfo {year}
  {2014})}\BibitemShut {NoStop}%
\bibitem [{\citenamefont {Vali}\ and\ \citenamefont
  {Khouzestani}(2015)}]{Vali_2015}%
  \BibitemOpen
  \bibfield  {author} {\bibinfo {author} {\bibfnamefont {R.}~\bibnamefont
  {Vali}}\ and\ \bibinfo {author} {\bibfnamefont {H.}~\bibnamefont
  {Khouzestani}},\ }\bibfield  {title} {\bibinfo {title} {Crossed andreev
  reflection in topological insulator f/d-wave sc/f junction},\ }\href
  {https://doi.org/https://doi.org/10.1016/j.physe.2014.12.018} {\bibfield
  {journal} {\bibinfo  {journal} {Physica E: Low-dimensional Systems and
  Nanostructures}\ }\textbf {\bibinfo {volume} {68}},\ \bibinfo {pages} {107}
  (\bibinfo {year} {2015})}\BibitemShut {NoStop}%
\bibitem [{\citenamefont {Islam}\ \emph {et~al.}(2017)\citenamefont {Islam},
  \citenamefont {Dutta},\ and\ \citenamefont {Saha}}]{Islam_2017}%
  \BibitemOpen
  \bibfield  {author} {\bibinfo {author} {\bibfnamefont {S.~F.}\ \bibnamefont
  {Islam}}, \bibinfo {author} {\bibfnamefont {P.}~\bibnamefont {Dutta}},\ and\
  \bibinfo {author} {\bibfnamefont {A.}~\bibnamefont {Saha}},\ }\bibfield
  {title} {\bibinfo {title} {Enhancement of crossed andreev reflection in a
  normal-superconductor-normal junction made of thin topological insulator},\
  }\href {https://doi.org/10.1103/PhysRevB.96.155429} {\bibfield  {journal}
  {\bibinfo  {journal} {Phys. Rev. B}\ }\textbf {\bibinfo {volume} {96}},\
  \bibinfo {pages} {155429} (\bibinfo {year} {2017})}\BibitemShut {NoStop}%
\bibitem [{\citenamefont {Zhang}\ and\ \citenamefont
  {Cheng}(2018)}]{Zhang_K_2018}%
  \BibitemOpen
  \bibfield  {author} {\bibinfo {author} {\bibfnamefont {K.}~\bibnamefont
  {Zhang}}\ and\ \bibinfo {author} {\bibfnamefont {Q.}~\bibnamefont {Cheng}},\
  }\bibfield  {title} {\bibinfo {title} {Electrically tunable crossed andreev
  reflection in a
  ferromagnet{\textendash}superconductor{\textendash}ferromagnet junction on a
  topological insulator},\ }\href {https://doi.org/10.1088/1361-6668/aac290}
  {\bibfield  {journal} {\bibinfo  {journal} {Superconductor Science and
  Technology}\ }\textbf {\bibinfo {volume} {31}},\ \bibinfo {pages} {075001}
  (\bibinfo {year} {2018})}\BibitemShut {NoStop}%
\bibitem [{\citenamefont {Ikegaya}\ \emph {et~al.}(2019)\citenamefont
  {Ikegaya}, \citenamefont {Asano},\ and\ \citenamefont
  {Manske}}]{Ikegaya_2019}%
  \BibitemOpen
  \bibfield  {author} {\bibinfo {author} {\bibfnamefont {S.}~\bibnamefont
  {Ikegaya}}, \bibinfo {author} {\bibfnamefont {Y.}~\bibnamefont {Asano}},\
  and\ \bibinfo {author} {\bibfnamefont {D.}~\bibnamefont {Manske}},\
  }\bibfield  {title} {\bibinfo {title} {Anomalous nonlocal conductance as a
  fingerprint of chiral majorana edge states},\ }\href
  {https://doi.org/10.1103/PhysRevLett.123.207002} {\bibfield  {journal}
  {\bibinfo  {journal} {Phys. Rev. Lett.}\ }\textbf {\bibinfo {volume} {123}},\
  \bibinfo {pages} {207002} (\bibinfo {year} {2019})}\BibitemShut {NoStop}%
\bibitem [{\citenamefont {Mackenzie}\ \emph {et~al.}(1998)\citenamefont
  {Mackenzie}, \citenamefont {Haselwimmer}, \citenamefont {Tyler},
  \citenamefont {Lonzarich}, \citenamefont {Mori}, \citenamefont {Nishizaki},\
  and\ \citenamefont {Maeno}}]{Mackenzie_1998}%
  \BibitemOpen
  \bibfield  {author} {\bibinfo {author} {\bibfnamefont {A.~P.}\ \bibnamefont
  {Mackenzie}}, \bibinfo {author} {\bibfnamefont {R.~K.~W.}\ \bibnamefont
  {Haselwimmer}}, \bibinfo {author} {\bibfnamefont {A.~W.}\ \bibnamefont
  {Tyler}}, \bibinfo {author} {\bibfnamefont {G.~G.}\ \bibnamefont
  {Lonzarich}}, \bibinfo {author} {\bibfnamefont {Y.}~\bibnamefont {Mori}},
  \bibinfo {author} {\bibfnamefont {S.}~\bibnamefont {Nishizaki}},\ and\
  \bibinfo {author} {\bibfnamefont {Y.}~\bibnamefont {Maeno}},\ }\bibfield
  {title} {\bibinfo {title} {Extremely strong dependence of superconductivity
  on disorder in ${\mathrm{sr}}_{2}{\mathrm{ruo}}_{4}$},\ }\href
  {https://doi.org/10.1103/PhysRevLett.80.161} {\bibfield  {journal} {\bibinfo
  {journal} {Phys. Rev. Lett.}\ }\textbf {\bibinfo {volume} {80}},\ \bibinfo
  {pages} {161} (\bibinfo {year} {1998})}\BibitemShut {NoStop}%
\bibitem [{\citenamefont {Anwar}\ and\ \citenamefont
  {Robinson}(2021)}]{Anwar_2021}%
  \BibitemOpen
  \bibfield  {author} {\bibinfo {author} {\bibfnamefont {M.~S.}\ \bibnamefont
  {Anwar}}\ and\ \bibinfo {author} {\bibfnamefont {J.~W.~A.}\ \bibnamefont
  {Robinson}},\ }\bibfield  {title} {\bibinfo {title} {A review of electronic
  transport in superconducting sr2ruo4 junctions},\ }\bibfield  {journal}
  {\bibinfo  {journal} {Coatings}\ }\textbf {\bibinfo {volume} {11}},\ \href
  {https://doi.org/10.3390/coatings11091110} {10.3390/coatings11091110}
  (\bibinfo {year} {2021})\BibitemShut {NoStop}%
\bibitem [{\citenamefont {Balatsky}\ \emph {et~al.}(2006)\citenamefont
  {Balatsky}, \citenamefont {Vekhter},\ and\ \citenamefont
  {Zhu}}]{Balatsky_2006}%
  \BibitemOpen
  \bibfield  {author} {\bibinfo {author} {\bibfnamefont {A.~V.}\ \bibnamefont
  {Balatsky}}, \bibinfo {author} {\bibfnamefont {I.}~\bibnamefont {Vekhter}},\
  and\ \bibinfo {author} {\bibfnamefont {J.-X.}\ \bibnamefont {Zhu}},\
  }\bibfield  {title} {\bibinfo {title} {Impurity-induced states in
  conventional and unconventional superconductors},\ }\href
  {https://doi.org/10.1103/RevModPhys.78.373} {\bibfield  {journal} {\bibinfo
  {journal} {Rev. Mod. Phys.}\ }\textbf {\bibinfo {volume} {78}},\ \bibinfo
  {pages} {373} (\bibinfo {year} {2006})}\BibitemShut {NoStop}%
\bibitem [{\citenamefont {De~Gennes}(1999)}]{DeGennes}%
  \BibitemOpen
  \bibfield  {author} {\bibinfo {author} {\bibfnamefont {P.~G.}\ \bibnamefont
  {De~Gennes}},\ }\href@noop {} {\emph {\bibinfo {title} {{Superconductivity of
  Metals and Alloys}}}},\ Advanced book classics\ (\bibinfo  {publisher}
  {Perseus},\ \bibinfo {address} {Cambridge, MA},\ \bibinfo {year}
  {1999})\BibitemShut {NoStop}%
\bibitem [{\citenamefont {Zhu}(2016)}]{Zhu2016}%
  \BibitemOpen
  \bibfield  {author} {\bibinfo {author} {\bibfnamefont {J.-X.}\ \bibnamefont
  {Zhu}},\ }\bibinfo {title} {Transport across normal-metal/superconductor
  junctions},\ in\ \href {https://doi.org/10.1007/978-3-319-31314-6_6} {\emph
  {\bibinfo {booktitle} {Bogoliubov-de Gennes Method and Its Applications}}}\
  (\bibinfo  {publisher} {Springer International Publishing},\ \bibinfo
  {address} {Cham},\ \bibinfo {year} {2016})\ pp.\ \bibinfo {pages}
  {141--167}\BibitemShut {NoStop}%
\bibitem [{\citenamefont {Zhang}\ \emph {et~al.}(2009)\citenamefont {Zhang},
  \citenamefont {Liu}, \citenamefont {Qi}, \citenamefont {Dai}, \citenamefont
  {Fang},\ and\ \citenamefont {Zhang}}]{Zhang_2009}%
  \BibitemOpen
  \bibfield  {author} {\bibinfo {author} {\bibfnamefont {H.}~\bibnamefont
  {Zhang}}, \bibinfo {author} {\bibfnamefont {C.-X.}\ \bibnamefont {Liu}},
  \bibinfo {author} {\bibfnamefont {X.-L.}\ \bibnamefont {Qi}}, \bibinfo
  {author} {\bibfnamefont {X.}~\bibnamefont {Dai}}, \bibinfo {author}
  {\bibfnamefont {Z.}~\bibnamefont {Fang}},\ and\ \bibinfo {author}
  {\bibfnamefont {S.-C.}\ \bibnamefont {Zhang}},\ }\bibfield  {title} {\bibinfo
  {title} {Topological insulators in bi2se3, bi2te3 and sb2te3 with a single
  dirac cone on the surface},\ }\href {http://dx.doi.org/10.1038/nphys1270}
  {\bibfield  {journal} {\bibinfo  {journal} {Nature Physics}\ }\textbf
  {\bibinfo {volume} {5}},\ \bibinfo {pages} {438 EP } (\bibinfo {year}
  {2009})}\BibitemShut {NoStop}%
\bibitem [{\citenamefont {Silvestrov}\ \emph {et~al.}(2012)\citenamefont
  {Silvestrov}, \citenamefont {Brouwer},\ and\ \citenamefont
  {Mishchenko}}]{Silvestrov_2012}%
  \BibitemOpen
  \bibfield  {author} {\bibinfo {author} {\bibfnamefont {P.~G.}\ \bibnamefont
  {Silvestrov}}, \bibinfo {author} {\bibfnamefont {P.~W.}\ \bibnamefont
  {Brouwer}},\ and\ \bibinfo {author} {\bibfnamefont {E.~G.}\ \bibnamefont
  {Mishchenko}},\ }\bibfield  {title} {\bibinfo {title} {Spin and charge
  structure of the surface states in topological insulators},\ }\href
  {https://doi.org/10.1103/PhysRevB.86.075302} {\bibfield  {journal} {\bibinfo
  {journal} {Phys. Rev. B}\ }\textbf {\bibinfo {volume} {86}},\ \bibinfo
  {pages} {075302} (\bibinfo {year} {2012})}\BibitemShut {NoStop}%
\bibitem [{\citenamefont {Cuevas}\ \emph {et~al.}(1996)\citenamefont {Cuevas},
  \citenamefont {Mart\'{\i}n-Rodero},\ and\ \citenamefont
  {Yeyati}}]{Yeyati_1996}%
  \BibitemOpen
  \bibfield  {author} {\bibinfo {author} {\bibfnamefont {J.~C.}\ \bibnamefont
  {Cuevas}}, \bibinfo {author} {\bibfnamefont {A.}~\bibnamefont
  {Mart\'{\i}n-Rodero}},\ and\ \bibinfo {author} {\bibfnamefont {A.~L.}\
  \bibnamefont {Yeyati}},\ }\bibfield  {title} {\bibinfo {title} {Hamiltonian
  approach to the transport properties of superconducting quantum point
  contacts},\ }\href {https://doi.org/10.1103/PhysRevB.54.7366} {\bibfield
  {journal} {\bibinfo  {journal} {Phys. Rev. B}\ }\textbf {\bibinfo {volume}
  {54}},\ \bibinfo {pages} {7366} (\bibinfo {year} {1996})}\BibitemShut
  {NoStop}%
\bibitem [{\citenamefont {Cuevas}\ and\ \citenamefont {Scheer}(2010)}]{Cuevas}%
  \BibitemOpen
  \bibfield  {author} {\bibinfo {author} {\bibfnamefont {J.~C.}\ \bibnamefont
  {Cuevas}}\ and\ \bibinfo {author} {\bibfnamefont {E.}~\bibnamefont
  {Scheer}},\ }\href@noop {} {\emph {\bibinfo {title} {Molecular electronics :
  an introduction to theory and experiment}}},\ World scientific series in
  nanoscience and nanotechnology;1\ (\bibinfo  {publisher} {New Jersey [u.a.] :
  World Scientific},\ \bibinfo {year} {2010})\BibitemShut {NoStop}%
\bibitem [{\citenamefont {McMillan}(1968)}]{McMillan_1968}%
  \BibitemOpen
  \bibfield  {author} {\bibinfo {author} {\bibfnamefont {W.~L.}\ \bibnamefont
  {McMillan}},\ }\bibfield  {title} {\bibinfo {title} {Theory of
  superconductor---normal-metal interfaces},\ }\href
  {https://doi.org/10.1103/PhysRev.175.559} {\bibfield  {journal} {\bibinfo
  {journal} {Phys. Rev.}\ }\textbf {\bibinfo {volume} {175}},\ \bibinfo {pages}
  {559} (\bibinfo {year} {1968})}\BibitemShut {NoStop}%
\bibitem [{\citenamefont {Herrera}\ \emph {et~al.}(2010)\citenamefont
  {Herrera}, \citenamefont {Burset},\ and\ \citenamefont
  {Yeyati}}]{Herrera_2010}%
  \BibitemOpen
  \bibfield  {author} {\bibinfo {author} {\bibfnamefont {W.~J.}\ \bibnamefont
  {Herrera}}, \bibinfo {author} {\bibfnamefont {P.}~\bibnamefont {Burset}},\
  and\ \bibinfo {author} {\bibfnamefont {A.~L.}\ \bibnamefont {Yeyati}},\
  }\bibfield  {title} {\bibinfo {title} {A green function approach to
  graphene–superconductor junctions with well-defined edges},\ }\href
  {http://stacks.iop.org/0953-8984/22/i=27/a=275304} {\bibfield  {journal}
  {\bibinfo  {journal} {Journal of Physics: Condensed Matter}\ }\textbf
  {\bibinfo {volume} {22}},\ \bibinfo {pages} {275304} (\bibinfo {year}
  {2010})}\BibitemShut {NoStop}%
\bibitem [{\citenamefont {Casas}\ \emph
  {et~al.}(2019{\natexlab{b}})\citenamefont {Casas}, \citenamefont
  {G\'omez~P\'aez}, \citenamefont {Levy~Yeyati}, \citenamefont {Burset},\ and\
  \citenamefont {Herrera}}]{Casas_2019_1}%
  \BibitemOpen
  \bibfield  {author} {\bibinfo {author} {\bibfnamefont {O.~E.}\ \bibnamefont
  {Casas}}, \bibinfo {author} {\bibfnamefont {S.}~\bibnamefont
  {G\'omez~P\'aez}}, \bibinfo {author} {\bibfnamefont {A.}~\bibnamefont
  {Levy~Yeyati}}, \bibinfo {author} {\bibfnamefont {P.}~\bibnamefont
  {Burset}},\ and\ \bibinfo {author} {\bibfnamefont {W.~J.}\ \bibnamefont
  {Herrera}},\ }\bibfield  {title} {\bibinfo {title} {Subgap states in
  two-dimensional spectroscopy of graphene-based superconducting hybrid
  junctions},\ }\href {https://doi.org/10.1103/PhysRevB.99.144502} {\bibfield
  {journal} {\bibinfo  {journal} {Phys. Rev. B}\ }\textbf {\bibinfo {volume}
  {99}},\ \bibinfo {pages} {144502} (\bibinfo {year}
  {2019}{\natexlab{b}})}\BibitemShut {NoStop}%
\bibitem [{\citenamefont {Herrmann}\ \emph
  {et~al.}(2010{\natexlab{b}})\citenamefont {Herrmann}, \citenamefont
  {Portier}, \citenamefont {Roche}, \citenamefont {Yeyati}, \citenamefont
  {Kontos},\ and\ \citenamefont {Strunk}}]{Alfredo_2010}%
  \BibitemOpen
  \bibfield  {author} {\bibinfo {author} {\bibfnamefont {L.~G.}\ \bibnamefont
  {Herrmann}}, \bibinfo {author} {\bibfnamefont {F.}~\bibnamefont {Portier}},
  \bibinfo {author} {\bibfnamefont {P.}~\bibnamefont {Roche}}, \bibinfo
  {author} {\bibfnamefont {A.~L.}\ \bibnamefont {Yeyati}}, \bibinfo {author}
  {\bibfnamefont {T.}~\bibnamefont {Kontos}},\ and\ \bibinfo {author}
  {\bibfnamefont {C.}~\bibnamefont {Strunk}},\ }\bibfield  {title} {\bibinfo
  {title} {Carbon nanotubes as cooper-pair beam splitters},\ }\href
  {https://doi.org/10.1103/PhysRevLett.104.026801} {\bibfield  {journal}
  {\bibinfo  {journal} {Phys. Rev. Lett.}\ }\textbf {\bibinfo {volume} {104}},\
  \bibinfo {pages} {026801} (\bibinfo {year} {2010}{\natexlab{b}})}\BibitemShut
  {NoStop}%
\bibitem [{\citenamefont {Löfwander}\ \emph {et~al.}(2001)\citenamefont
  {Löfwander}, \citenamefont {Shumeiko},\ and\ \citenamefont
  {Wendin}}]{Lofwander_2001}%
  \BibitemOpen
  \bibfield  {author} {\bibinfo {author} {\bibfnamefont {T.}~\bibnamefont
  {Löfwander}}, \bibinfo {author} {\bibfnamefont {V.~S.}\ \bibnamefont
  {Shumeiko}},\ and\ \bibinfo {author} {\bibfnamefont {G.}~\bibnamefont
  {Wendin}},\ }\bibfield  {title} {\bibinfo {title} {Andreev bound states in
  high- t c superconducting junctions},\ }\href
  {http://stacks.iop.org/0953-2048/14/i=5/a=201} {\bibfield  {journal}
  {\bibinfo  {journal} {Superconductor Science and Technology}\ }\textbf
  {\bibinfo {volume} {14}},\ \bibinfo {pages} {R53} (\bibinfo {year}
  {2001})}\BibitemShut {NoStop}%
\bibitem [{\citenamefont {Casas}(2019)}]{Casas_2019}%
  \BibitemOpen
  \bibfield  {author} {\bibinfo {author} {\bibfnamefont {O.~E.}\ \bibnamefont
  {Casas}},\ }\bibfield  {title} {\bibinfo {title} {Transporte el\'ectrico en
  nanoestructuras topológicas}} (\bibinfo {year} {2019}),\ \bibinfo {note}
  {www.bdigital.unal.edu.co/73258/}\BibitemShut {NoStop}%
\bibitem [{\citenamefont {Takagaki}\ and\ \citenamefont
  {Ploog}(1999)}]{Takagaki_1999}%
  \BibitemOpen
  \bibfield  {author} {\bibinfo {author} {\bibfnamefont {Y.}~\bibnamefont
  {Takagaki}}\ and\ \bibinfo {author} {\bibfnamefont {K.~H.}\ \bibnamefont
  {Ploog}},\ }\bibfield  {title} {\bibinfo {title} {Quantum point contact
  spectroscopy of d-wave superconductors},\ }\href
  {https://doi.org/10.1103/PhysRevB.60.9750} {\bibfield  {journal} {\bibinfo
  {journal} {Phys. Rev. B}\ }\textbf {\bibinfo {volume} {60}},\ \bibinfo
  {pages} {9750} (\bibinfo {year} {1999})}\BibitemShut {NoStop}%
\bibitem [{\citenamefont {Herrera}\ \emph {et~al.}(2009)\citenamefont
  {Herrera}, \citenamefont {Yeyati},\ and\ \citenamefont
  {Mart\'{\i}n-Rodero}}]{Herrera_2009b}%
  \BibitemOpen
  \bibfield  {author} {\bibinfo {author} {\bibfnamefont {W.~J.}\ \bibnamefont
  {Herrera}}, \bibinfo {author} {\bibfnamefont {A.~L.}\ \bibnamefont
  {Yeyati}},\ and\ \bibinfo {author} {\bibfnamefont {A.}~\bibnamefont
  {Mart\'{\i}n-Rodero}},\ }\bibfield  {title} {\bibinfo {title} {Long-range
  crossed andreev reflections in high-temperature superconductors},\ }\href
  {https://doi.org/10.1103/PhysRevB.79.014520} {\bibfield  {journal} {\bibinfo
  {journal} {Phys. Rev. B}\ }\textbf {\bibinfo {volume} {79}},\ \bibinfo
  {pages} {014520} (\bibinfo {year} {2009})}\BibitemShut {NoStop}%
\bibitem [{\citenamefont {Sigrist}(2005)}]{Sigrist_2005}%
  \BibitemOpen
  \bibfield  {author} {\bibinfo {author} {\bibfnamefont {M.}~\bibnamefont
  {Sigrist}},\ }\bibfield  {title} {\bibinfo {title} {Introduction to
  unconventional superconductivity},\ }\href
  {https://doi.org/10.1063/1.2080350} {\bibfield  {journal} {\bibinfo
  {journal} {AIP Conference Proceedings}\ }\textbf {\bibinfo {volume} {789}},\
  \bibinfo {pages} {165} (\bibinfo {year} {2005})},\ \Eprint
  {https://arxiv.org/abs/https://aip.scitation.org/doi/pdf/10.1063/1.2080350}
  {https://aip.scitation.org/doi/pdf/10.1063/1.2080350} \BibitemShut {NoStop}%
\end{thebibliography}%

\end{document}